\pdfoutput=1
\documentclass[a4paper,11pt,notitlepage,amssymb]{article}
\usepackage{graphicx}
\usepackage[margin=20mm]{geometry}
\usepackage{setspace}
\usepackage{amsmath}
\usepackage[font={small}, labelfont={bf,small}]{caption}
\usepackage{siunitx}
\usepackage{cite}
\usepackage{hyperref}
\usepackage{multirow}

\usepackage[dvipsnames]{xcolor}
\usepackage{latexsym}
\usepackage{cite}

\usepackage{float}

\usepackage{xspace}
\usepackage{amssymb}
\usepackage{amsmath}
\usepackage{bm}
\usepackage{siunitx}
\usepackage[capitalise]{cleveref}
\usepackage{xparse}
\usepackage{booktabs}
\usepackage{setspace} 

\usepackage{enumitem}
\usepackage{alphalph}
\usepackage{longtable}

\newcommand{\sifigref}[1]{Fig~\ref{#1}}                 
\newcommand{\sitabref}[1]{Table~\ref{#1}}
\begin{document}
 \begin{center}
   \LARGE{\textbf{A multilevel network approach to revealing  patterns of online political selective exposure}}
 
\vspace{1cm}

\large Yuan Zhang$^{1,\ast}$, Laia Castro Herrero$^{2}$, Frank Esser$^{1}$ and Alexandre Bovet$^{3,4,\ast}$

\vspace{0.2cm} \normalsize 
$^{1}$Department of Communication and Media Research, University of Zurich, Switzerland\\
$^{2}$Department of Political Science, University of Barcelona, Spain\\
$^{3}$Department of Mathematical Modeling and Machine Learning, University of Zurich, Switzerland\\
$^{4}$Digital Society Initiative, University of Zurich, Switzerland.\\
$\ast$ y.zhang@ikmz.uzh.ch, alexandre.bovet@uzh.ch

\end{center}

\begin{abstract}
Selective exposure, individuals' inclination to seek out information that supports their beliefs while avoiding information that contradicts them, plays an important role in the emergence of polarization and echo chambers. In the political domain, selective exposure is usually measured on a left-right ideology scale, ignoring finer details. To bridge the gap, this work introduces a multilevel analysis framework based on a multi-scale community detection approach. To test this approach, we combine survey and Twitter/X data collected during the 2022 Brazilian Presidential Election and investigate selective exposure patterns among survey respondents in their choices of whom to follow. We construct a bipartite network connecting survey respondents with political influencers and project it onto the influencer nodes. Applying multi-scale community detection to this projection uncovers a hierarchical clustering of political influencers, where each cluster is more frequently co-followed by a specific subgroup of survey respondents compared to others. Different indices of selective exposure, such as Community Overlap, Identity Diversity, Information Diversity, Structural Integration, and Connectivity Inequality, suggest that the characteristics of the influencer communities engaged by survey respondents vary with the level of community resolution. This finding indicates that online political selective exposure exhibits a more complex structure than a mere left-right dichotomy. Moreover, depending on the resolution level we consider, we find different associations between network indices of exposure patterns and 189 individual attributes of the survey respondents. For example, at finer levels, survey respondents' Community Overlap is associated with several factors, such as ideological position, demographics, news consumption frequency, and incivility perception. In comparison, only their ideological position is a significant factor at coarser levels. Our work demonstrates that measuring selective exposure at a single level, such as left and right, misses important information necessary to capture this phenomenon correctly.\\
\end{abstract}


\section*{Introduction}
The internet, especially social media, has facilitated the online consumption of political information in recent decades \cite{rainie2012social}. According to a national survey we conducted during the 2022 Brazilian Presidential Election, 82 percent of the Brazilian population acquired political information via social media, with approximately 51 percent of them consuming or engaging with political information on Twitter (now known as X) at least once a week. On Twitter/X, ordinary users mainly receive political information by following certain political influencers, such as politicians, media outlets, journalists, and individual opinion leaders. Despite social networks like Twitter/X providing a high-choice environment for information consumption and the possibility of deliberative communication, it is a concern that users on social networks predominantly follow information aligned with their own opinions, leading to political fragmentation, polarization, and the creation of echo chambers \cite{bright2018explaining, pildes2021age, terren2021echo, cinelli2021echo, GonzalezBailon2023}. This phenomenon, known as selective exposure, is attributed to bounded rationality—humans have limited capacity for information processing and prefer to be exposed to information they find agreeable \cite{van2005global} and to the influence of personalized recommendation algorithms on social media platforms \cite{garcia2023influence, santos2021link, huszar2022algorithmic, haroon2023auditing, guess2023social, allcott2024effects}. Selective exposure online can, therefore, limit people's access to diverse opinions and even lead to the spread of extreme views within certain groups \cite{ohara2015echo}.

Many studies have explored selective exposure and related topics such as political polarization and echo chambers, e.g., \cite{messing2014selective, delvicario2016echo, bessi2016users, bail2018exposure, yang2020exposure, levin2021dynamics, cinelli2021echo, balietti2021reducing, tornberg2022how}. Most of them focus on the U.S. case, where the political system has traditionally been divided between two major parties - the Democrats and the Republicans \cite{abramowitz2008is}. A similar left–right division is evident in users' behaviorally selective exposure, where individuals are more likely to interact with members of their own party \cite{barbera2015birds}. This phenomenon deepens polarization both ideologically and behaviorally through the reinforcement of echo chambers \cite{johnson2020issues}. However, the dichotomous model of selective exposure, polarization, or echo chambers might be not fully applicable to countries like Brazil, which has a multi-party system and weak party affiliation. Scholars often observe non-ideological cleavage fragmentation in multi-party systems, although the left–right ideological divide persists \cite{zucco2021fragmentation}. Therefore, this study focuses on Brazil as a case study, where both non-ideological fragmentation and ideological division may contribute to the complex patterns of selective exposure among Brazilian online users.  

Furthermore, user-generated content on social media has reshaped the dichotomous nature of online political dynamics globally\cite{cinelli2020selective, flamino2023political, efstratiou2023non}. It is not only elite accounts such as media outlets, political parties, and politicians that promote political propaganda, but also individual opinion leaders who create more targeted political information to represent specific social groups or achieve business attention \cite{perez2019political, riedl2021rise}. In light of this background, political division is more complex in the contemporary era than ever. Evidence from the 2016 U.S. presidential election presented by Bartels \textit{et al.} \cite{bartels2018partisanship} highlights that within the Democratic group, various cultural subgroups exist where social identities and other demographic characteristics play a role in forming closely connected communities. Similarly, Brazil's party systems are also intertwined with complex socio-cultural groups that could even transcend party lines \cite{garcia2023political}. 

That said, traditional network community detection algorithms that perform well in identifying left–right selective exposure, polarization, and echo chambers in the U.S. are often inadequate for uncovering fragmentation patterns in countries like Brazil or in the current social media age. Modularity-based quality functions suffer from a resolution limit, making them ineffective at detecting smaller communities within large networks and at multiple resolution levels \cite{fortunato2007resolution}. Hence, in this study, we apply a multi-scale community detection framework based on Markov stability to uncover patterns of political selective exposure among Brazilian online users sampled from a representative national survey across varying levels of resolution \cite{delvenne2010stability}. While other network-based approaches can also measure polarization in multipolar online systems, our approach offers a distinct advantage in uncovering the hierarchical structure of political systems \cite{martin2023multipolar, peralta2024multidimensional}. Additionally, we propose five novel indices of selective exposure and examine their associated factors at various resolution levels. For each surveyed individual, they are computed based on the communities that their followed influencers form.
They are: (i) \emph{Community Overlap}, capturing the bridging capacity of the surveyed individuals; (ii) \emph{Identity Diversity}, measuring the diversity of the identities of  influencers that surveyed individuals follow; (iii) \emph{Information Diversity}, capturing the information provision diversity of their influencers; (iv) \emph{Structural Integration}, measuring the network integration of their influencers; and (v) \emph{Connectivity Inequality}, capturing the influence imbalance of their influencers. Overall, our research is guided by two main questions:
\begin{itemize}
\item[RQ1:] How can we capture the complexity of individuals’ selective exposure, reflected in Community Overlap, Identity Diversity, Information Diversity, Structural Integration, and Connectivity Inequality, within a fragmented political context?
\item[RQ2:] Which individual attributes are associated with selective exposure patterns across different resolution levels?
\end{itemize}

To answer the two research questions, we combine a representative national survey and Twitter/X political influencers the survey participants followed, and construct consumer-supplier relations. Consumers are represented by 204 respondents from a national survey who provided their Twitter/X handles, while the suppliers are 2,307 political influencers (including politicians, media, and individual influencers) followed by consumers. 
The distribution of demographic attributes of the 204 survey participants does not show significant statistical differences compared to the overall survey sample, which is representatively sampled from Brazil's population. Our analysis, therefore, reports on selective exposure from the point of view of a representative sample of users. Moreover, political influencers are annotated based on dimensions of political ideology (Left/Right/Center), campaign support (Lula camp/Bolsonaro camp), social identity (Women/Religious/LGBTQ/Black), and account type (Politician/Media/Individual). Then, we uncover the complex structures of Brazilian users' selective exposure on Twitter/X using a multi-scale community detection approach. The results suggest that ordinary users tend to expose themselves to political influencers based on ideological positions, however, selective exposure patterns are also evident along other dimensions, such as electoral supports, social identities, and categories of accounts. Therefore, Brazilian users' selective exposure on Twitter/X forms a multi-level and hierarchical structure. 

By examining a comprehensive set of individuals' survey attributes with the five selective exposure indices at various levels, we find that, at the lowest level, a wide range of variables, including ideological position, demographics, news consumption frequency, and incivility perception, are significantly associated with the survey individuals' Community Overlap. As the level increases, the ideological position becomes more prominent. The other indices are also associated with different individual attributes at various levels. These findings suggest that the multilevel framework based on multi-scale community detection is an effective approach for revealing individuals' selective exposure across different levels of resolution, and it can be also applied to other contexts and larger datasets.

This study significantly contributes to the application of multi-scale community detection in social network research, demonstrating that different resolution levels can reveal distinct patterns of individuals’ selective exposure. This approach is particularly valuable for studying selective exposure, polarization, and echo chambers in multi-party political systems or societies with complex social structures.

\section*{Materials and methods}
\subsection*{\textbf{Data collection and preprocessing}}
During the 2022 Brazilian presidential election (from October to November 2022), a national survey was conducted by NetQuest, an international survey company. The data were gathered using NetQuest’s exclusive online panel, using stratified quota sampling aligned with national distributions of age, gender, and geographic area to ensure national demographic representativeness. Statistical tests indicate that the survey participants constitute a representative sample of the national electorate (see Fig B in Supplementary Information). Participants were enrolled through a double opt-in process and completed the survey online in return for incentives. Approximately 40 questions and 189 variables regarding socio-demographics, news consumption behaviors, and political predispositions were asked in the survey (see the questionnaire in section 2 in Supplementary Information). We also requested respondents' consent to provide their Twitter/X handles. Of the 1,018 respondents, 403 consented, and 271 were verified as existing accounts on Twitter/X, accounting for 26.62\% of the whole panel. 
 
Based on the 271 survey respondents, we collected the Twitter/X accounts followed by them (57,645 accounts and 73,755 following pairs were collected) using the Twitter/X API. We use ``following'' relations to represent individuals' preference to be exposed to information posted by the accounts they follow. We then identify political influencers from the 57,645 accounts. We define political influencers as a composition of both ordinary citizens and celebrities (e.g., politicians, parties, media outlets, journalists, and individual opinion leaders) who satisfy two conditions: 1) with at least more  1,000 followers and 2) indicate political related information in their profiles or are known political figures \cite{khamis2017self, harff2023influencers} (see section 3 in Supplementary Information for more details about the identification of political influencers). After filtering out the survey respondents who did not follow any political influencers, we obtained 204 survey respondents (also referred to as political consumers in the text), 2,307 political influencers (also referred to as political suppliers in the text), and 4,107 following pairs between the two groups.
Pearson’s chi-squared tests (for categorical variables) and Mann–Whitney U tests (for discrete variables) comparing the demographic distributions of the 204 subsample with the 1,018 survey respondents do not reject the null hypothesis that the 204 samples are drawn from the same distribution as the survey respondents at the 5\% significance level (see Fig C in Supplementary Information). Therefore, the survey respondents who provided their handles are very likely to form a representative sample of the Brazilian electorate.

This study was reviewed and approved by the Ethics Committee of the Faculty of Arts and Social Sciences, University of Zurich. Informed electronic consent was obtained prior to participation via the NetQuest survey platform. Participants were informed of the study purpose, data processing, voluntary participation, and their right to withdraw at any time. Twitter/X data were collected in accordance with the platform’s Terms of Service and API policies and were limited to publicly available content.

\subsection*{\textbf{Multi-scale community detection}}
 To analyze the relationships between political consumers and political influencers, we construct a bipartite network \( \mathcal{G} = (\mathcal{C}, \mathcal{S}, \mathcal{E}) \), where \( \mathcal{C} \) is the set of ordinary users who primarily consume political information online, and \( \mathcal{S} \) is the set of political influencers who produce such content. The set \( \mathcal{E} \) contains edges that indicate which users follow which influencers. We then project this bipartite network onto the influencer side, resulting in a new graph \( \mathcal{G}^\mathcal{S} = (\mathcal{S}, \mathcal{E}^\mathcal{S}) \), where edges in \( \mathcal{E}^\mathcal{S} \) indicate that the connected influencers are followed by at least one common user. The weight \( w_{ij} \) of an edge between influencers \( s_i \) and \( s_j \) in the projected graph is equal to the number of shared consumers following both influencers. Similarly, we create a consumer network $\mathcal{G}^\mathcal{C} = (\mathcal{C}, \mathcal{E}^\mathcal{C})$ where edges between two consumer nodes represent the number of common influencers they follow. 

We then apply a multi-scale community detection approach to the projected influencer graph \(\mathcal{G}^\mathcal{S}\). Community detection is the process of identifying groups of nodes that are more densely connected to each other than to the rest of the network. It is a common strategy to study phenomena like selective exposure, polarization, and echo chambers \cite{himelboim2013tweeting, cinelli2020selective, cinelli2021echo, conover2011political}. However, most studies detect communities using single-scale approach, such as modularity optimization \cite{newman2004fast}. But real-world social networks are usually multi-scale and hierarchical, for example, smaller class groups are embedded within the broader structure of a school \cite{clauset_hierarchical_2008,ahn2010link}. Multi-scale community detection approaches allow for the discovery of communities at different resolution scales. This flexibility also enables selecting the scales closest to the real situation among all the experimental scales, depending on the research goals \cite{luo2019multiscale}.  

A variety of techniques have been designed for multi-scale community detection, such as methods with adjustable resolution parameters (e.g., \cite{reichardt2006statistical,lancichinetti2009detecting, delvenne2010stability}) and methods which automatically detect relevant scales (e.g., \cite{Rosvall2011,peixoto_hierarchical_2014,arnaudon_algorithm_2024}). Here, we use Markov Stability\cite{delvenne2010stability,lambiotte2014random} with automatic scale selection\cite{arnaudon_algorithm_2024} and optimization using the Leiden algorithm \cite{Traag2019}. We note that our approach is applicable to the results of any multiscale community detection approach. The rationale of the Markov Stability method is to identify communities as groups of nodes where a random walk remains confined over varying time scales, thus revealing community structures at different resolutions. We give a detailed mathematical explanation of the Markov Stability method in section 5 in Supplementary Information.

\subsection*{\textbf{Multidimensional annotation of identities}}
We then annotate four identity attributes of political influencer nodes in the network: political ideology, campaign support, social identity, and account type. The data are manually annotated based on the self-presentation of political and social identities in users' Twitter/X profiles, combined with a formal list of politicians and parties participating in the election. It is shown that users who disclose their political identities publicly are generally more active than those who do not \cite{phillips2024why}. Almost all political influencers have profile descriptions. 

Political ideology is the most traditional and straightforward indicator of individuals' political opinions and comprises categories \{Left, Right, Center\}. During the 2022 Brazilian Presidential Election, Lula and Bolsonaro are the two candidates who represent leftist and rightist leaders, respectively. Nevertheless, it is not necessarily the case that all left-wing individuals support Lula and all right-wing individuals support Bolsonaro, due to increasing cleavages on issues and socio-cultural factors, even though there is still significant overlap. To differentiate between these situations, we also annotate them as "Pro-Lula" or "Pro-Bolsonaro" if they support certain candidates during the election campaign.

Many users also indicate their social identities, as cultural issues have become increasingly related to politics. We identify four social identities most frequently revealed in the profiles: Women (in the context of feminism or supporting women' rights), Religious, Black, and LGBTQ. For religious, Black, and LGBTQ identities, annotation is based on either users' self-identification with the group or their expression of support for its rights. Note that, unlike the fixed options of political ideology, social identity encompasses more categories; accordingly, we include only those observed in our samples. The same situation applies to account type, which indicates what the accounts represent. We consider media (including media outlets and journalists), politicians, and individual opinion leaders, as they are the most common accounts that produce political information. 

As not all political influencers reveal their identities at all dimensions, we assign those unrevealed profiles as ``Unlabeled''. The annotation results on political ideology, campaign support, social identity, and account types for influencers in communities of all scale levels are included in section 4 in Supplementary Information.  

\subsection*{\textbf{Measurements of selective exposure indices}}
We introduce five indices – Community Overlap, Identity Diversity, Information Diversity, Structural Integration, and Connectivity Inequality to measure individuals' selective exposure patterns at different levels. We consider a followship bipartite network between political influencers and survey respondents, and project the network onto both the survey participants’ side and the political influencers’ side. From the projection on the survey participants, we capture the number of influencer communities in which they participate. From the projection on the political influencers, we detect communities and identify the characteristics of influencer communities followed by each survey participant. In this context, we introduce the following measures:

\paragraph{Community overlap.} This index is derived from the network's projection on the survey participants and reflects the number of influencer communities they engage with. It captures how online users are divided into fragmented communities in which individuals primarily interact with others who share similar interests, and how they could act as bridges between multiple communities. This index is based on the phenomenon of cyberbalkanization introduced by Alstyne and Brynjolfsson in their 2005 study \cite{van2005global} which showed that some internet users are more inclined to have diverse interests and engage with multiple communities, while others prefer to remain within a single community. Survey users who connect with multiple communities have a higher potential to act as bridgers—individuals who consume information from a variety of sources \cite{bakshy2015exposure}. To quantify survey respondents' capacity to bridge multiple communities, we introduce the Community Overlap Index.

Community Overlap is measured for each consumer node and at each level and corresponds to the number of influencer communities a consumer belongs to. Each consumer node can follow influencers either in a single community or from multiple communities. 

At scale $s$, for a consumer node \( v \in \mathcal{C} \), let \( I(v, E_i^s) \) be an indicator function that is 1 if node \( v \) belongs to the hyperedge \( E_i^s \), and 0 otherwise. The number of hyperedges node $v$ belongs to, i.e., the number of influencer communities that the users it follows belong to, is expressed as

\begin{equation}
\text{CO}^s(v) = \sum_{i=1}^{K_s} I(v, E_i^s).
\label{eq:1}
\end{equation}

\paragraph{Identity diversity.} This index is measured from the network's projection on the political influencers and indicates the diversity of identities among the influencers engaged by survey participants. According to social identity theory and social categorization theory, people may act based on inter-group behaviors arising from a sense of belonging to certain social groups \cite{tajfel1979integrative, turner1987rediscovering}. It can be applied to the political domain, where political identities advance social identities, such as national identities, partisan affiliations, and political ideologies. \cite{huddy2001from}. The diversity of influencers' political identity in a community is measured by the Identity Diversity Index. 

We annotate identities for influencers on four dimensions: Political Ideology, Campaign Support, Social Identity, and Account Type, as indicated in their profile descriptions. Influencers who strongly exhibit certain attributes may lead individuals to filter into specific community structures. However, not all influencers disclose identities in all dimensions. To address unlabeled accounts, we apply probability assignment, as described below (also see the percentages of unlabeled identities in Figs E–H in Supplementary Information). We only measure the diversity of labels on the Political Ideology dimension as it is the most direct indicator of political identity. 

We measure this index using the Gini-Simpson Diversity Index, used in fields such as ecology, sociology, and psychology to measure the diversity of types in a dataset\cite{simpson1949measurement}. The Gini-Simpson Diversity Index gives the probability that two entities taken at random from the dataset of interest represent different labels. It ranges from 0 to 1, where 0 represents a totally homogeneous distribution of ideology labels, and 1 represents totally diversified labels. For an influencers community $C_i^s$ at scale $s$, the Gini-Simpson Diversity is computed as
\begin{equation}
\text{IdD}(C_i^s) = 1 - \sum_{\ell=1}^{L} \left( \frac{n_\ell (n_\ell - 1)}{N (N - 1)} \right)
\label{eq:2}
\end{equation}

Where \( L \) is the number of ideological labels (here:``Left'', ``Right'' and ``Center''), \( n_\ell \) is the number of influencers of label \( \ell \) in community $C_i^s$, and \( N \) is the total number of influencers in community $C_i^s$. As not all the influencers are labeled, we apply two approaches to deal with the missing values: in one approach, we assign probabilities of labels to the unlabeled influencer accounts based on the proportions of labels present in each local community; in the other approach, we only consider the influencer accounts that are labeled. The two approaches yield similar results (see section 6 in Supplementary Information).

\paragraph{Information diversity.} Likewise, this index is computed based on the communities found on the network's projection on the influencers. Instead of being exposed to certain political identity groups, ordinary users may focus on the information retrieved from specific communities. People may have a preference for certain information sources or opinions due to personal interests or confirmation bias \cite{del2017modeling, knobloch2020confirmation}. Here, we measure the extent to which individuals are exposed to diverse information using the website domain links shared by political influencers.

Similar to Identity Diversity, we use the Gini-Simpson Diversity Index to measure the diversity of website domains shared by the influencers in that community, which is defined as Information Diversity. The Information Diversity, $\text{InfD}(C_i^s)$ of community $C_i^s$ at scale $s$ is used to measure exposure at the information level and is computed with \eqref{eq:2}, with \( L \) as the number of different website domains, \( n_\ell \) the number of times a website from domain \( \ell \) is shared by an influencer of community $C_i^s$, and \( N \) is the total number of links shared by influencers in community $C_i^s$.

This index measures how diversified the information sources are witnessed by individual consumers who follow that community instead of just the diversity of influencers' profiles. To achieve better reliability, we remove the communities that share fewer than 100 website domains and have less than 50 percent of influencer sharing links (see section 6 in Supplementary Information).

\paragraph{Structural integration.} This is a structural measure of the influencer communities engaged by survey participants. Each influencer community occupies a distinctive structural position within the network. Some communities are relatively isolated from others, while some are more integrated into the overall structure. In the co-following influencer network, a more isolated community indicates that it caters to a more targeted consumer group, whereas greater integration suggests exposure to a broader audience.

To quantify this, we introduce the Structural Integration Index. This index is derived from the Normalized Cut, a widely used metric in image segmentation and graph partitioning that evaluates the connectivity between two disjoint subsets of nodes in a network \cite{shi2000normalized}. We employ the Normalized Cut to assess the degree of integration of a given community by comparing its internal and external connectivity. For a community $C_i^s$ at scale $s$, the Structural Integration is defined as:

\begin{equation}
\text{SI}(C_i^s) = \frac{c_i^s}{2m_i^s + c_i^s} + \frac{c_i^s}{2(m - m_i^s) + c_i^s}
\label{eq:3}
\end{equation}

where \(c_i^s\) is the size of the cut, i.e., the sum of the weights of the edges crossing the boundary of the community $c_i^s=\sum_{u\in C_i^s, v \notin C_i^s} w_{uv}$,  \(m_i^s\) is the sum of the weights of the edges inside of the community, i.e. $m_i^s=\sum_{u,v\in C_i^s} w_{uv}$, and \(m\) is the sum total of the edge weights in the influencer network, $m=\sum_{u,v\in \mathcal{S}} w_{uv}$. The Structural Integration Index reaches a minimum value of 0 and a maximum value of 2 for two degenerate cuts: the empty cut, and the cut containing all edges. In this work, all the values of the index were between 0 and 1. Lower values indicate stronger isolation with the rest of the network, and larger values reflect greater integration, meaning the community has relatively more or stronger connections to other parts of the network.
This measure enables us to evaluate whether survey participants are primarily exposed to influencers from structurally peripheral communities or from those that are more centrally embedded in the network.

\paragraph{Connectivity inequality.} This index is also based on the influencer communities but focuses on their internal structures. Some communities exhibit an unequal distribution of (weighted) degrees, with certain nodes having a larger number of connections and potentially playing a more central role, while others display a more even distribution. 

We quantify the inequality in the distribution of influencer degree weights—defined as the number of shared survey consumers each influencer has with others within a community—using the Gini Index \cite{gastwirth1972estimation}.
It captures the skewness of the degree distribution. A highly skewed distribution may arise due to the presence of ``super-influencers'', such as prominent politicians, who share survey consumers with many other influencers. The metric is computed as

\begin{equation}
\text{CI}(C_i^s) = \frac{\sum_{u,v\in C_i^s} |k_u - k_v|}{2{N_i^s}^2 \left< k_i^s \right>}
\label{eq:4}
\end{equation}
where \( N_i^s \) is the number of nodes in the community $C_i^s$, $k_u=\sum_{v\in C_i^s} w_{uv}$ is the weighted internal degree of node $u$ and $\left< k_i^s \right>$ is the average weighted internal degree of the nodes in community $C_i^s$.

Finally, all influencer community-level indices—Identity Diversity, Information Diversity, Structural Integration, and Connectivity Inequality—are aggregated at the user level for the regression analysis by calculating the weighted average across all influencer communities with which the individual is connected.

Specifically, let \( \mathcal{C}_u \) denote the set of influencer communities followed by survey participant \( u \), and let \( w_{uC_j} \) represent the number of influencers in community \( C_j \in \mathcal{C}_u \) followed by user \( u \). For each index \( I \in \{\text{IdD}, \text{InfD}, \text{SI}, \text{CI}\} \), the aggregated individual-level score for user \( u \) is computed as a weighted average of the community-level values:

\begin{equation}
I_u = \frac{\sum_{C_j \in \mathcal{C}_u} w_{uC_j} \cdot I(C_j)}{\sum_{C_j \in \mathcal{C}_u} w_{uC_j}}
\label{eq:user_level_aggregation}
\end{equation}

This step ensures that communities with more connections to a given user contribute proportionally more to their overall exposure, reflecting the strength of affiliation and potential information influence. In contrast, the Community Overlap Index (\( \text{CO}^s(v) \)) is inherently computed at the individual level based on the number of influencer communities followed, and thus does not require further aggregation.

\subsection*{\textbf{Variable dimension reduction and regression analysis}}

Having measured the patterns of consumers' selective exposure, we continue to examine the association of ten groups of individual attributes obtained from the survey: Demographics, News Consumption, Political Communication, Political Identification, Political Engagement, Perceptions of Incivility, Perceptions of Disinformation, Authority Trust, Populism, and Attitudes toward Democracy, with the five selective exposure indices (see section 7 in Supplementary Information for detailed description of the ten groups of individual attributes). 

We perform a regression analysis to explore factors related to the five indices of selective exposure patterns.
The independent variables are consumers' individual attributes. To examine a wide range of consumers' attributes, we construct 10 groups of 189 variables obtained from the National Survey in Brazil, including Demographics, News Consumption, Political Communication, Political Identification, Political Engagement, Perceptions of Incivility, Perceptions of Disinformation, Authority Trust, Populism, and Attitudes to Democracy. These groups encompass individuals' social, political, and media attributes. We then reduce the dimensions of variables in each group by applying PCA, retaining only one variable for each significant dimension and 25 variables in total (see section 7 in Supplementary Information). 

The dependent variables are measurements of five indices of selective exposure patterns. As mentioned before, except for Community Overlap, the other four indices—Identity Diversity, Information Diversity, Structural Integration, and Connectivity Inequality—are initially measured based on influencer communities. We transform these four indices onto individual levels by calculating the average values of influencer communities each consumer interacts with, weighted by the number of links between consumers and influencers in each community.

Finally, we conduct a regression analysis between the reduced attribute variables and the transformed measurements of five indices. We employ a forward selection approach with BIC to select the appropriate regression models for all the selective exposure indices; variables are added iteratively until the BIC does not improve anymore. We use a Zero-truncated Negative Binomial regression (implemented with \textit{vglm} R package, family = \textit{posnegbinomial}) for Community Overlap, since it accepts values in positive integers, and Beta regression with a logit link function for the other indices, which all range between zero and one (implemented with \textit{betareg} R package \cite{cribari2010beta}).

For a more intuitive understanding of our methodological design, please refer to Fig A in Supplementary Information.
\section*{Results}
\subsection*{\textbf{Multidimensional profiles of political influencers and their interconnections}}

We build a bipartite network between 204 political consumers and 2,307 political influencers by combining datasets of surveys and Twitter/X. Political consumers are sampled from respondents of a national survey conducted during the 2022 Brazilian Presidential Election. Political influencers are identified based on the Twitter/X accounts followed by these survey respondents and can be politicians, media, and individual opinion leaders. Finally, 4,107 following pairs are found in the followship network.  

\begin{figure}[h]
\centering
\includegraphics[width=0.9\linewidth]{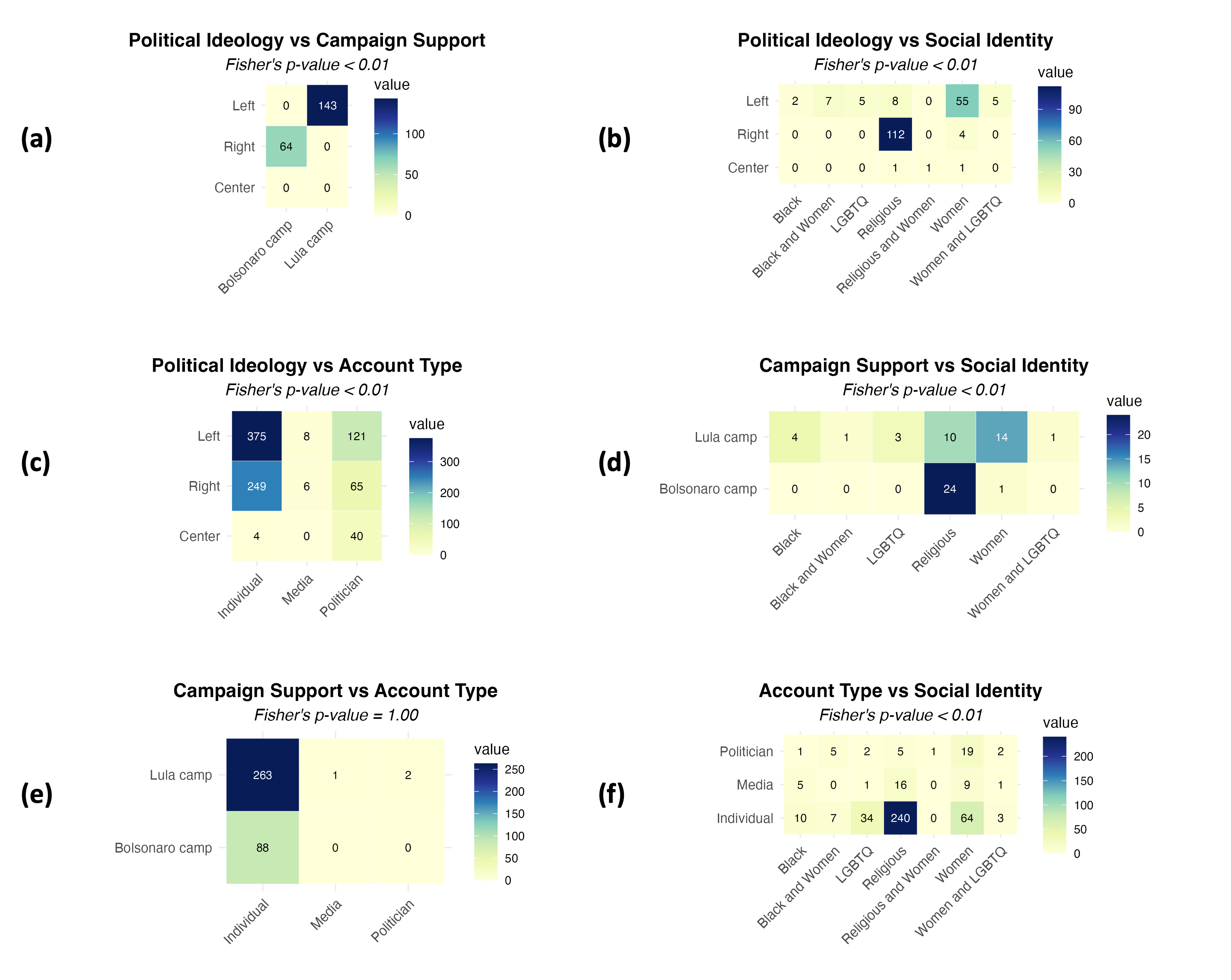}
\caption{\textbf{Contingency tables of multidimensional annotations of political influencer Twitter/X accounts.} Influencer accounts are annotated based on the dimensions of Political Ideology ($N$ = 886), Campaign Support ($N$ = 356), Social Identity ($N$ = 438), and Account Type ($N$ = 2,129). Correlations among categories are examined for each dimension pair (a–f). A multivariate Fisher’s exact test is applied to each contingency table to assess dependencies between dimensions, with significance levels indicated. In Social Identity, \emph{Women} denotes feminism or support for women’s rights; \emph{Religious}, \emph{Black}, and \emph{LGBTQ} refer to self-identification or expressed support for the rights of these groups.}
\label{fig:correlations}
\end{figure}

Based on our annotation of political influencers on four dimensions - Political Ideology, Campaign Support, Social Identity, and Account Type, We find that influencers’ attributes across different dimensions are correlated to some degree (see Fig \ref{fig:correlations}). A Fisher's exact test confirms that accounts with left-wing ideologies are associated with a support of Lula - the leftist leader and current president of Brazil and advocate for women's rights (Fisher's Exact Test $p$-value $<0.01$). Conversely, right-wing accounts are more likely to support Bolsonaro - the rightist leader, former president of Brazil, and be associated with religious groups ($p<0.01$). Most accounts that express support for electoral candidates are from individual opinion leaders, showing only weak dependency between dimensions of Campaign Support and Account Type ($p=1.00$). However, left-wing accounts demonstrate higher proportions of politician accounts compared to others, and religious accounts show higher correlations with individual opinion leaders compared to other identities ($p<0.01$).

These correlation patterns can suggest a dependency relation or potentially hierarchical relation between dimensions. To understand whether there is a more complex structure of individuals' selective exposure to political influencers, we investigate how the political influencers are organized in the network with a multi-level analysis.

\subsection*{\textbf{Hierarchical structure of online political selective exposure}}

We then project the bipartite network onto the influencer side and conduct multi-scale community detection on the influencer-projected network to measure the complexity of individuals' selective exposure structures (see the schematic diagram in Fig \ref{fig:community detection} (a) \& (b) and the Materials and methods section for more details.). 

To capture the multi-scale and hierarchical organization of real-world social networks\cite{clauset_hierarchical_2008,ahn2010link},
we conduct multi-scale community detection\cite{delvenne2010stability,arnaudon_algorithm_2024} on the influencer network (see details in the Materials and methods section and section 5 in Supplementary Information). We detect five significant levels (or scales). The number of influencer communities detected at five levels is shown in Fig \ref{fig:community detection} (c), ranging from 46 communities at the lowest level and 2 communities at the highest level. In this setting, a community represents a set of political influencers more likely to be co-followed by the same set of consumers. 

\begin{figure}[h]
\centering
\includegraphics[width=0.9\linewidth]{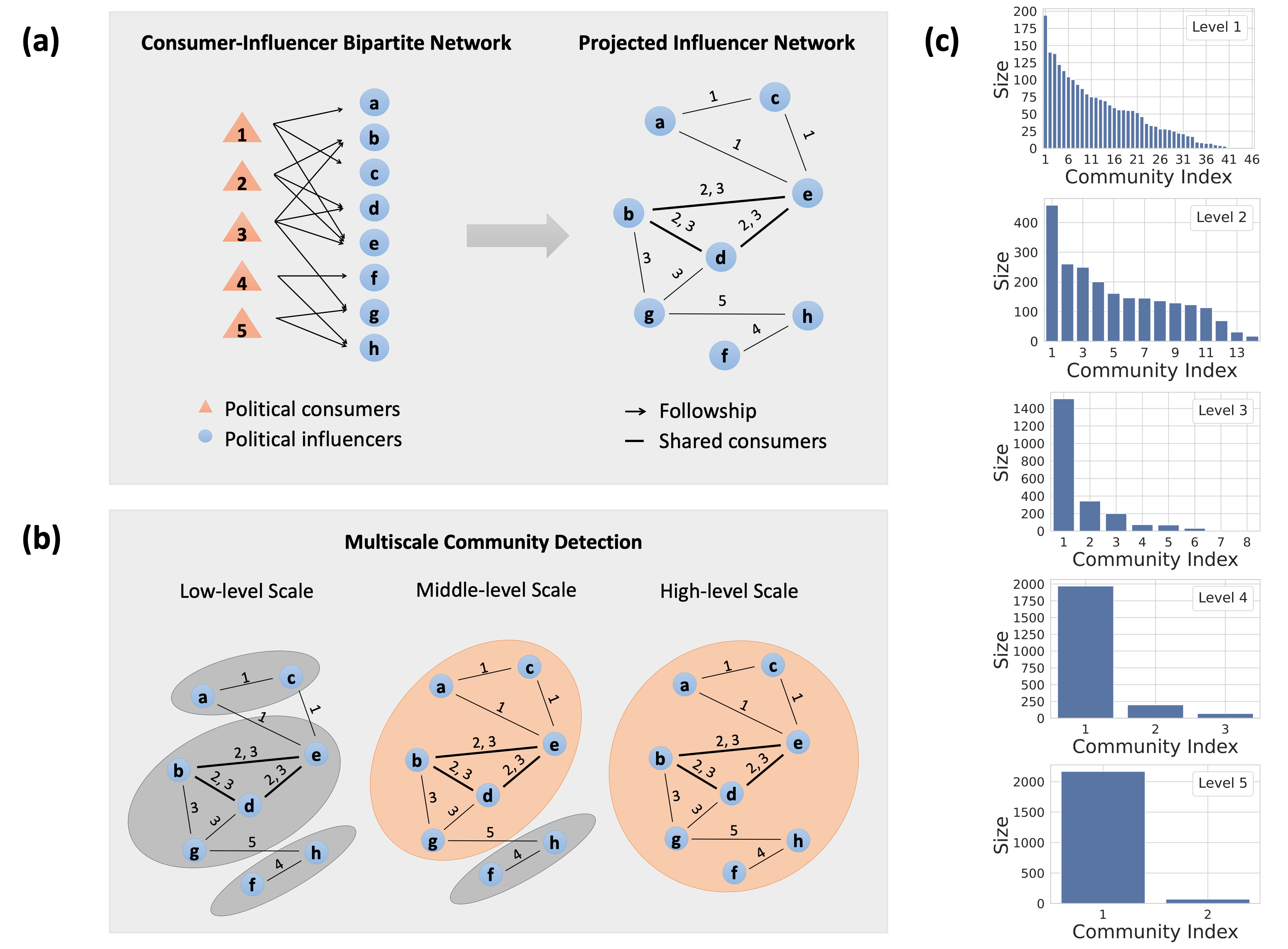}
\caption{\textbf{Schematic diagram of network construction and community detection.} (a) The procedure used to construct the bipartite network between political consumer nodes (orange triangles) and political influencers nodes (blue circles). We project the bipartite network onto political influencers. The directed arrows between consumers and influencers represent the following relationships on Twitter/X. The undirected lines between influencers indicate their shared consumers, weighted by the number of shared consumers. (b) Illustration of multi-scale community detection in the projected influencer network. At the lower level, there are more communities, and the size of each community is smaller (shown in grey). At the higher level, the smaller communities merge, resulting in fewer, larger communities (shown in orange). (c) Histograms of community sizes at the five detected levels.}
\label{fig:community detection}
\end{figure}

\begin{figure}[h]
\centering
\includegraphics[width=0.9\linewidth]{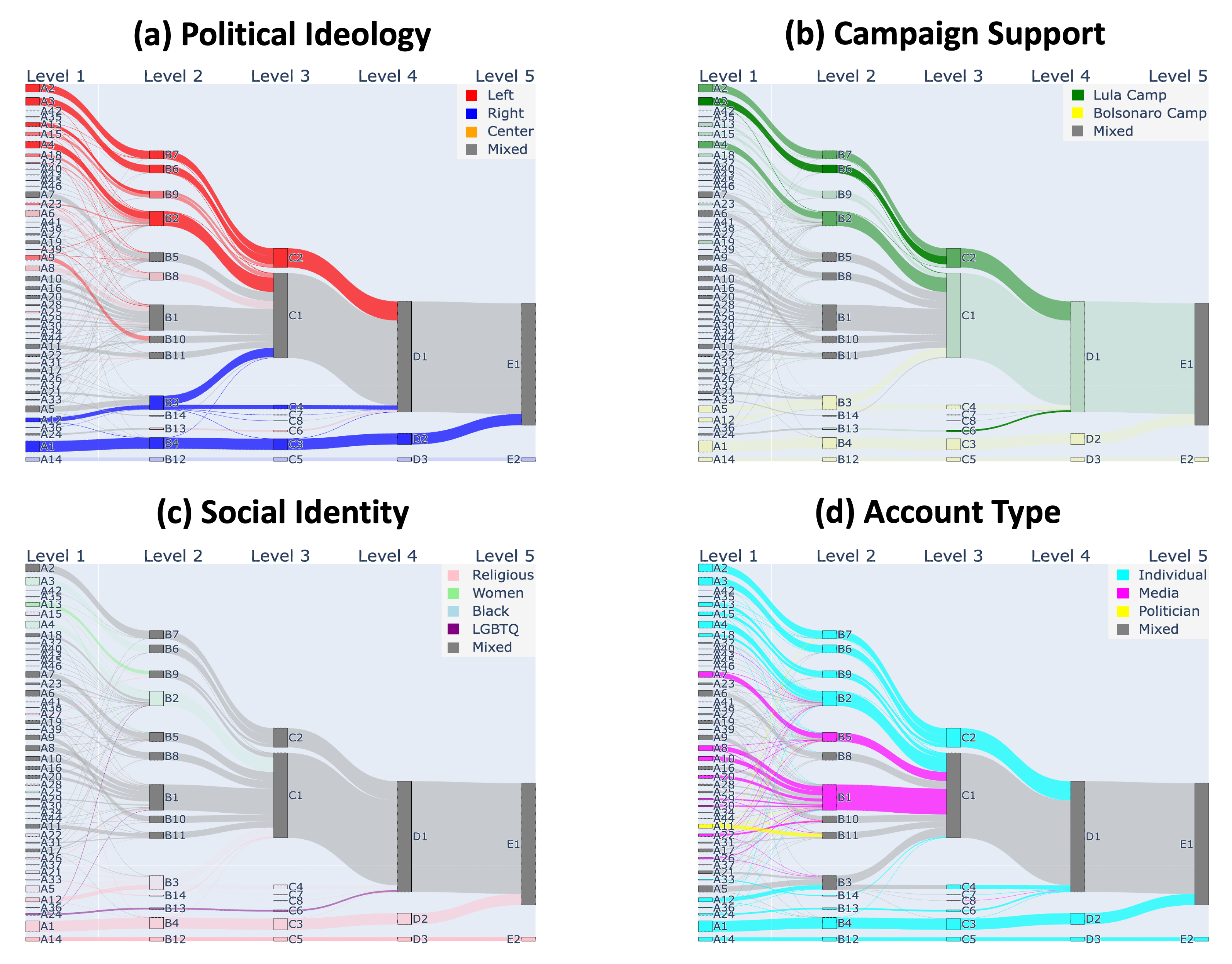}
\caption{\textbf{Alluvial diagram showing the hierarchical organization of political influencer communities across five levels.} The different panels display the labeling of political influencer communities according to the following dimensions: Political Ideology (a), Campaign Support (b), Social Identity (c), and Account Type (d). The partition of the influencers in communities from a granular level to coarser levels is shown by the different grouping and arrangement of the bundles. We assign a color to a community if more than half of the influencer accounts within the community belong to one category. The transparency of the colors is proportional to the homogeneity of the labels; the higher the proportion of the majority category, the more salient the color. We only color communities when more than 30\% of influencer accounts are annotated. In Social Identity, \emph{Women} denotes feminism or support for women’s rights; \emph{Religious}, \emph{Black}, and \emph{LGBTQ} refer to self-identification or expressed support for the rights of these groups.}
\label{fig:sanky diagram}
\end{figure}

Fig \ref{fig:sanky diagram} shows the results of the multi-scale community detection of the influencer network as an alluvial diagram, revealing its hierarchical structure. As we can see, at the most granular level (Level 1), diverse communities are formed based not only on ideologies but also on political support, social group affiliation, and account types. At the second scale (Level 2), communities with mostly the same categories merge, resulting in more unified left-wing and right-wing communities, as well as distinct Lula and Bolsonaro camps. From level 3 onward, the left-wing and right-wing communities, along with Lula and Bolsonaro supporters, merge together. At the same time, many categories based on social identity and account types become less prominent. By levels 4 and 5, we observe a large mixed community and predominantly right-wing communities and Bolsonaro supporters on the side, who are more isolated and mainly come from religious groups and individual opinion leaders.

The results show that a single scale in community detection may not always detect the polarized selective exposure pattern between left-wing and right-wing groups. Lower-level scales can uncover more fragmented communities nested within ideological groups, while higher-level scales may capture more integrated ideological communities. Our findings also suggest that the higher-level ideological segmentation in Brazil is asymmetric, with some right-leaning communities being more isolated while others are integrated into larger, more mixed communities. This aligns with findings on political news consumption in the United States \cite{GonzalezBailon2023}.

\subsection*{\textbf{Measurements of online political selective exposure patterns}}

\begin{figure}[h]
\centering
\includegraphics[width=0.9\linewidth]{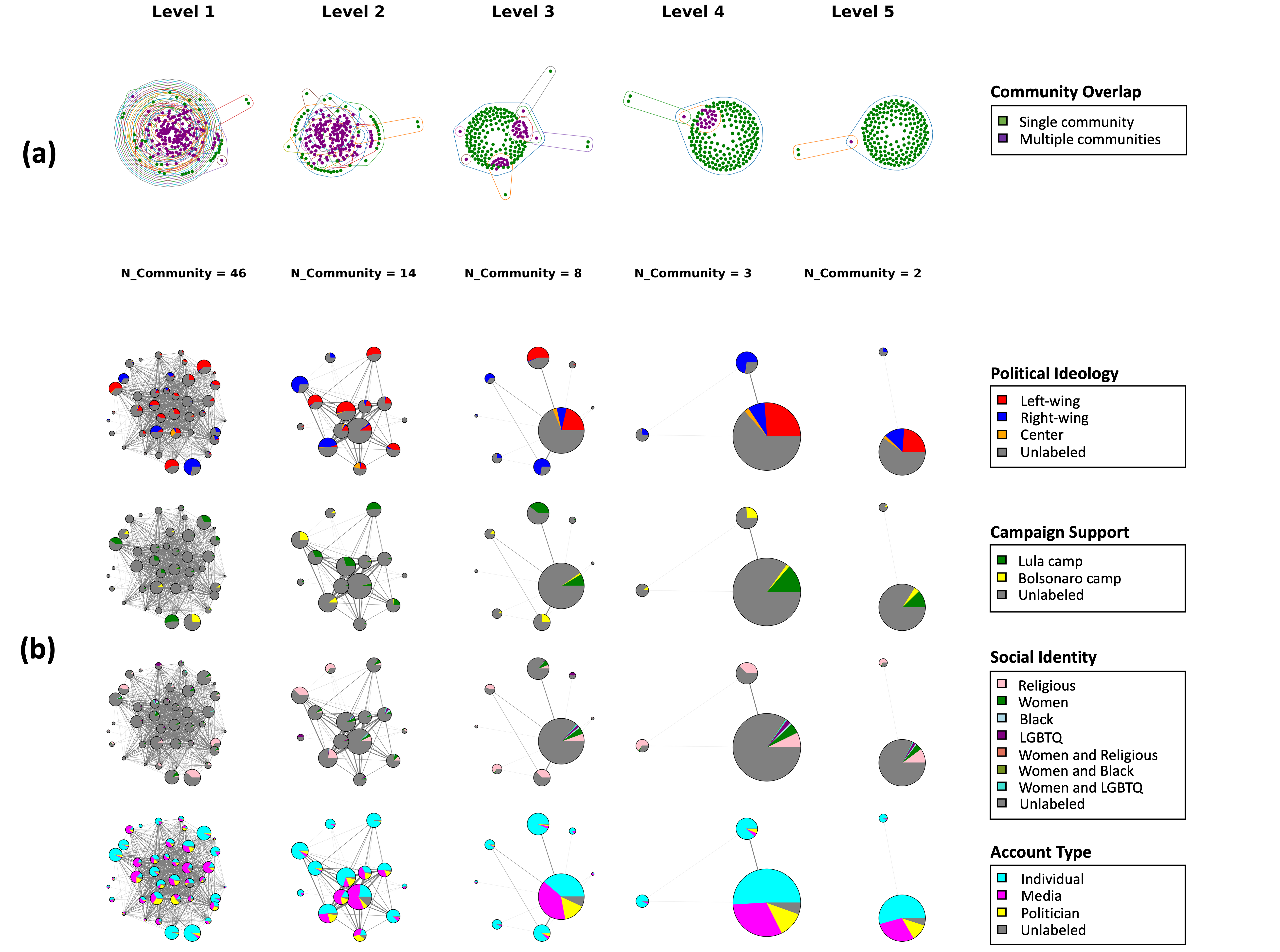}
\caption{\textbf{Projection of the consumer-influencer bipartite network onto both the consumer and influencer sides.} (a) Influencer communities projected on the consumer network. Each node represents a consumer who can follow single or multiple influencer communities. Consumers following the same communities are encircled. Consumer nodes that follow a single community are annotated in green, while those that follow more than one community are annotated in purple. All five levels are displayed. (b) Projected pie-chart graph of the influencer community network. The nodes are influencer communities, indicated by a pie chart of labels on four dimensions: Political Ideology, Campaign Support, Social Identity, and Account Type. Proportions of categories in each dimension are shown inside the pie chart. Edges between two nodes denote the consumers who follow both influencer communities. In Social Identity, \emph{Women} denotes feminism or support for women’s rights; \emph{Religious}, \emph{Black}, and \emph{LGBTQ} refer to self-identification or expressed support for the rights of these groups.}
\label{fig:community visualization}
\end{figure}

We also project the consumer-influencer bipartite network onto the consumer side. We group consumers together if they follow political influencers in the same influencer community. The consumers can, therefore, be involved in multiple or just a single influencer community (see Fig \ref{fig:community visualization} (a)). We observe that as the level increases, the number of individuals involved in multiple influencer communities decreases. In contrast, the number of individuals who belong to only a single influencer community increases.

Consumers who are involved in multiple influencer communities create closer connections between these influencer communities. We visualize the influencer community network connected by shared consumers in Fig \ref{fig:community visualization} (b), where each pie chart represents an influencer community node and shows the proportions of different identity categories across four dimensions. The edges represent consumers who follow both communities. At Level 1 and 2, we observe that influencer communities closely connected by shared consumers exhibit high proportions of left-wings and media categories. Consumers following these categories of influencer accounts are more versatile and fragmented than others, and this pattern can be mainly observed at the lower level.

\begin{figure}[h]
\centering
\includegraphics[width=0.9\linewidth]{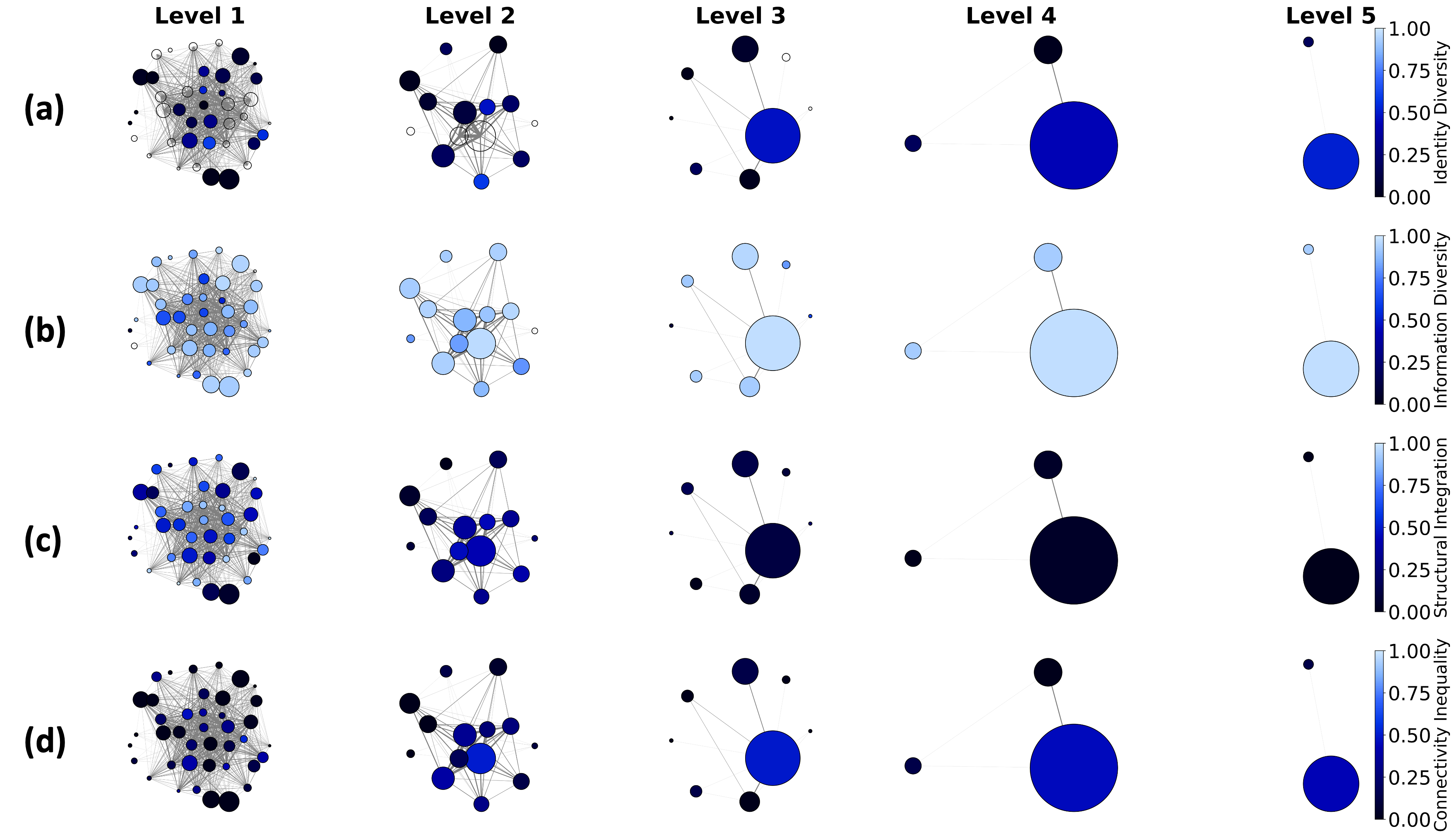}
\caption{\textbf{Visualization of measurements of four community-level indices in the influencer community network.} (a) Identity Diversity, (b) Information Diversity, (c) Structural Integration, and (d) Connectivity Inequality. The nodes in each network represent detected influencer communities, and the edges represent the shared consumers between two influencer communities. The area of each node is proportional to the number of political influencers in the community. The color spectrum ranges from black to blue, with a darker color indicating more homogeneous ideological identities and domain information, less isolation from other communities, and a more equal distribution of degree weights.}
\label{fig:local measurements}
\end{figure}

We also show the values of the four community-level indices in Fig \ref{fig:local measurements}. The granular levels (Level 1 and Level 2) demonstrate diverse values across the four indices. As the levels increase, there is a large merged influencer community where the political ideologies are more diverse and the (weighted) degree distribution is more unequal compared to other communities. Information Diversity and Structural Integration show little distinction among communities at higher levels.

\subsection*{\textbf{Individual attributes related to multi-level selective exposure patterns}}

We continue to investigate how certain consumer attributes are related to selective exposure patterns at various levels with a regression analysis. Individual attributes are conceptualized into 10 groups, encompassing 189 variables, including Demographics, News Consumption, Political Communication, Political Identification, Political Engagement, Perceptions of Incivility, Perceptions of Disinformation, Authority Trust, Populism, and Attitudes toward Democracy. We conduct Principal Component Analysis (PCA) within each variable group to address the issue of multicollinearity. This results in 25 variables after PCA dimension reduction, which are then regressed against the five indices. It is worth noting that we convert the four community-level indices—Identity Diversity, Information Diversity, Structural Integration, and Connectivity Inequality—into individual-level indices by calculating the average values of the influencer communities each consumer interacts with, weighted by the number of influencers each consumer follows (see mathematical definitions in the Materials and methods section).

We identify significant explanatory variables among the 25 initial variables for each index at each level using a forward selection approach based on the Bayesian Information Criterion (BIC). This model selection procedure aims to find the statistical model with the fewest explanatory variables by iteratively adding them until the statistical significance of the model, as measured by the BIC, no longer improves (see section 7 in Supplementary Information for more details).

\begin{figure}[H]
\centering
\includegraphics[width=0.9\linewidth]{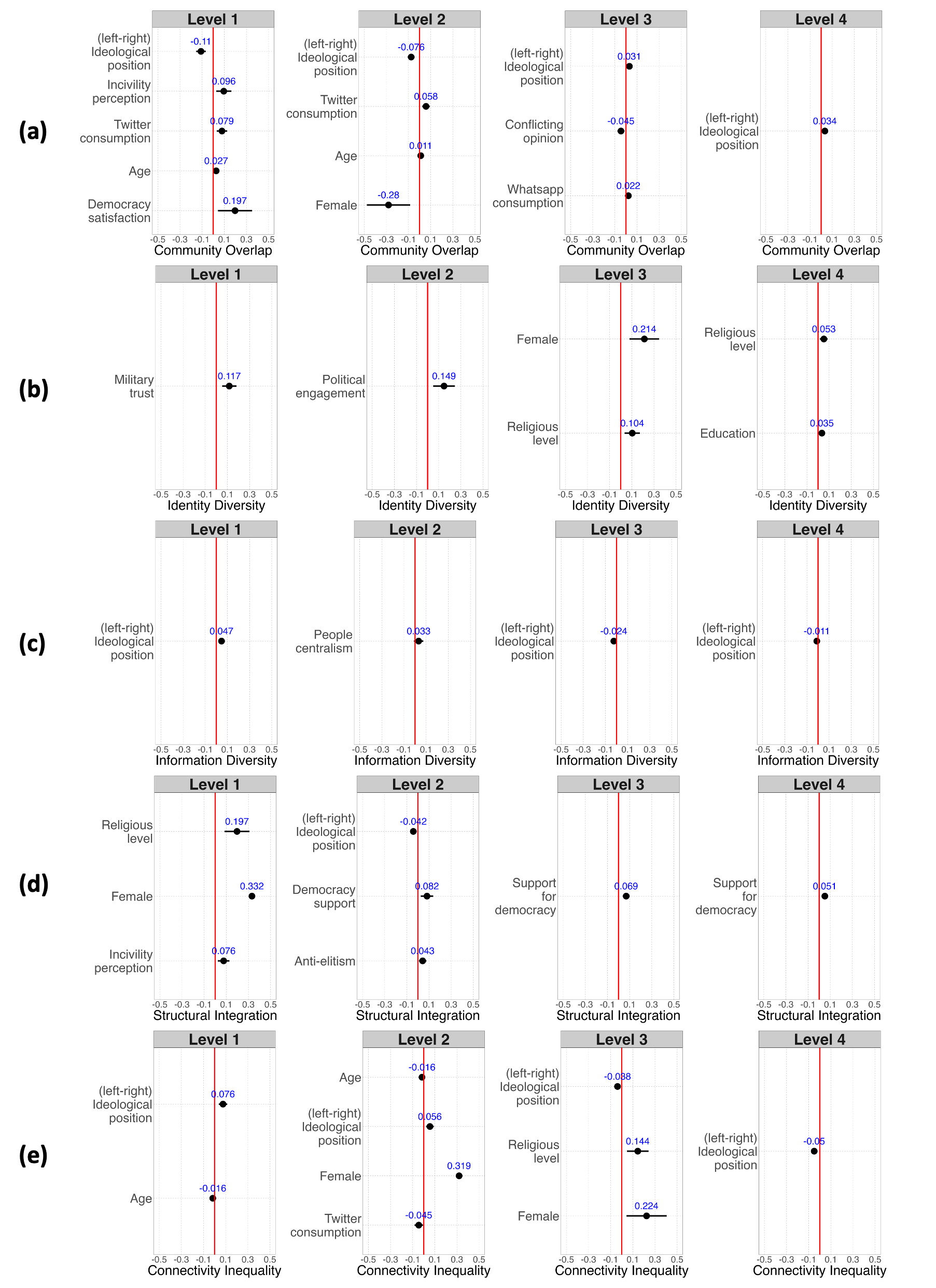}
\caption{\textbf{Regression plot of significant individual attributes related to selective exposure patterns.} Regression analyses of consumers' attributes on Demographics, News Consumption, Political Communication, Political Identification, Political Engagement, Perceptions of Incivility, Perceptions of Disinformation, Authority Trust, Populism, and Attitudes toward Democracy are conducted across five indices of selective exposure patterns: (a) Community Overlap, (b) Identity Diversity, (c) Information Diversity, (d) Structural Integration, and (e) Connectivity Inequality at four network community levels. The variables displayed are significant factors resulting from a forward selection based on BIC. The coefficient values are displayed in blue, and the confidence intervals at the 95\% confidence level are shown with error bars. The red lines indicate the $0$ value.}
\label{fig:regression}
\end{figure}

We fit the minimal explanatory variables into the regression model for each index and each level. After conducting distribution checks for the values of dependent variables, the zero-truncated Negative Binomial Regression model is chosen as the basic model for the Community Overlap Index. For the indices of Identity Diversity, Information Diversity, Structural Integration, and Connectivity Inequality, the Beta Regression model is employed. For more details, refer to the Materials and methods section and to section 7 in Supplementary Information. The regression results for the first four levels are shown in Fig \ref{fig:regression}. 

For the Community Overlap Index, we can observe that the ideological position (measured on a scale from 0 to 10, where 0 means “Very left-wing” and 10 means “Very right-wing”) of individuals plays a significant role in explaining their involvement in multiple communities at all levels. However, it is not the only factor; at lower levels of communities, more factors contribute to the versatile engagement of consumers in influencer communities (see Fig \ref{fig:regression} (a)). The impact of ideological position corresponds to the fragmentation pattern displayed in Fig \ref{fig:regression} (b), suggesting that left-wing influencer communities are more likely to be shared by consumers with high bridging capacity. Another two significant factors that remain at the first two levels are the frequency of news consumption on Twitter/X and age. At Level 3, the sign of the ideological position coefficient changes, demonstrating a turning point from which right-wing consumers are more inclined to be involved in multiple influencer communities. This is because as the left-wing influencers merge, most of the remaining isolated communities are right-wings. The ideological position plays an increasingly important role at higher levels, especially at Level 4, where only the ideological position remains significant.

The other community-level indices show correlations with different attributes, but similarly to the Community Overlap, they demonstrate various associations across levels. For instance, trust to police, military, and government and the frequency of participating political activities primarily explain Identity Diversity at Level 1 and Level 2 (see Fig \ref{fig:regression} (b)). However, demographic variables play a more important role at Levels 3 and 4, where the religious level is consistent at both levels.

The ideological position of individuals plays a more significant role in Information Diversity compared to other indices. As shown in Fig \ref{fig:regression} (c), the ideological position is the only significant factor at Levels 1, 3, and 4, and a sub-dimension of populist ideology—people centralism, which is the demand for prioritizing the rights of the majority, is a significant factor at Level 2.

For Structural Integration, there are many explanatory factors at the finer levels, while the degree of integration of communities is mainly related to consumers' support for democracy at Levels 3 and 4 (see Fig \ref{fig:regression} (d)). Regarding Connectivity Inequality, similar to Community Overlap, at lower scales, ideological position, as well as a variety of other factors such as age, gender, religious level, and the frequency of news consumption on Twitter/X, have explanatory power. Only the ideological position remains significant at Level 4 (see Fig \ref{fig:regression} (e)).

\section*{Discussion}

The findings of this study reveal the intricate and hierarchical nature of individuals' political selective exposure in the Brazilian Twitter/X space. By employing a multidimensional annotation design, we discover that attributes of political influencers are correlated across different dimensions. For instance, left-wing accounts are correlated with attributes of supporting Lula and advocating for women's rights. In contrast, right-wing accounts are more likely to be associated with Bolsonaro supporters and religious believers. This highlights political influencers involved in homogeneous left-wing and right-wing communities or Lula-camp and Bolsonaro-camp also reassemble each other on demographic grounds, even though it does not apply for all the communities. 


In the long history of political studies, demographics such as gender, race, ethnicity, and religion have been related to ideological positions and voting behaviors \cite{Manza1999}. In recent years, demographic cleavages have become more significant due to the emergence of far-right and populist political leaders in political campaigns \cite{golder2016far}. Layton et al. (2021) demonstrated a pattern of new alignments in Brazil’s electorate since the 2018 presidential election, showing that political groups are now split more based on demographics and issues rather than purely partisanship \cite{layton2021demographic}. In particular, the far-right candidate Bolsonaro has attracted new supporters by creating demographic polarization beyond the traditional left-versus-right division. For instance, his misogynistic views on gender, racially offensive remarks to Black activists, and evangelical propositions have played a role in this polarization \cite{layton2021demographic}. The strong correlations between political ideology and other dimensions suggest that a new trend of political division might be ongoing.

In light of this background, media scholars and political scientists have gradually shifted from the traditional focus on selective exposure, polarization, and echo chambers of political ideology and partisanship to a wider range of divisions such as issues, demographics, and beliefs \cite{yarchi2021political}. However, effective methods to detect such selective exposure at various levels are still lacking. This study provides a novel perspective by utilizing a multi-scale community detection approach to identify different patterns of selective exposure. This approach can also be applied to research on similar topics, such as polarization and echo chambers. 

The hierarchical patterns observed through multi-scale community detection demonstrate the importance of levels in understanding the complexity of political selective exposure (RQ1). As community detection moves to coarser levels, the communities in which survey participants are involved change accordingly. These changes can be captured by various measurements, including Community Overlap, Identity Diversity, Information Diversity, Structural Integration, and Connectivity Inequality. For instance, participants may shift from engaging with multiple communities to engaging with only one as the level becomes more aggregated (Community Overlap), and the communities they engage with may become more homogeneous or more diverse in terms of identity (Identity Diversity). These measurements offer a quantitative perspective for understanding how selective exposure happens at various resolution levels. At the most granular level, diverse communities form based on various factors, including political ideology, candidate support, social group affiliation, and account type. As the scale increases, these communities gradually merge, forming more unified—albeit asymmetric—ideological groups. This process continues until higher scales predominantly reveal right-wing communities and Bolsonaro supporters as homogeneous and isolated communities, which are mainly associated with religious groups and individual opinion leaders.

We find that Community Overlap is a useful measure of a consumer's selective exposure patterns. Many variables, such as demographics, news consumption frequency, incivility perception, and ideological position, are found to be related to individuals' versatility in connecting to influencer communities. However, only ideological position is significant at the coarsest level. Our investigation into the related factors of selective exposure patterns reveals that different factors play a role in multi-level selective exposure (RQ2). We reveal a more nuanced picture of political information filtering than traditional left–right ideological models alone can capture. Specifically, the emergence of fine-grained communities nested within broader ideological camps suggests that demographic attributes (e.g., age, gender, religion), social media experience (e.g., Twitter/X consumption frequency, incivility perception), as well as other factors, may exert greater influence at localized scales. For example, we find that the gender female emerges as a significant factor at the second level, associated with engagement in fewer communities. Indeed, Fig \ref{fig:sanky diagram}(c) shows that most influencers advocating for women's rights form a single, homogeneous community at this level, implying that individuals' exposure to political content is not only shaped by ideological alignment but also by socio-demographic proximity and information consumption habits, which may reinforce specific opinions or beliefs within subgroups. For instance, the alignment of social and political identities might lead to increased polarization and group-based animosity, potentially resulting in toxicity and severe attacks. This multilevel approach thus offers a powerful lens for understanding how political attitudes are shaped in complex media ecosystems, where micro-level affiliations and macro-level ideologies intersect. Moreover, this framework has broader applicability beyond the Brazilian context, offering a valuable tool for studying political communication dynamics in other multi-party democracies or conducting cross-system comparison. By uncovering the layered mechanisms underlying selective exposure, our approach advances understanding of the complex drivers of fragmentation and mitigates the risk of overgeneralizing conclusions to entire ideological groups. Methodologically, it also suggests that online communities detected at a single level might miss important information; therefore, various levels should be examined in relevant studies.

This study builds both theoretical and methodological foundations for exploring multilevel selective exposure and related concepts such as polarization, and echo chambers. Although the survey sample size is limited in this study, this does not influence the validity of the insights brought by our novel multi-scale approach, which would also apply to larger samples. Additionally, we did not screen for potential Twitter/X bots. However, as influencers who are potentially bots, or followed by many bots, have been willingly followed by survey users and participate in their information exposure, including them is necessary to correctly capture the selective exposure of survey users.
We also only used following relations rather than other types of interactions. This is because we focus on information exposure rather than engagement. It is worth noting that recommendation algorithms can also influence information exposure independently of whom the survey participants follow. Future research could explore other forms of interaction and assess the role of recommendation algorithms to complement this approach. Methodologically, when projecting the bipartite network, this work retains all link information in the projection. We note that alternative projection methods for bipartite networks, such as the fixed degree sequence model or the stochastic degree sequence model, that extract a backbone preserving only the most significant links \cite{neal2021comparing} could be used in future work. Additionally, for research aimed at uncovering nested structures and fine-grained subgroups within political systems—where actors may belong to different communities at different scales, as in this study—the Markov stability multiscale approach offers distinct advantages over modularity-based methods. However, alternative approaches may be more suitable depending on the research goals. For instance, overlapping community detection methods are well suited for capturing multifaceted affiliations, where nodes can simultaneously belong to multiple communities at the same scale. If the main goal is to reveal both hierarchical structures and multidimensional affiliations of users at each scale, alternative methodologies such as multi-scale overlapping community detection can be explored. In summary, this study demonstrates the necessity of employing multi-level approaches to effectively capture the complexity of political selective exposure in countries with multi-party systems and complex social structures. Our results highlight that political communities are not monolithic; rather, they are characterized by intricate interconnections across various dimensions and scales. Single-scale analyses may overlook critical aspects of community formation and evolution, potentially leading to incomplete or biased understandings of online political patterns. Future research should continue to refine such multi-scale methodology, potentially combining with content and issues, to interpret the multi-level behaviors of minority groups and radical ideological groups.

\section*{Acknowledgements}
 
The authors thank Reinhard Furrer, Dorian Quelle, Yasaman Asgari, and Samuel Koovely for helpful advice and discussions.

\section*{Data Availability}
All relevant data and code are available from the Harvard Dataverse repository (\url{https://doi.org/10.7910/DVN/46ZS2Y}). To safeguard participant privacy, all data have been fully anonymized prior to publication.

\section*{Funding}

This work was supported in part by funds from the Swiss National Science Foundation (grant: 100017\_204483). The funders had no role in study design, data collection and analysis, decision to publish, or preparation of the manuscript.

\section*{Author Declaration} No authors have competing interests.

\clearpage
\setcounter{figure}{0} 
\renewcommand{\thefigure}{\AlphAlph{\value{figure}}}

\begin{center}
   \LARGE{\textbf{Supplementary Information for ``A multilevel network approach to revealing  patterns of online political selective exposure''}}
 
\vspace{1cm}

\large Yuan Zhang$^{1,\ast}$, Laia Castro$^{2}$, Frank Esser$^{1}$ and Alexandre Bovet$^{3,4,\ast}$

\vspace{0.2cm} \normalsize 
$^{1}$Department of Communication and Media Research, University of Zurich, Switzerland\\
$^{2}$Department of Political Science, University of Barcelona, Spain\\
$^{3}$Department of Mathematical Modeling and Machine Learning, University of Zurich, Switzerland\\
$^{4}$Digital Society Initiative, University of Zurich, Switzerland.\\
* y.zhang@ikmz.uzh.ch, alexandre.bovet@uzh.ch

\end{center}

\section*{1. Overview of Methodological Design}

\begin{figure}[H]
\centering
\includegraphics[width=1.0\linewidth]{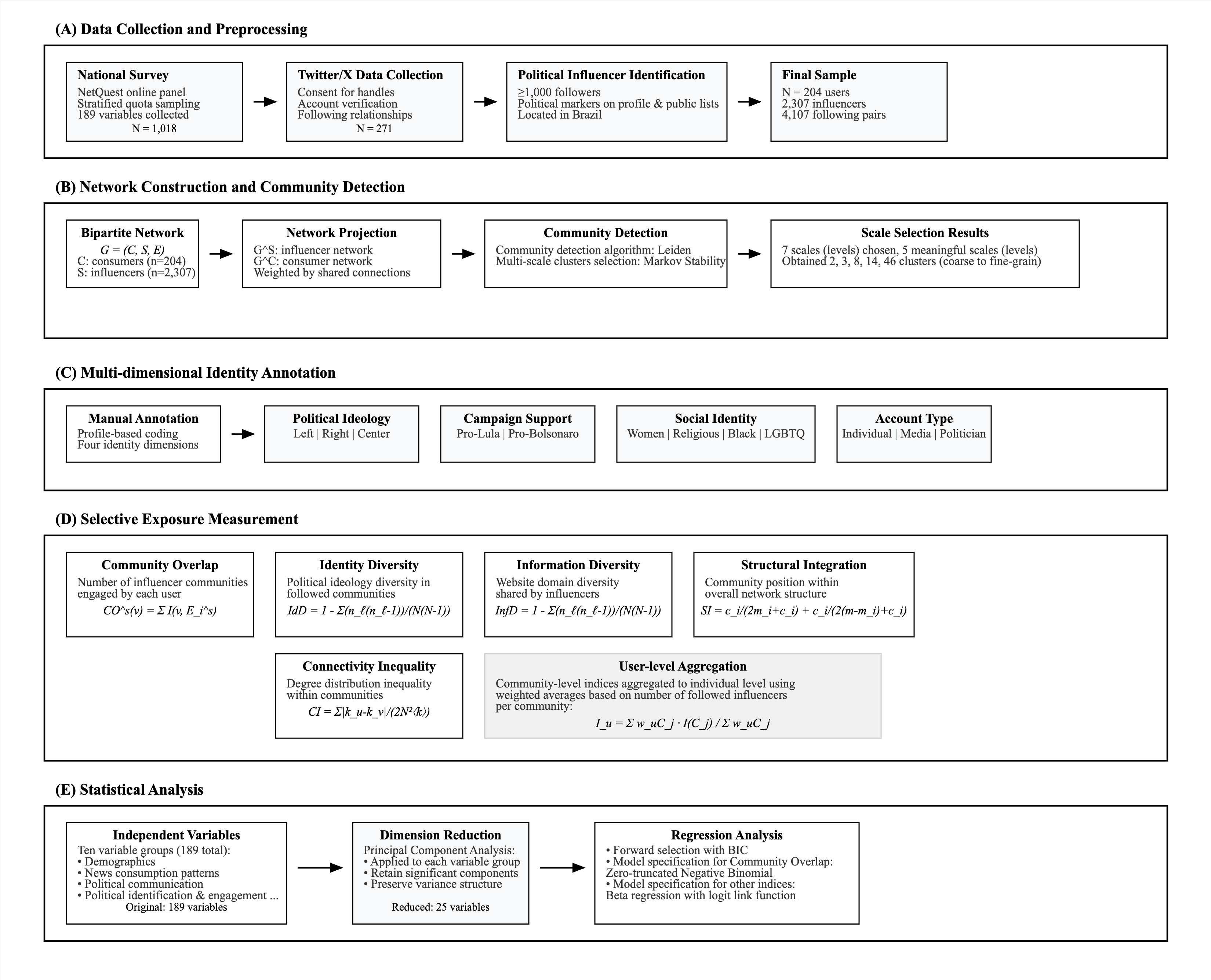}
\caption{Flowchart of the Methodological Design Overview. The analytical framework comprises five stages: (A) Data Collection and Preprocessing. (B) Network Construction and Community Detection. (C) Multi-dimensional Identity Annotation. (D) Selective Exposure Measurement. (E) Statistical Analysis.}
\label{fig:methodology_flowchart}
\end{figure}

The following flowchart illustrates the overall methodological design of this study (see \sifigref{fig:methodology_flowchart}). 
 
\section*{2. Survey Questionnaire}

\subsection*{Section I. Demographics}\hfill\\

\noindent\textbf{(Age)} How old are you?\hfill\\

\noindent\textbf{(Gender)} Are you…? 
\begin{enumerate}[label=\arabic*.]
    \item Male
    \item Female
    \item Other
\end{enumerate}

\noindent\textbf{(Ethnic)} How do you describe yourself? (select all that apply)

\begin{enumerate}[label=\arabic*.]
    \item White
    \item Black
    \item Mixed
    \item Asia
    \item Indigenous
    \item Other (\textit{write in}): \underline{\hspace{5cm}}
    \item Prefer not to answer
\end{enumerate}

(PT version) \noindent\textbf{(Etnia)} A sua cor ou raça é? (RU)
Nesta pergunta é possível assinalar somente uma alternativa. 

\begin{enumerate}[label=\arabic*.]
    \item Branca
    \item Preta
    \item Parda
    \item Amarela
    \item Indígena
    \item Outro (\textit{anote}): \underline{\hspace{5cm}}
    \item Prefiro não responder
\end{enumerate}

\noindent\textbf{(Education)} What is the highest degree or level of school you have completed? (If you’re currently enrolled in school, please indicate the highest degree you have received.)

\begin{enumerate}[label=\arabic*.]
    \item Até a Pré Escola
    \item Até a 4ª série/ 5º ano do Ensino Fundamental
    \item Até a 8ª série/ 9º ano do Ensino Fundamental
    \item Até o 1º ano do Ensino Médio
    \item Até o 2º ano do Ensino Médio
    \item Até o 3º ano do Ensino Médio
    \item Superior Incompleto
    \item Superior Completo
    \item Pós-graduação ou Mestrado
    \item Doutorado
\end{enumerate}

(PT version) \noindent\textbf{(Educação)} Até que ano da escola você cursou? (RU)
Nesta pergunta é possível assinalar somente uma alternativa.

\begin{enumerate}[label=\arabic*.]
    \item Até a Pré Escola
    \item Até a 4ª série/ 5º ano do Ensino Fundamental
    \item Até a 8ª série/ 9º ano do Ensino Fundamental
    \item Até o 1º ano do Ensino Médio
    \item Até o 2º ano do Ensino Médio
    \item Até o 3º ano do Ensino Médio
    \item Superior Incompleto
    \item Superior Completo
    \item Pós-graduação ou Mestrado
    \item Doutorado
\end{enumerate}

\noindent\textbf{(Religion)} What is your present religion, if any?

\begin{enumerate}[label=\arabic*.]
    \item Roman Catholic
    \item Protestant evangelical or pentecostal
    \item Protestant non-evangelical
    \item Non-christian Oriental Religions (Islamism, Hinduism, Buddhism)
    \item Jeova's Witnesss
    \item Afro-brazilian religions (Umbanda, Candomble)
    \item Kardecist
    \item Jewish
    \item Other religions
    \item Is religious but doesn't follow any religion / Agnostic
    \item Atheist
    \item Prefer not to answer
\end{enumerate}

(PT version) \noindent\textbf{(Religião)} Qual a sua religião ou culto? (RU)
Nesta pergunta é possível assinalar somente uma alternativa.

\begin{enumerate}[label=\arabic*.]
    \item Católica Apostólica Romana
    \item Protestante evangélico e pentecostal (Igreja Universal, Quadrangular, Batista, Adventista,
etc.)
    \item Protestante não evangélico (Calvinista, Luterano, Metodista, Anglicano, etc)
    \item Religiões orientais não cristãs (Islamismo, Budismo, Hinduísmo)
    \item Testemunha de Jeová
    \item Religiões afro-brasileiras (Candomblé, Umbanda)
    \item Kardecista, espírita
    \item Judeu
    \item Outras religiões
    \item É religioso mas não segue nenhuma / Agnóstico
    \item Ateu
    \item Prefiro não responder
\end{enumerate}

\noindent\textbf{(Religious Level)} Regardless of whether you belong to a particular religion, how religious would you say you are?

\begin{enumerate}[label=\arabic*.]
    \item Very religious
    \item Moderately religious
    \item Somewhat religious
    \item Not at all religious
    \item Prefer not to answer
\end{enumerate}

\noindent\textbf{(Living)} Which description best describes the area where you live?

\begin{enumerate}[label=\arabic*.]
    \item A big city
    \item The suburbs or outskirts of a big city
    \item A town or a small city
    \item A country village
    \item A farm or a home on the countryside
    \item Don’t know 
\end{enumerate}

\noindent\textbf{(Income)} Which of the following income brackets applies to your household income, including you and everyone who lives with you? Please include all sources of income, such as salaries, pensions etc.

\begin{enumerate}[label=\arabic*.]
    \item Up to R\$ 1.212,00
    \item R\$ 1.212,01 to R\$ 2.424,00
    \item R\$ 2.242,01 to R\$ 3.636,00
    \item R\$ 3.636,01 to R\$ 6.060,00
    \item R\$ 6.060,01 to R\$ 12.120,00
    \item R\$ 12.120,01 to R\$ 24.240,00
    \item R\$ 24.240,01 to R\$ 36.360,00
    \item More than R\$ 36.360,00
    \item I don't know / Prefer
\end{enumerate}

(PT version) \noindent\textbf{(Renda)} Em qual das faixas abaixo estava a renda total da sua família no
mês passado, somando as rendas de todas as pessoas que moram com você, inclusive a
sua?
Nesta pergunta é possível assinalar somente uma alternativa.

\begin{enumerate}[label=\arabic*.]
    \item Até R\$ 1.212,00
    \item De R\$ 1.212,01 até R\$ 2.424,00
    \item De R\$ 2.242,01 até R\$ 3.636,00
    \item De R\$ 3.636,01 até R\$ 6.060,00
    \item De R\$ 6.060,01 até R\$ 12.120,00
    \item De R\$ 12.120,01 até R\$ 24.240,00
    \item De R\$ 24.240,01 até R\$ 36.360,00
    \item Mais de R\$ 36.360,00
    \item Não sei/ Prefiro não responder
\end{enumerate}

\subsection*{Section II. News Consumption}\hfill\\

\noindent\textbf{(News)} In general, how often would you say you read or watch news and get information on the following platforms?

\begin{enumerate}[label=\alph*.]
    \item Television news 
    \begin{enumerate}[label=]
        \item (- Never, Less often than once a month, Once a month, Once every 2 to 3 weeks, 1-2 days a week, 3-4 days a week, 5-6 days a week, Once a day, More often than once a day +)
    \end{enumerate}
    
    \item National newspapers 
    \begin{enumerate}[label=]
        \item (- Never, Less often than once a month, Once a month, Once every 2 to 3 weeks, 1-2 days a week, 3-4 days a week, 5-6 days a week, Once a day, More often than once a day +)
    \end{enumerate}
    
    \item Regional newspapers 
    \begin{enumerate}[label=]
        \item (- Never, Less often than once a month, Once a month, Once every 2 to 3 weeks, 1-2 days a week, 3-4 days a week, 5-6 days a week, Once a day, More often than once a day +)
    \end{enumerate}
    
    \item Radio news 
    \begin{enumerate}[label=]
        \item (- Never, Less often than once a month, Once a month, Once every 2 to 3 weeks, 1-2 days a week, 3-4 days a week, 5-6 days a week, Once a day, More often than once a day +)
    \end{enumerate}
    
    \item Online news sources (e.g., online newspaper, online magazines, etc.) 
    \begin{enumerate}[label=]
        \item (- Never, Less often than once a month, Once a month, Once every 2 to 3 weeks, 1-2 days a week, 3-4 days a week, 5-6 days a week, Once a day, More often than once a day +)
    \end{enumerate}

    \item News via social media 
    \begin{enumerate}[label=]
        \item (- Never, Less often than once a month, Once a month, Once every 2 to 3 weeks, 1-2 days a week, 3-4 days a week, 5-6 days a week, Once a day, More often than once a day +)
    \end{enumerate}  
\end{enumerate}

\noindent\textbf{(News Social Media)} In a typical week, which of the following sites or mobile apps do you use to find, read, watch, share, or comment news? (Please, select all that apply)

\begin{enumerate}[label=\arabic*.]
    \item WhatsApp
    \item YouTube
    \item Telegram
    \item Twitter
    \item Facebook
    \item Other (please specify)
\end{enumerate}

\noindent\textbf{(Campaign News)} Now let’s talk about your political news habits to gain knowledge about the Presidential election campaign. How often would you say you read or watched political and campaign news and information on the following platforms in the past week? (0 days - 7 days) 

\begin{enumerate}[label=\alph*.]
    \item Television news
    \item National newspapers
    \item Regional newspapers 
    \item Radio news 
    \item Online news sources (e.g., online newspaper, online magazines, etc.) 
    \item News via social media
\end{enumerate}

\noindent\textbf{(Social Campaign Information Engagement)} How often would you say you read or watched political information \& campaign information on the following social media platforms over the past week? (0 days - 7 days - Don’t know)\hfill\\

\noindent\textbf{(Social Campaign Information Sharing)} How often would you say you shared or liked (on Twitter: also retweet) political information \& campaign information on the following social media platforms over the past week? (0 days - 7 days - Don’t know)\hfill\\

\noindent\textbf{(Social Campaign Information Commenting)} How often would you say you commented or posted (on Twitter: tweet or reply) political information \& campaign information on the following social media platforms over the past week? (0 days - 7 days - Don’t know)\hfill\\

\textbf{Platforms:}
\begin{enumerate}[label=\arabic*.]
    \item Twitter
    \item YouTube
    \item Facebook
    \item Whatsapp
    \item Telegram
    \item Other (please specify)
\end{enumerate}

\noindent\textbf{(Election Information Received on WhatsApp/Telegram)} Have you received information about the election in your WhatsApp or Telegram groups from:

\begin{enumerate}[label=\arabic*.]
    \item Family and friends
    \item People I don't know very well (e.g., colleagues, acquaintances, neighbors)
    \item People I don't know personally
    \item I have not received information about the election in my WhatsApp or Telegram groups (97)
\end{enumerate}

\noindent\textbf{(Election Discussion Participation on WhatsApp/Telegram)} Have you participated in a discussion about the elections in your WhatsApp or Telegram groups with:

\begin{enumerate}[label=\arabic*.]
    \item Family and friends
    \item People I don't know very well (e.g., colleagues, acquaintances, neighbors)
    \item People I don't know personally
    \item I have not participated in a discussion about the election in my WhatsApp or Telegram groups (97)
\end{enumerate}

\noindent\textbf{(Public Political Information Group)} Do you belong to any public group that shares information about politics and/or the Presidential election (i.e., public groups are accounts anyone can join by using a URL link)? 

\begin{enumerate}[label=\arabic*.]
    \item Yes
    \item No
\end{enumerate}

\subsection*{Section III. Political Communication}\hfill\\

\noindent\textbf{(Online Political Discussion Fatigue)} Thinking about the posts you see on social media about politics, to what extent do you agree or disagree with the following statements? Strongly disagree (1) - Strongly agree (7)

\begin{enumerate}[label=\alph*.]
    \item I like seeing lots of posts and political discussions on social media.
    \item I am worn-out by how many political posts and discussions I see on social media.
    \item I don’t feel strongly about these discussions one way or another.
\end{enumerate}

\noindent\textbf{Conflict Orientation)} Please indicate your level of agreement with the following statements. Strongly disagree (1) - Strongly agree (7)

\begin{enumerate}[label=\alph*.]
    \item I enjoy challenging the opinions of others.
    \item I find conflicts exciting.
    \item I hate arguments.
    \item I feel upset after an argument.
    \item Arguments don’t bother me.
\end{enumerate}

\noindent\textbf{(Political Discussion Frequency)} In general, how often do you discuss politics with…

\begin{enumerate}[label=\alph*.]
    \item your family and/or friends?
    \item colleagues, acquaintances, and/or neighbors?
    \item people online whom I know well (e.g., on social media)?
    \item people online whom I don’t know well or I don’t know personally (e.g., on social media)?
\end{enumerate}

\noindent\textbf{(Encounter of Diverse Opinions)} When you talk to people in your surrounding about the 2022 Brazilian Presidential Election, how often do you encounter opinions that are NOT in line with your own opinion? Strongly disagree (1) - Strongly agree (7)

\begin{enumerate}[label=\alph*.]
    \item your family and/or friends
    \item colleagues, acquaintances, and/or neighbors
    \item People online whom I know well (e.g., on social media)?
    \item People online whom I don’t know well or I don’t know personally (e.g., on social media)?
\end{enumerate}

\subsection*{Section IV. Political Identification}\hfill\\

\noindent\textbf{(Party Affiliation)} Do you consider yourself to be close to any particular political party? If so, which party do you feel close to?

\begin{enumerate}[label=\arabic*.]
    \item MDB
    \item PT
    \item PDT
    \item Novo
    \item PL
    \item PCB
    \item PSTU
    \item Democracia Cristã
    \item Unidade Popular
    \item Pros
    \item PTB
    \item União Brasil
    \item Other party (please specify): \rule{5cm}{0.4pt}
    \item No, I do not feel close to any particular party.
\end{enumerate}

\noindent\textbf{(Degree of Closeness to Party)} How close do you feel to this party?

\begin{enumerate}[label=]
    \item (- Not very close ... Very close +)
\end{enumerate}

\noindent\textbf{(Political Position)} In political matters people talk of “the left” and “the right”. What is your position? (Please indicate your views using any number on a scale from 0 to 10, where 0 means “Very left-wing” and 10 means “Very right-wing”)

\begin{enumerate}[label=]
    \item (0 Very left-wing ... Very right-wing 10)
\end{enumerate}

\noindent\textbf{(Candidate Likeability)} To what extent do you like or dislike each of the following party candidates? (Please use the scale below, where 0 is ‘strongly dislike’ and 10 is ‘strongly like’. Rate just the leaders that you know)

\begin{enumerate}[label=]
    \item (0 Strongly dislike ... Strongly like 10)
\end{enumerate}

\textbf{Candidates:}
\begin{enumerate}[label=\alph*.]
    \item Simone Tebet (MDB)
    \item Lula (PT)
    \item Ciro Gomes (PDT)
    \item Felipe D’Ávila (Novo)
    \item Jair Bolsonaro (PL)
    \item Sofia Manzano (PCB)
    \item Vera Lúcia Salgado (PSTU)
    \item Constituinte Eymael (Democracia Cristã)
    \item Léo Péricles (Unidade Popular)
    \item Pablo Marçal (Pros)
    \item Padre Kelmon (PTB)
    \item Soraya Thronicke (União Brasil)
\end{enumerate}

\noindent\textbf{(Party Likeability)} To what extent do you like or dislike each of the following parties? (Please use the scale below, where 0 is ‘strongly dislike’ and 10 is ‘strongly like’. Rate just the parties that you know)

\begin{enumerate}[label=]
    \item (0 Strongly dislike ... Strongly like 10)
\end{enumerate}

\textbf{Parties:}
\begin{enumerate}[label=\alph*.]
    \item MDB
    \item PT
    \item PDT
    \item Novo
    \item PL
    \item PCB
    \item PSTU
    \item Democracia Cristã
    \item Unidade Popular
    \item Pros
    \item PTB
    \item União Brasil
\end{enumerate}

\subsection*{Section V. Political Engagement}\hfill\\

\noindent\textbf{(Civic and Political Engagement)} During the past 12 months, how often have you done any of the following (Never, rarely, from time to time, frequently, very often):

\textbf{Institutional/Electoral campaign}

\begin{enumerate}[label=\alph*)]
    \item Contacted an elected official (by letter, telephone or email)
    \item Donated money to a political party, a political organization, or a candidate running for public office
    \item Volunteered for a political party or campaign (like distributing leaflets)
    \item Participated in a political meeting
    \item Discussed social and political issues with others
    \item Encouraged others to take action about political issues
    \item Encouraged others to vote
\end{enumerate}

\textbf{Protest}

\begin{enumerate}[label=\alph*)]
    \item Signed a petition
    \item Participated in a march or street demonstration
    \item Refused to buy, or boycotted, certain products or services because of the social or political values of the company
    \item Joined unofficial strikes
\end{enumerate}

\textbf{Civic engagement}

\begin{enumerate}[label=\alph*)]
    \item Volunteered for a non-profit organization or charity (like environmental organization or Red Cross)
    \item Donated money to a non-profit or charity organization (like environmental organization or Red Cross)
\end{enumerate}

\textbf{Online participation}

\begin{enumerate}[label=\alph*)]
    \item Posted your own political opinion on social media
    \item Commented on a political post on social media
    \item Shared a political post on social media
    \item Followed a political party, a candidate or a politician on social media
    \item Changed your profile picture on social media to support a social cause or in response to a current event
\end{enumerate}

\noindent\textbf{(Interest in Politics)} Generally speaking, how interested are you in politics? 

\begin{enumerate}[label=]
    \item (- Not at all interested ... Very interested +) 
\end{enumerate}

\noindent\textbf{(Following Brazilian Politicians or Political Parties)} Do you regularly follow any Brazilian politician or political party on...

\textbf{Rows}

\begin{enumerate}[label=\arabic*.]
    \item No, I don’t follow any politicians or parties
    \item Yes, I follow 1 politician or party.
    \item Yes, I follow some politicians or parties.
    \item Yes, I follow many politicians or parties.
\end{enumerate}

\textbf{Platforms}

\begin{enumerate}[label=\alph*]
    \item Facebook
    \item Twitter
    \item Other social media (please specify)
\end{enumerate}

\noindent\textbf{(Type of Politicians or Political Parties Followed)} What type of politicians or political parties do you follow?

\begin{enumerate}[label=\arabic*.]
    \item Those whom I share similar views
    \item I follow political figures and parties with a diversity of views
    \item I do not share their views
\end{enumerate}

\subsection*{Section VI. Incivility Perception}\hfill\\

\noindent\textbf{(Platform-Specific Incivility)} Thinking about the posts and political discussions around the campaign and the Presidential election you witness these days on social media, how likely or unlikely would it be for you to come across the following messages on each of the following social media platforms? Scale – to + : highly unlikely / unlikely / somewhat unlikely / neither likely nor unlikely / somewhat likely / likely / highly likely

\begin{enumerate}[label=]
    \item (Exclusion) Messages that deny political or social groups the right to participate in politics.
    
    \begin{enumerate}[label=\alph*., align=left]
        \item Whatsapp
        \item YouTube
        \item Twitter
        \item Facebook
        \item Telegram
    \end{enumerate}
    
    \item (Impoliteness) Messages containing name-calling (such as “traitor”, “idiot”), offensive and/or pejorative language.
    
    \begin{enumerate}[label=\alph*., align=left]
        \item Whatsapp
        \item YouTube
        \item Twitter
        \item Facebook
        \item Telegram
    \end{enumerate}
    
    \item (Physical Harm/Violence) Messages that threaten others with physical harm or incite others to inflict harm to other individuals.
    
    \begin{enumerate}[label=\alph*., align=left]
        \item Whatsapp
        \item YouTube
        \item Twitter
        \item Facebook
        \item Telegram
    \end{enumerate}
    
    \item (Negativity) Messages that evoke negative emotions such as hatred anger or anxiety.
    
    \begin{enumerate}[label=\alph*., align=left]
        \item Whatsapp
        \item YouTube
        \item Twitter
        \item Facebook
        \item Telegram
    \end{enumerate}
    
    \item (Personal Attack) Verbal attacks to political opponents’ and their family members based on their physical characteristics, character or personal beliefs.
    
    \begin{enumerate}[label=\alph*., align=left]
        \item Whatsapp
        \item YouTube
        \item Twitter
        \item Facebook
        \item Telegram
    \end{enumerate}
    
    \item (Stereotype/Hate Speech/Discrimination) Messages negatively associating a person with a specific societal group by using offensive labels related to their sexuality, gender or race.
    
    \begin{enumerate}[label=\alph*., align=left]
        \item Whatsapp
        \item YouTube
        \item Twitter
        \item Facebook
        \item Telegram
    \end{enumerate}
    
    \item (Threat to Democratic Freedoms) Messages threatening democratic freedoms such as proposing to stage a coup to overthrow a democratically-elected government.
    
    \begin{enumerate}[label=\alph*., align=left]
        \item Whatsapp
        \item YouTube
        \item Twitter
        \item Facebook
        \item Telegram
    \end{enumerate}
\end{enumerate}

\noindent\textbf{(Online-Offline Incivility)} Compared to your political discussions or those discussions you witness outside the Internet, are the political discussions you engage in or witness online:

\begin{enumerate}[label=\alph*.]
    \item Very negative - Very positive
    \item Not respectful at all - Very respectful
    \item Not inclusive at all - Very inclusive
    \item Very angry - Not angry at all
    \item Very intolerant - Very tolerant
    \item Very uncivil - Very civil
\end{enumerate}

\noindent\textbf{(Online Political Discussion Experiences)} Considering your conversations about politics and the elections on social media and messaging apps, did you experience any of the following situations in the past month? Scale – to +: Never / Less than few times a month / A few times a month / A few times a week / Every day or almost every day

\begin{enumerate}[label=\alph*.]
    \item Was attacked or insulted by someone else during a political discussion
    \item Attacked or insulted someone during a political discussion
    \item Felt offended by someone else during a political discussion
    \item Witnessed a tense or otherwise uncomfortable political discussion
\end{enumerate}

\noindent\textbf{(Political Self-Censorship)} When people have been rude or attacked you online, how often have you responded by? [- Never, Rarely, Sometimes, Often, All of the time +]

\begin{enumerate}[label=\alph*.]
    \item Telling them to stop
    \item Continue the argument and articulate further
    \item Leaving the whole conversation
    \item Just ignoring the person who attacked you
    \item Refrain from participating in other political conversations and/or sharing your political views
    \item Making insulting or attacking comments back
    \item Unfriended, unfollowed, reported or blocked the person
\end{enumerate}

\subsection*{Section VII. Disinformation Perception}\hfill\\

\textbf{(Perception of Disinformation)} How often do you come across political news or information that you suspect is false? 

\begin{enumerate}[label=]
    \item (- Never, Rarely, Occasionally, Frequently, Very often +) 
\end{enumerate}

\begin{enumerate}[label=\alph*.]
    \item On Facebook
    \item On Twitter
    \item On YouTube 
    \item On Whatsapp
    \item On Telegram
    \item On news media (e.g., newspapers, TV, radio, news websites) 
\end{enumerate}

\subsection*{Section VIII. Authority Trust}\hfill\\

\noindent\textbf{(Trust in Institutions)} Please indicate your level of agreement with the following statements: 

\begin{enumerate}[label=]
    \item (1 Fully disagree ... Fully agree 7)
\end{enumerate}

\begin{enumerate}[label=\alph*.]
    \item I trust the parliament
    \item I trust politicians
    \item I trust political parties
    \item I trust the media
    \item I trust the legal system
    \item I trust the police
    \item I trust the military/Armed Forces
    \item I trust the government
\end{enumerate}

\subsection*{Section IX. Populism}\hfill\\

\noindent\textbf{(Populism)} Please indicate your level of agreement with the following statements.

\begin{enumerate}[label=]
    \item (1 Fully disagree ... Fully agree 7)
\end{enumerate}

\begin{enumerate}[label=\alph*.]
    \item Politicians are not really interested in what people like me think.
    \item Politicians make decisions that harm the interests of the ordinary people.
    \item The people, not the politicians or experts, should make our most important policy decisions.
    \item Economic forces and economic interest groups should be brought under greater control.
    \item For being truly Brazilian, it is important to have been born here.
    \item Achieving compromise among differing viewpoints is important in politics.
    \item The will of the majority should be exercised instead of constantly measuring the rights of minorities.
\end{enumerate}

\subsection*{Section X. Attitudes to Democracy}\hfill\\

\noindent\textbf{(Support for Democracy)} Please indicate how much you agree/disagree with the following statement:

\begin{enumerate}[label=]
        \item Democracy is the best system for a country like Brazil.
\end{enumerate}
    
\begin{enumerate}[label=]
    \item (1 Fully disagree ... Fully agree 7)
\end{enumerate}

\noindent\textbf{(Satisfaction with democracy)} On the whole, how satisfied or dissatisfied are you with the way democracy works in Brazil? Are you ...
\begin{enumerate}[label=\arabic*.]
    \item Not at all satisfied
    \item Not very satisfied
    \item Fairly satisfied
    \item Very satisfied
\end{enumerate}

\subsection*{Consent for data linkage}\hfill\\
Additionally, we asked the survey respondents to provide their Twitter handles given their consent:\hfill\\ 

\noindent\textbf{(Twitter handle)} Do you have a personal Twitter account?
\begin{enumerate}[label=\alph*.]
    \item Yes
    \item No
\end{enumerate}

\noindent\textbf{(If Yes)} We are interested in being able to linking people's answers to
this survey to the ways in which they use Twitter. This survey is part of a research project
about who and how people use Twitter conducted by a team of researchers at University of
Zurich.
Are you willing to provide us with your personal Twitter account and for this to be passed to
researchers at University of Zurich, along with your answers to this survey? Your Twitter
name would not be published. In compliance with Twitter terms of service, we will neither
provide any private user’s identifying information nor the full text of the tweets used.

\begin{enumerate}[label=\alph*.]
    \item Yes
    \item No
\end{enumerate}

\noindent\textbf{(If Yes)} Please enter your Twitter name here: Open Question (Maximum of 100 characters)\hfill\\

The survey participants are sampled by the survey company NetQuest using stratified quota sampling based on a soft quota of national statistics for age, gender, and geographical area (see \sitabref{tab:softquota}). The demographic distribution of the final survey participants is reported in \sitabref{tab:survey_demographics}. Results from Pearson’s chi-squared tests and Mann–Whitney U tests for age, gender, and geographical area indicate that the survey participants and the national electorate are drawn from the same distribution (see \sifigref{fig:sample validation_1}). We reject the null hypothesis when the p-value is smaller than 0.05. Additional demographic variables of the survey participants are presented in \sitabref{tab:survey_demographics}.

\section*{3. Identification of Political Influencers}

We use a heuristic strategy of identifying political influencers from the 57,645 accounts followed by survey respondents. We define political influencers as a composition of both ordinary citizens and celebrities (e.g., politicians, parties, media outlets, journalists, and ordinary opinion leaders) who satisfy two conditions: 1) have a comparatively large followers base and 2) are likely to produce political content \cite{khamis2017self, harff2023influencers}. According to this definition, we select political influencers in three steps. Firstly, we identify influential accounts that have a number of followers exceeding a certain threshold. We experiment with various thresholds and find that more generous thresholds can reveal significant multi-level clusters, but we also avoid using a threshold that is too low. We choose 1,000 as the ideal threshold for achieving this goal.
\sifigref{fig:community detection} (A) displays the Complementary Cumulative Distribution Function (CCDF) plots for the number of followers of accounts followed by survey respondents. We establish a threshold of 1,000 followers, and only accounts exceeding this threshold are retained, accounting for 63\% of the total accounts who are followed by survey respondents.

Second, from the accounts with more than 1,000 followers, we further select Brazilian accounts based on the location information displayed in their profile (Brazil or Brazilian cities). 

Third, we filter the accounts that might produce political information from the Brazilian influencers, covering categories of politicians, parties, media outlets, journalists, and ordinary opinion leaders. They are also popular categories used in other studies \cite{soares2018influencers, dubois2014multiple,flamino2023political}. We manually examine approximately 2,000 random profile examples and create a politically relevant keyword list based on these samples. Accounts potentially generating political content are identified by matching politics-related keywords in their profiles and supplemented by additional lists of politicians (based on 2022 presidential election candidates), parties, and media outlets (based on Digital News Report 2022 produced by Reuters Institute Oxford). See \sitabref{table:political_influencers} for more details.

The three steps result in identifying 2,307 Brazilian political influencers from the 57,645 followed accounts. Moreover, out of 271 individuals, 204 survey participants are found to follow political influencers. We examine the distribution of demographic variables among these 204 respondents and compare it to the demographic variables of 1,018 survey respondents (see \sifigref{fig:sample validation_2}). The Pearson's chi-squared test and Mann–Whitney U test results across all variables show that the null hypothesis of the same distribution between our sample and survey respondents cannot be rejected. Therefore, based on our evidence, this 204-person subsample does not introduce significant bias into either the overall group of survey respondents or the national electorate.

\section*{4. Multi-dimension Annotation of Identities of Political Influencers}

\begin{figure}[H]
\centering
\includegraphics[width=1.0\linewidth]{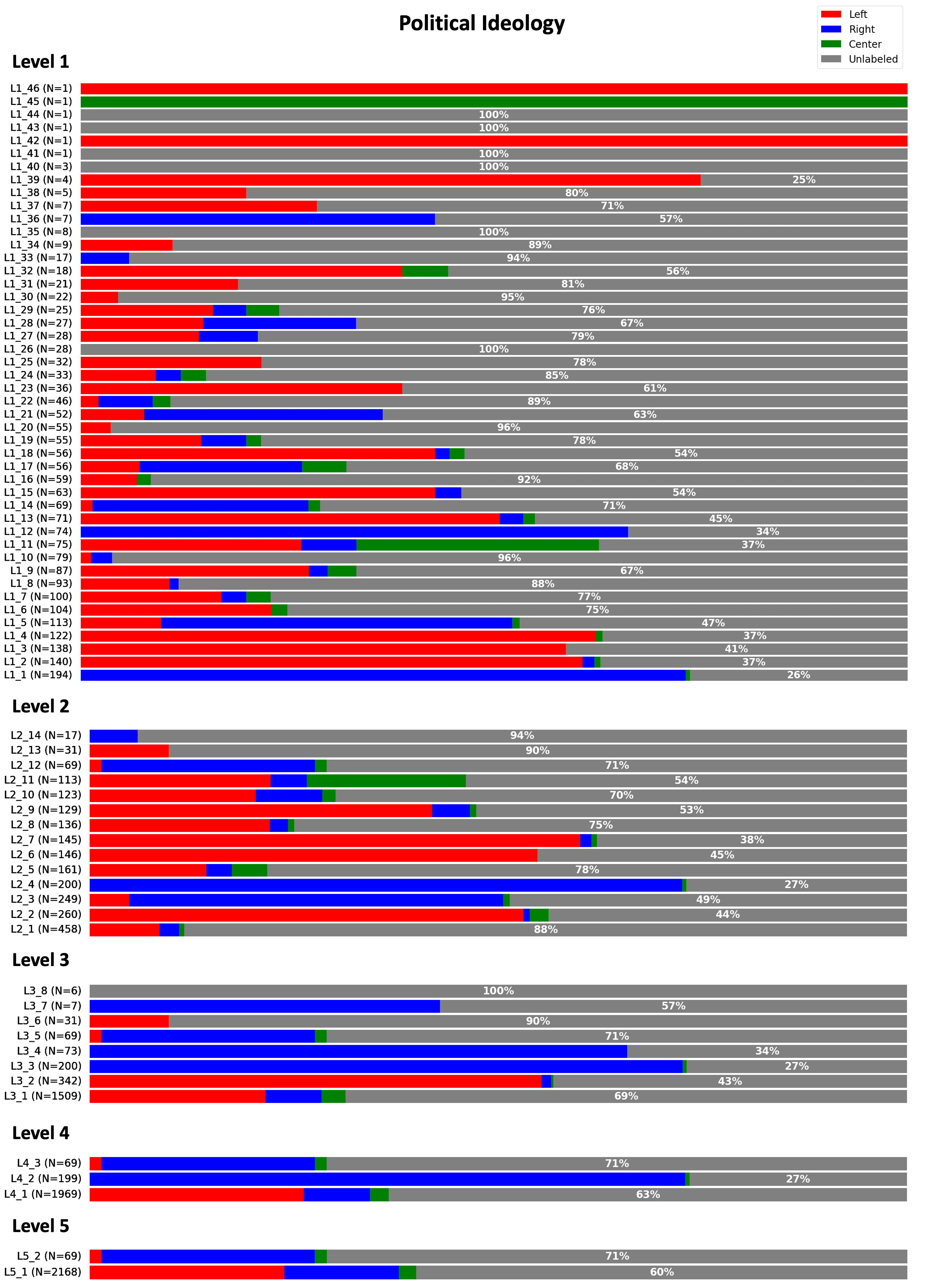}
\caption{Annotation of political ideology among political influencers in communities across five scale levels.}
\label{fig:annotation1}
\end{figure}

\begin{figure}[H]
\centering
\includegraphics[width=1.0\linewidth]{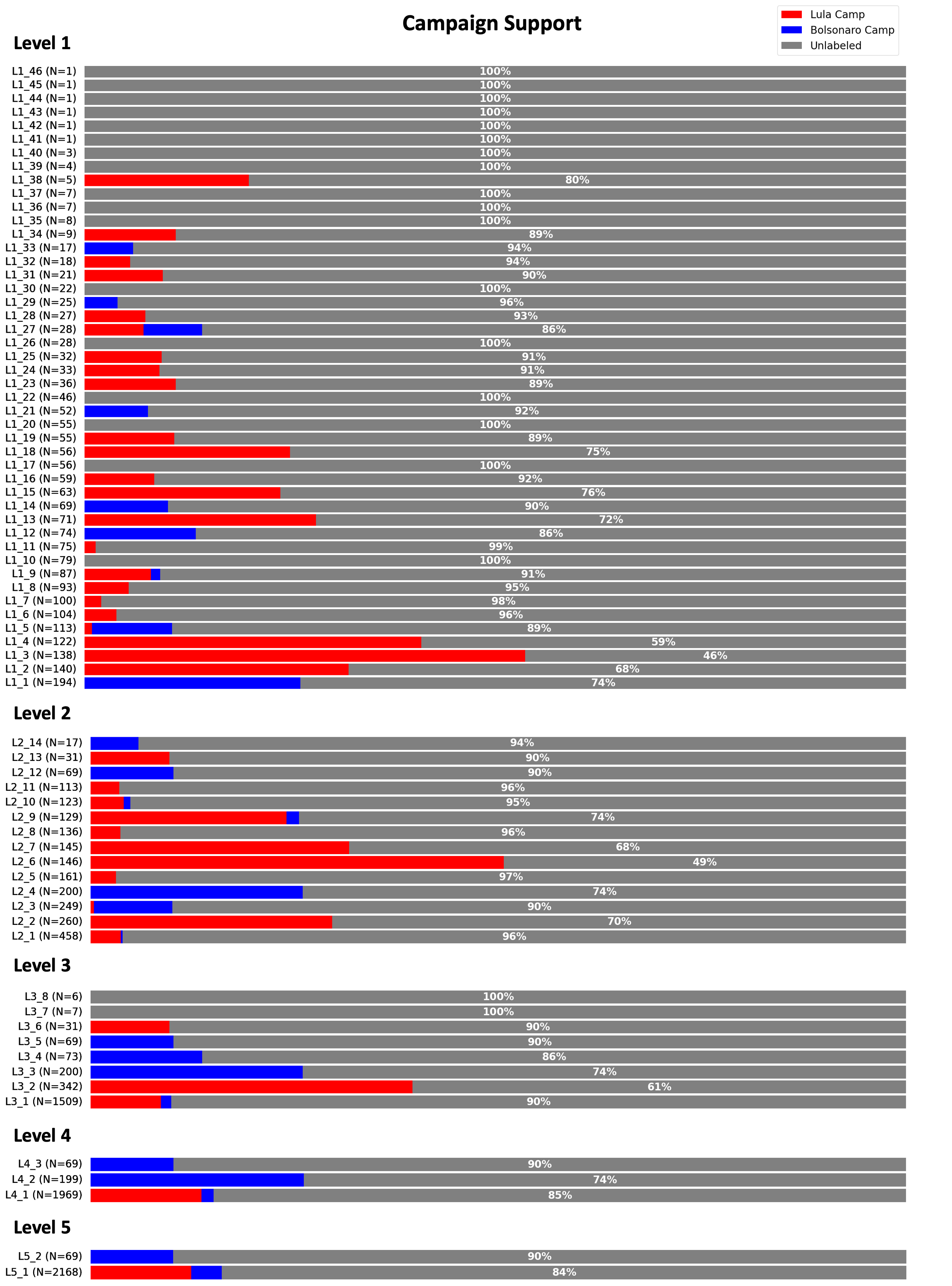}
\caption{Annotation of campaign support among political influencers in communities across five scale levels.}
\label{fig:annotation2}
\end{figure}

\begin{figure}[H]
\centering
\includegraphics[width=0.9\linewidth]{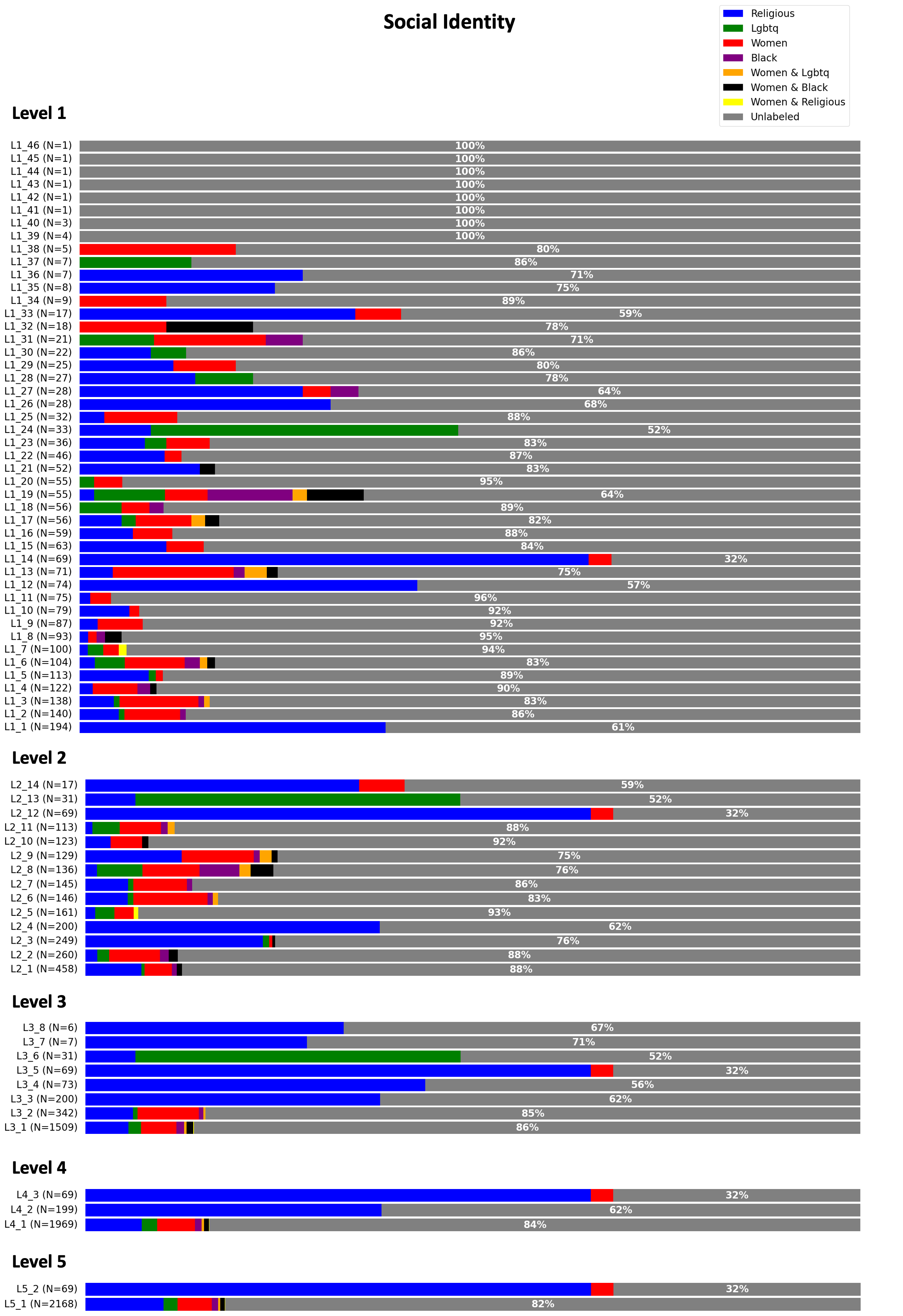}
\caption{Annotation of social identity among political influencers in communities across five scale levels.}
\label{fig:annotation3}
\end{figure}

\begin{figure}[H]
\centering
\includegraphics[width=0.9\linewidth]{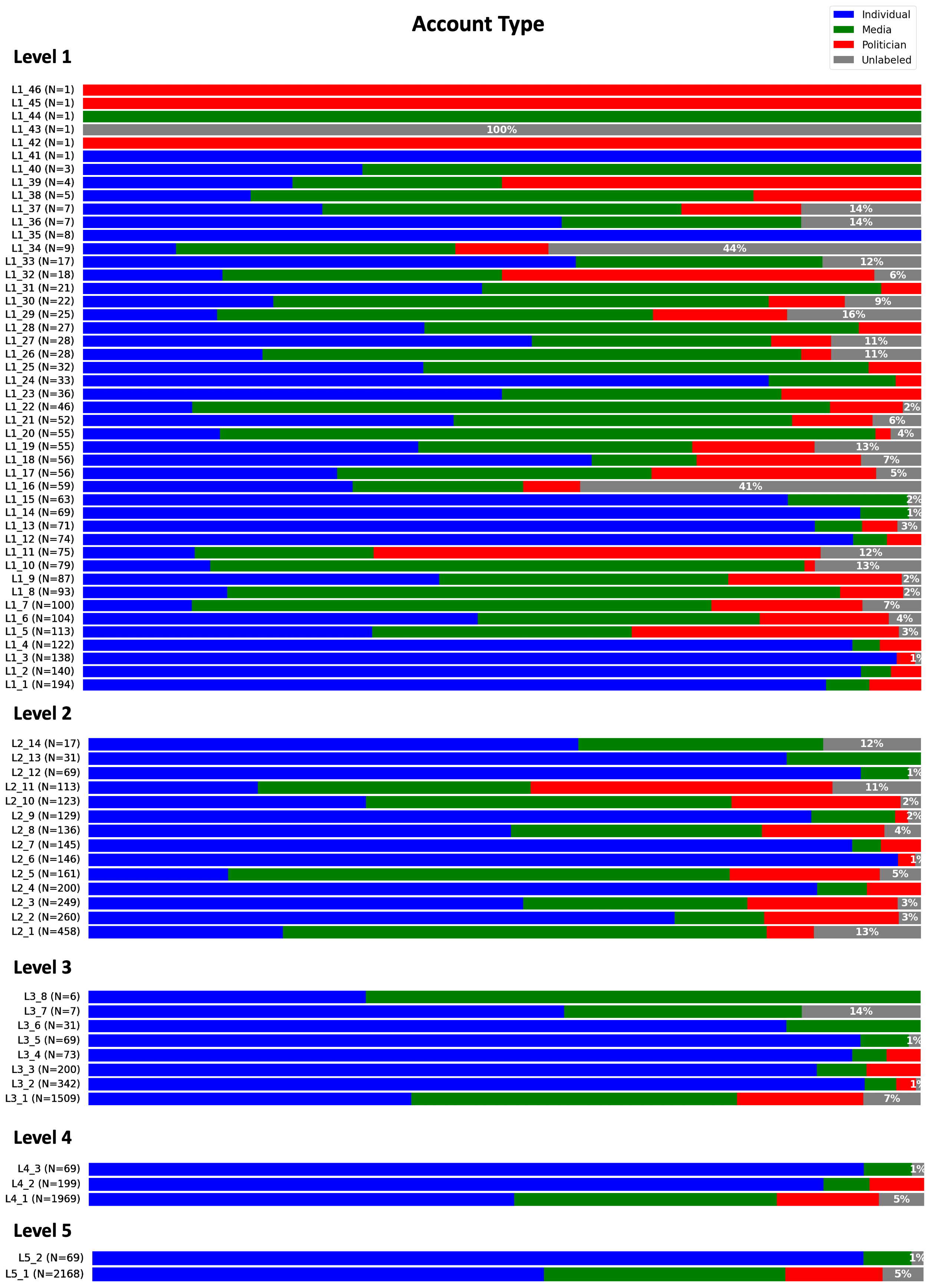}
\caption{Annotation of account type among political influencers in communities across five scale levels.}
\label{fig:annotation4}
\end{figure}

The annotation results on Political Ideology, Campaign Support, Social Identity, and Account Type for influencers in communities of all scale levels are displayed in \sifigref{fig:annotation1}, \sifigref{fig:annotation2}, \sifigref{fig:annotation3}, and \sifigref{fig:annotation4}.

\section*{5. Multi-scale Community Detection}

At the heart of the multi-scale community detection method proposed by \cite{arnaudon2023pygenstability} lies the generalized Markov Stability function, which quantifies the quality of a partition \( H \) of a graph \( G \) at a specific time scale \( t \). The goal is to find partitions that maximize this stability, indicating strong community structures.

The optimization problem is formulated as:

\[
H^*(t) = \arg\max_H Q_{\text{gen}}(t, H) = \arg\max_H \operatorname{Tr}\left[ H^\top \left( F(t) - \sum_{k=1}^m v_{2k-1} v_{2k}^\top \right) H \right]
\]

Where:

\begin{itemize}
  \item \( H \in \mathbb{R}^{N \times c} \) is the indicator matrix for the partitioning of \( N \) nodes into \( c \) communities.
  \item \( F(t) \in \mathbb{R}^{N \times N} \) is the node similarity matrix at time \( t \), capturing the probability of a random walker transitioning between nodes over time.
  \item \( \{v_k\}_{k=1}^{2m} \) are vectors defining the null model, representing expected connections in a randomized version of the graph.
\end{itemize}

This formulation ensures that the detected communities are not only cohesive but also statistically significant when compared to a null model.

The parameter \( t \) serves as a resolution factor:

\begin{itemize}
  \item Small \( t \): The random walker has limited time to move, leading to finer community structures.
  \item Large \( t \): The random walker explores more of the graph, revealing coarser community structures.
\end{itemize}

By analyzing the stability of partitions across different \( t \) values, it uncovers the multiscale community structure inherent in complex networks.

This method finds communities as groups of nodes that best retain the flow of a random walk process. Longer times used by the random walker correspond to larger scales. Optimal scales are found as minima of the variation of information of an ensemble partitions computed at each scale \cite{lambiotte2014random}. The process results in seven optimal scales in the Brazilian network. We discard the two first scales with 1767 and 1170 communities, respectively, as they have very low granularity, and more than 90 percent of the communities only include one node. Finally, we have five levels for analysis, with 46, 14, 8, 3, and 2 communities respectively.

For each scale $s$, the community detection yields a partition of the influencer network $\mathcal{G}^\mathcal{S}$ in $K_s$ non-overlapping communities \(\{C^s_i\}_{i=1}^{K_s}\).
By projecting the partitions on the consumer network $\mathcal{G}^\mathcal{C}$, we obtain, possibly overlapping, communities of consumers where two nodes are in the same community if they follow influencers belonging to the same community. 
This overlapping partitioning of the consumer network can also be seen as a hypergraph, where for each community $C_i^s$ at scale $s$ in the influencer network, corresponds a hyperedge, $E_i^s$ in the consumer network defined as the set of users following at least one influencer in $C_i^s$. At scale $s$, $\{E^s_i\}_{i=1}^{K_s}$ is therefore the set of $K_s$ hyperedges of the consumer network mapping the community partition $\{C^s_i\}_{i=1}^{K_s}$ of the influencer network.

The community detection in the influencer network is implemented with the Python package \textit{PyGenStability}\footnote{\url{https://barahona-research-group.github.io/PyGenStability/}}, which is designed for multi-scale community detection with Markov Stability \cite{arnaudon_algorithm_2024} and includes an automatic detection of significant scales. 

We run the code with the following parameters \textit{PyGenStability}:

\begin{verbatim}
    method = "leiden",  
    min_scale=-3,  
    max_scale=3,
    n_scale=1000,
    n_tries=100,
    constructor= "linearized", 
    n_workers=4, 
\end{verbatim}

For more information about the parameters, please refer to the documentation.

The result of \textit{PyGenStability} multi-scale community detection is shown in \sifigref{fig:community detection} (B). 
The Markov time $t$ (resolution parameter) is varied from $10^{-3}$ to $10^{3}$, with 1000 steps, and
robust partitions are detected as minima of the Normalized Variation of Information (NVI) of an ensemble of 100 partitions computed at each step. We obtain seven robust scales, and five are manually evaluated as effective partitions.

\section*{6. Evaluation of Measurements for Selective Exposure Indices}

\subsection*{Identity Diversity}
The index of Identity Diversity is calculated for each community and is based on the labels assigned to political influencers. As introduced in section 3, we annotate political influencers on four dimensions: Political Ideology, Campaign Support, Social Identity, and Account Type. We use Political Ideology labels as the main indicator of the political identity of political influencers since they are the most commonly used labels in politics. This approach also simplifies inferring the political identity of accounts with missing labels, as they can be assigned probabilities from the three options: \{Left, Right, Center\}. 

As we only label accounts that explicitly reveal their ideology on their profile, we have several accounts in each community that are unlabeled. We put forward two approaches to deal with the unlabeled values when calculating the diversity of Political Ideology of political influencers with the Gini-Simpson Index: First, 
we do not consider the unlabeled accounts and, second, we compute the Gini-Simpson Index, which is the probability of drawing two labels from a community that are different from one another, by assigning probabilities of political ideology labels to the unlabeled accounts equal to the proportions of political ideology labels in the community.
Using both strategies in the regression analysis did not significantly change the results.

\subsection*{Information Diversity}
\begin{figure}[H]
\centering
\includegraphics[width=1.0\linewidth]{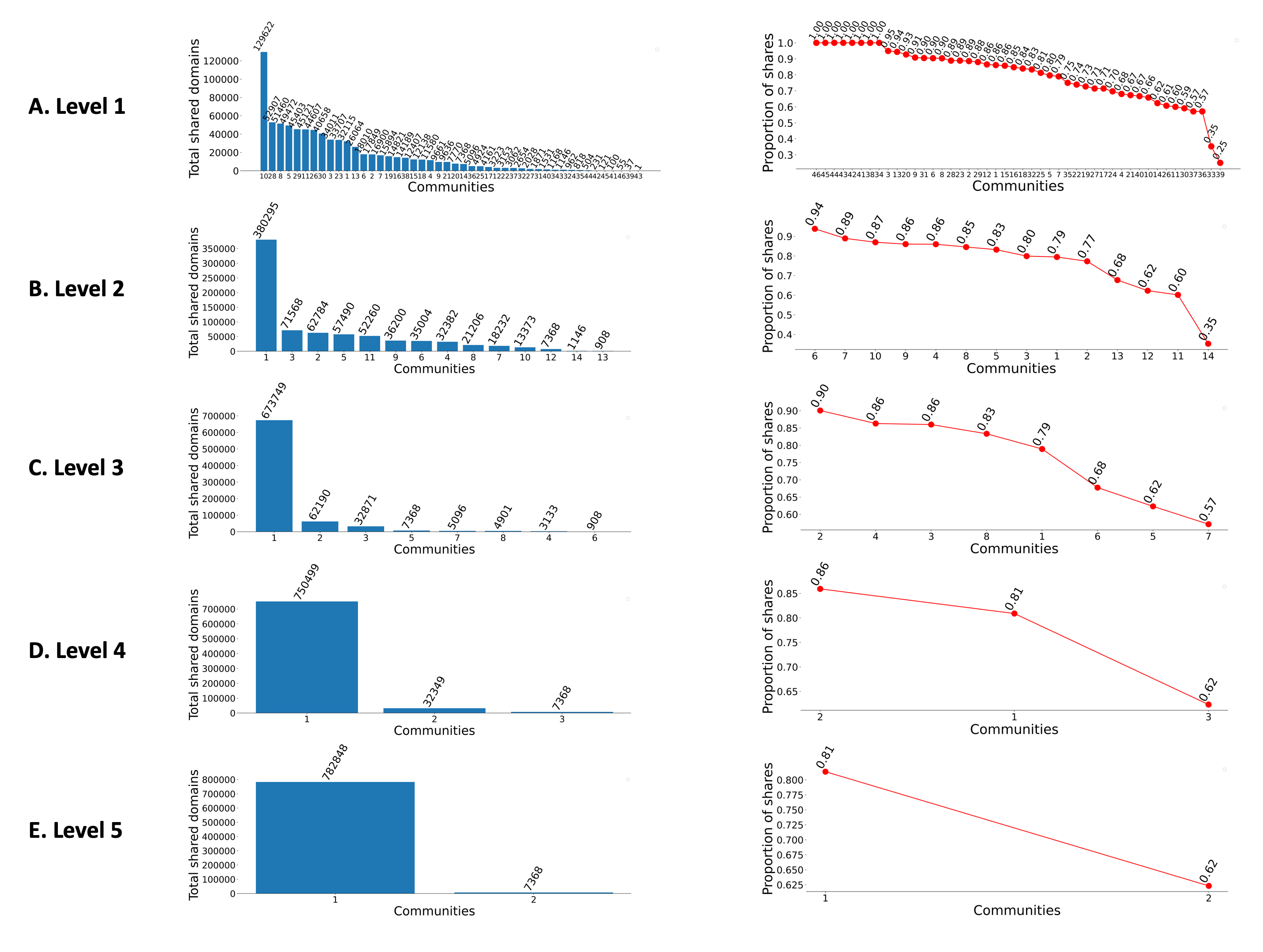}
\caption{The left column shows the total number of website domain links shared by political influencers in communities at (A) Level 1, (B) Level 2, (C) Level 3, (D) Level 4, (E) Level 5. The right column shows the proportion of website domain links shared by political influencers in communities at (A) Level 1, (B) Level 2, (C) Level 3, (D) Level 4, (E) Level 5.}
\label{fig:shared domains}
\end{figure}

The index of Information Diversity is calculated on each community and based on the website domain links posted by the political influencers. We unshorten the website links and extract the domain from the URLs with the Python package \textit{unshortenit}\footnote{\url{https://pypi.org/project/unshortenit/}}, and calculate the diversity of domains using the Gini-Simpson Diversity Index. Communities with fewer than 100 shared domain links or where less than 50\% of the influencers share domain links are excluded due to reliability concerns (see \sifigref{fig:shared domains}).

\section*{7. Statistical Regression Modeling}
\begin{figure}[H]
\centering
\includegraphics[width=1.0\linewidth]{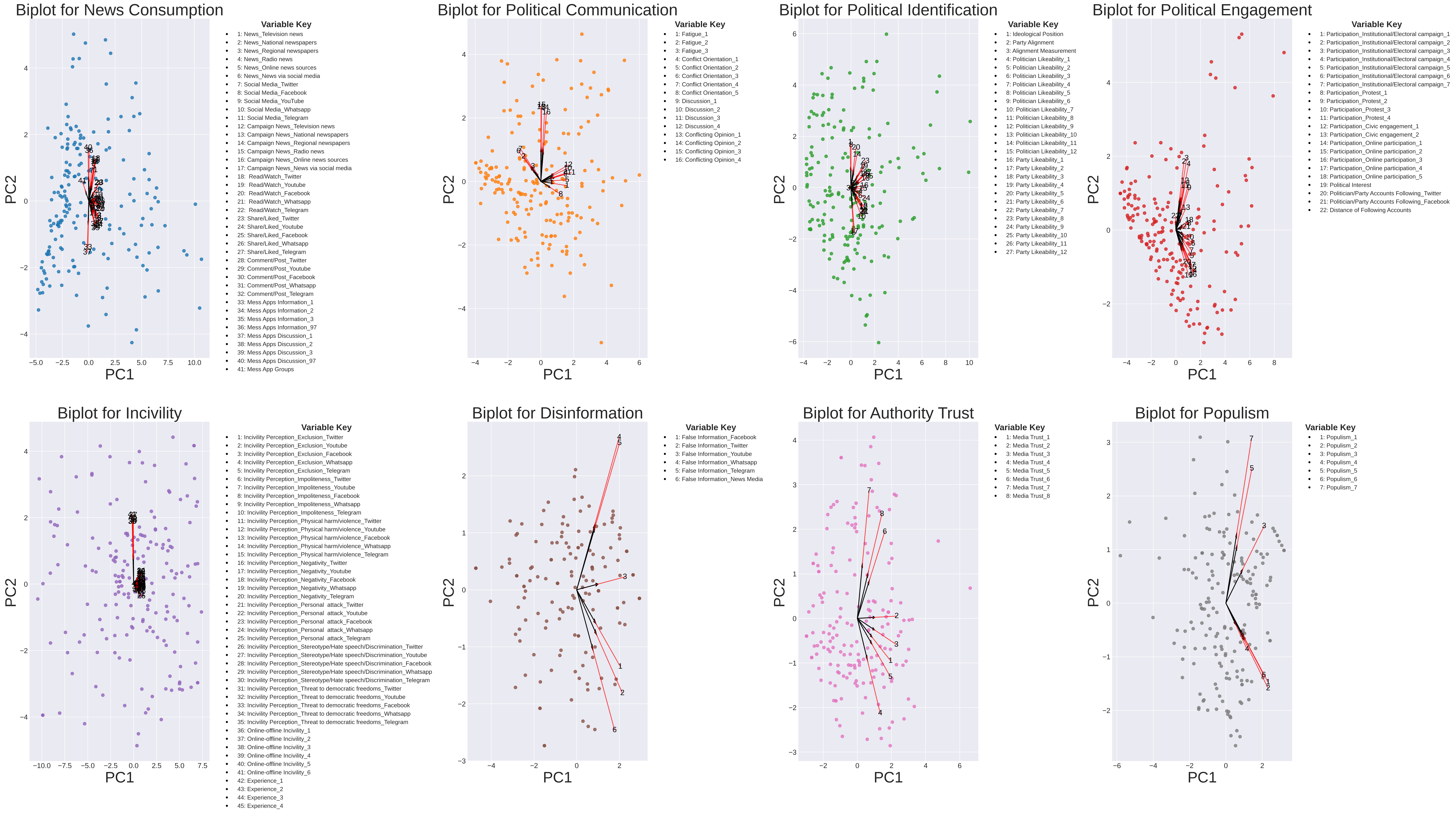}
\caption{Biplot of PCA projection of survey individuals' attribute variables. Clusters are shown within each pre-defined group: News Consumption, Political Communication, Political Identification, Political Engagement, Incivility, Disinformation, Authority Trust, and Populism.}
\label{fig:pca biplot}
\end{figure}

\begin{figure}[H]
\centering
\includegraphics[width=1.0\linewidth]{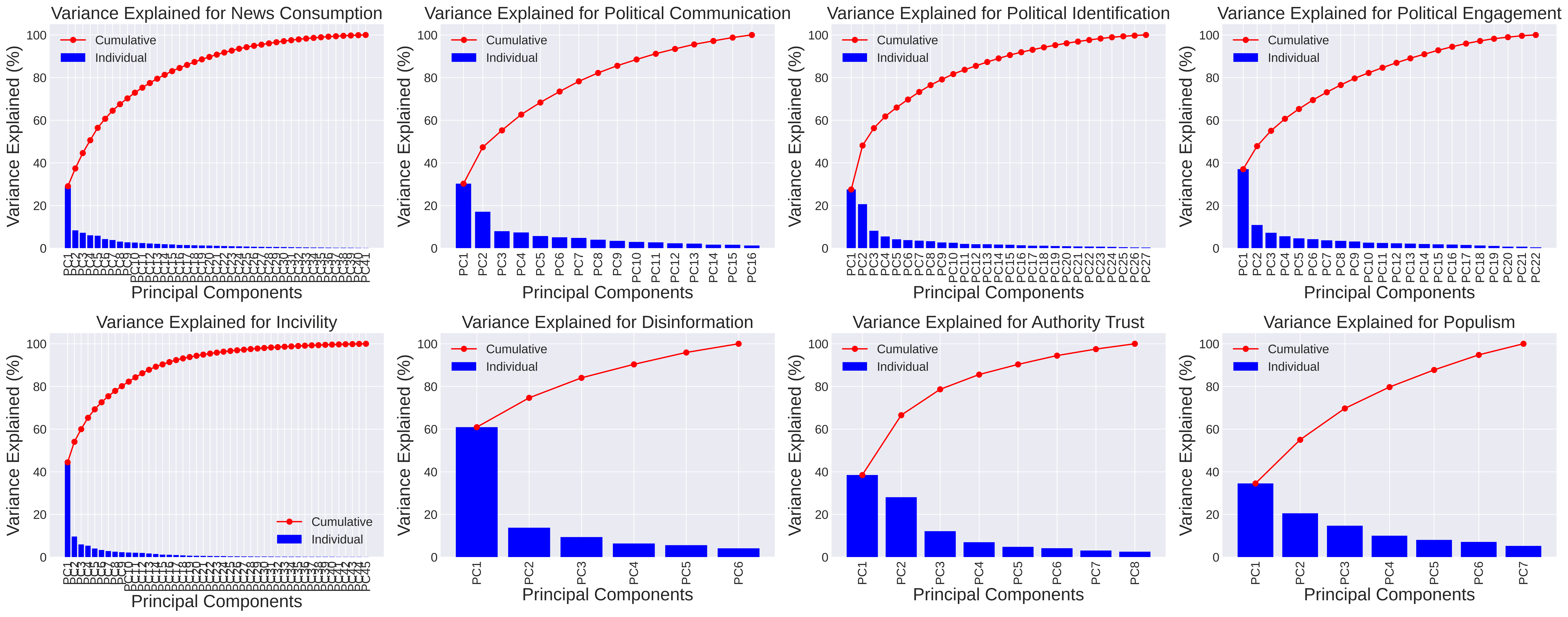}
\caption{Scree plot of PCA projection for survey individuals' attribute variables. The percentages of explained variance are displayed at different numbers of components.}
\label{fig:pca screeplot}
\end{figure}

The regression modeling between the individuals' attributes (independent variables) and selective exposure indices (dependent variables) consists of three steps:

First, we reduce the dimensions of the 189 independent variables obtained from ten thematic groups of survey questionnaires: Demographics, News Consumption, Political Communication, Political Identification, Political Engagement, Perceptions of Incivility, Perceptions of Disinformation, Authority Trust, Populism, and Attitudes toward Democracy (see section 2). PCA projection is implemented to observe the clusters or sub-dimensions formed by these variables. We define the sub-dimensions heuristically by combining the results from both the Biplot (\sifigref{fig:pca biplot}) and Screeplot (\sifigref{fig:pca screeplot}). All the sub-dimensions have been qualitatively interpreted (see \sitabref{table:survey attributes}). The variables included in each sub-dimension are shown in \sitabref{table:variable clusters}. We choose one representative variable from each sub-dimension for the following regression analysis. Finally, 25 variables are selected after the dimension reduction.

Second, as the values of Community Overlap are positive integers, we choose a zero-truncated negative binomial regression model, implemented with the \textit{vglm}\footnote{\url{https://www.rdocumentation.org/packages/VGAM/versions/1.1-11/topics/vglm}} package in R and the family is set with \textit{posnegbinomial}.
The other indexes take values in the real numbers between 0 and 1. We use a beta regression model implemented with the \textit{betareg} in R due to the flexibility of modeling diversity and concentration indices in the unit interval\cite{smithson2006better, cribari2010beta}. 
To deal with zeros and ones values in the indices, we utilize the following transformation approach from Ref. \cite{smithson2006better}, which transforms the variable $y\in[0,1]$ to $y'\in (0,1)$

\begin{equation}
y' = \frac{y(N - 1) + \frac{1}{2}}{N},
\label{eq:5}
\end{equation}
where \( N \) is the sample size.
We discard Level 5 because we only have values for 2 communities, and the communities detected at Level 5 have a similar pattern to Level 4.

\begin{figure}[H]
\centering
\includegraphics[width=1.0\linewidth]{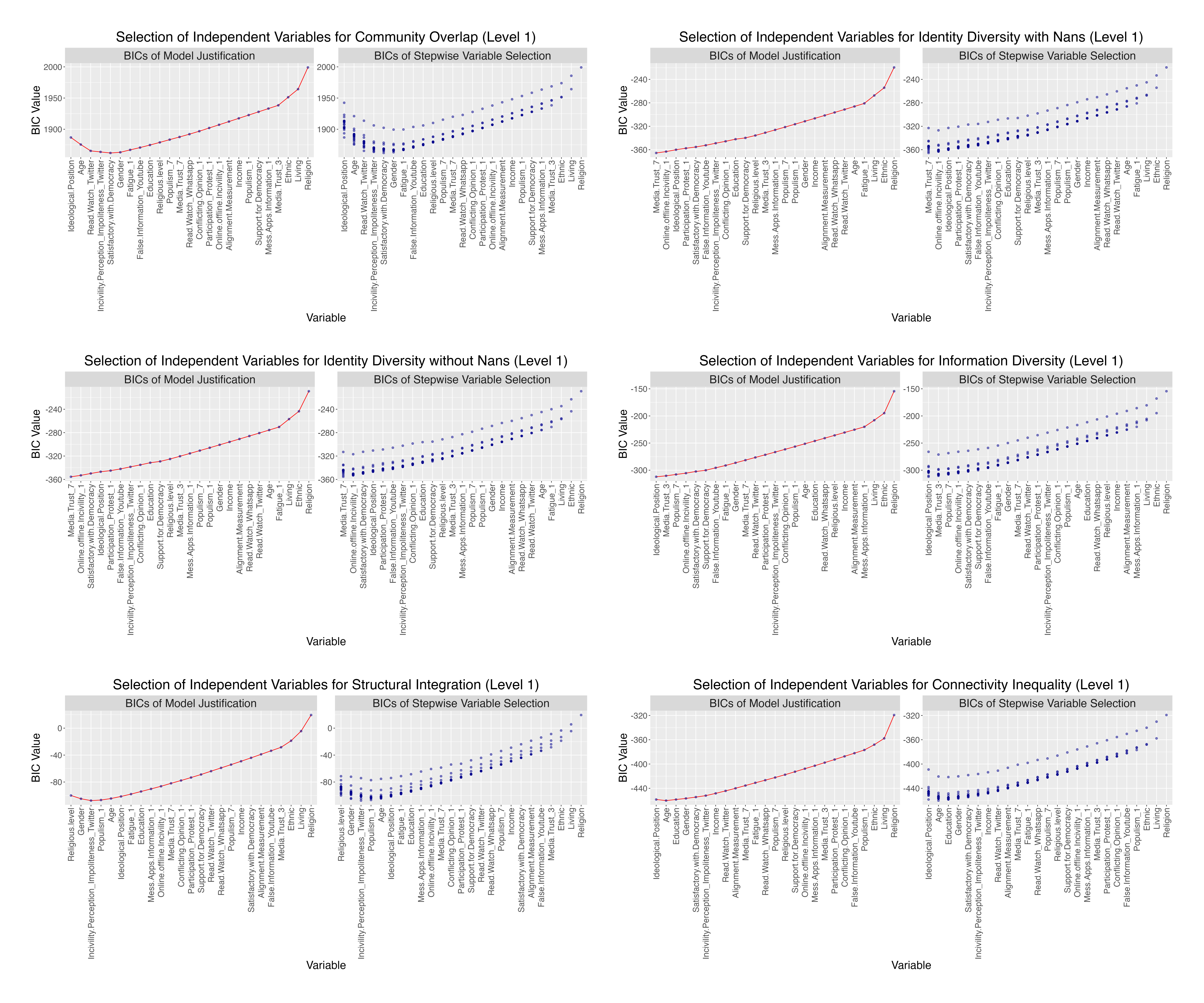}
\caption{BIC forward selection for regression models of six measurements: Community Overlap, Identity Diversity with unlabeled accounts, Identity Diversity without unlabeled accounts, Information Diversity, Structural Integration, and Connectivity Inequality, at scale level 1.}
\label{fig:bic_l1}
\end{figure}

\begin{figure}[H]
\centering
\includegraphics[width=1.0\linewidth]{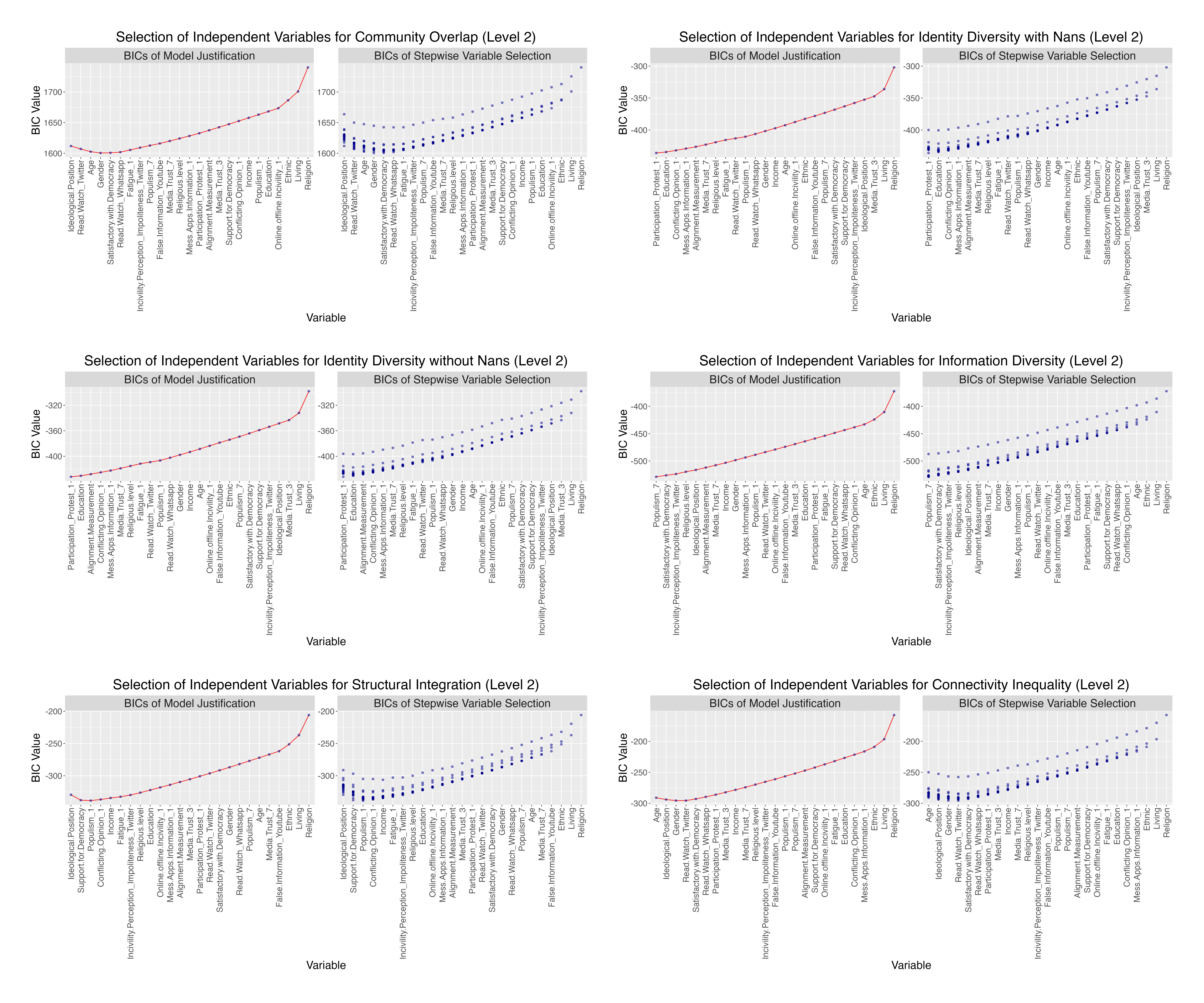}
\caption{BIC forward selection for regression models of six measurements: Community Overlap, Identity Diversity with unlabeled accounts, Identity Diversity without unlabeled accounts, Information Diversity, Structural Integration, and Connectivity Inequality, at scale level 2.}
\label{fig:bic_l2}
\end{figure}

\begin{figure}[H]
\centering
\includegraphics[width=1.0\linewidth]{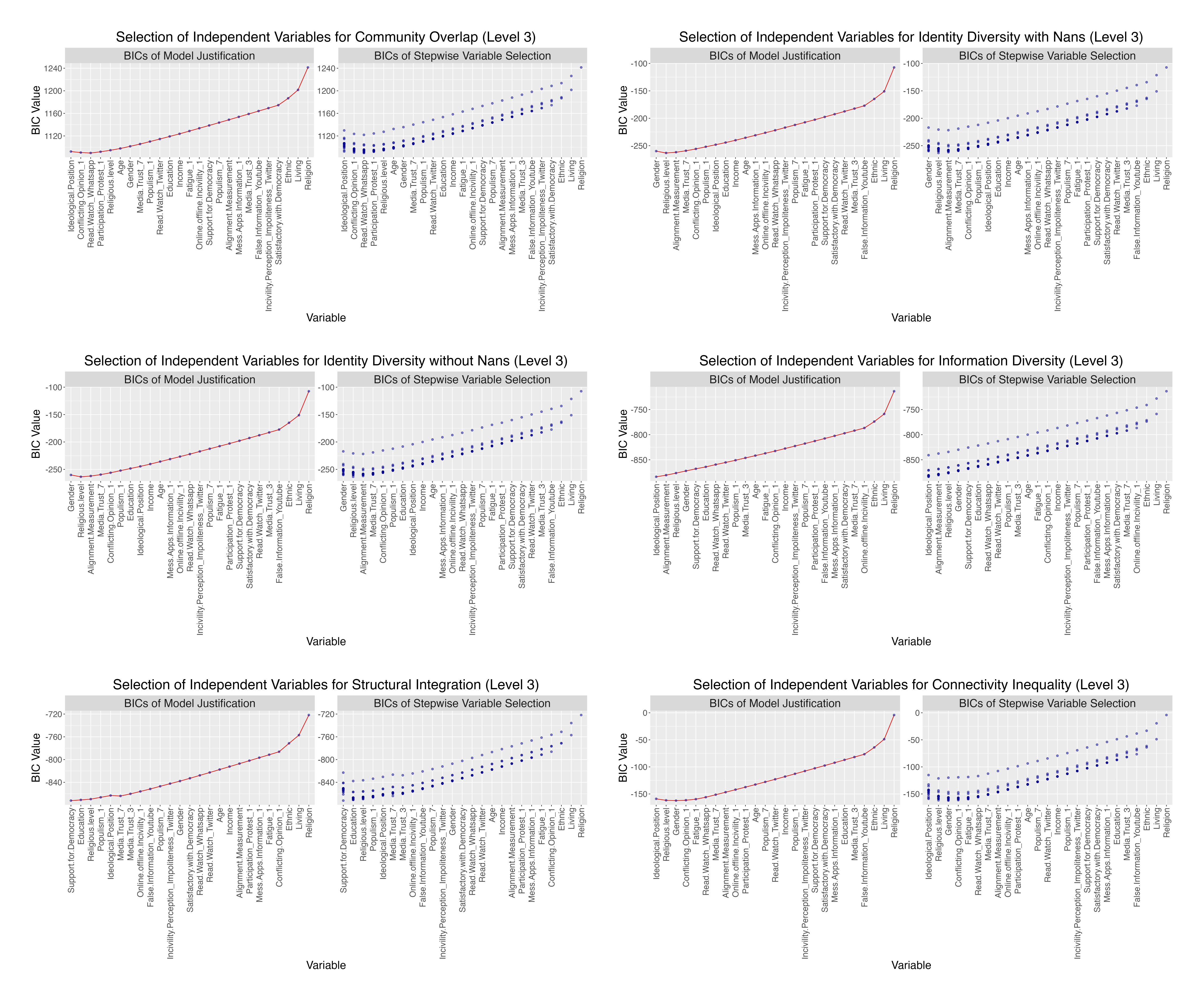}
\caption{BIC forward selection for regression models of six measurements: Community Overlap, Identity Diversity with unlabeled accounts, Identity Diversity without unlabeled accounts, Information Diversity, Structural Integration, and Connectivity Inequality, at scale level 3.}
\label{fig:bic_l3}
\end{figure}

\begin{figure}[H]
\centering
\includegraphics[width=1.0\linewidth]{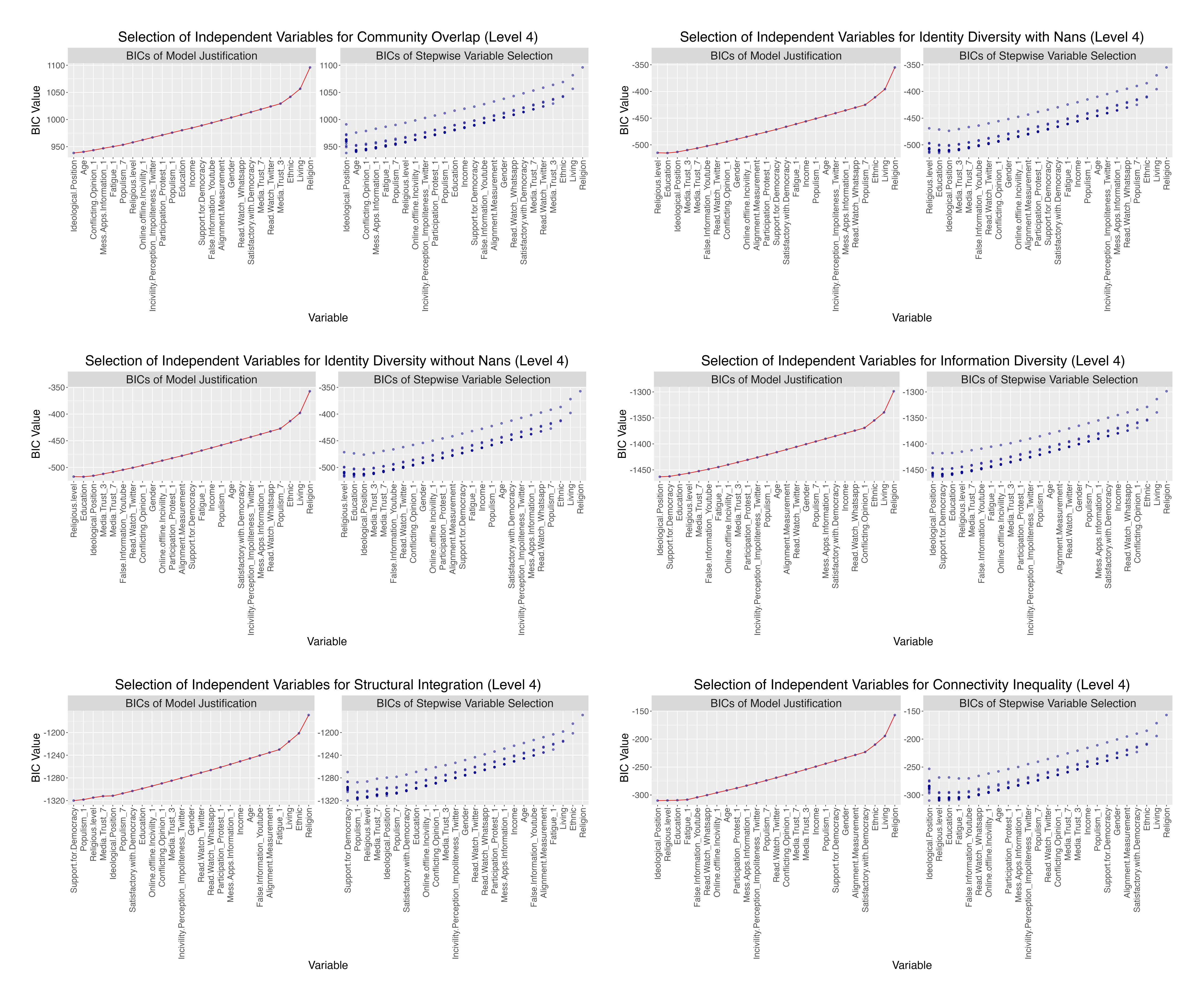}
\caption{BIC forward selection for regression models of six measurements: Community Overlap, Identity Diversity with unlabeled accounts, Identity Diversity without unlabeled accounts, Information Diversity, Structural Integration, and Connectivity Inequality, at scale level 4.}
\label{fig:bic_l4}
\end{figure}

Third, we select the optimal regression models using Bayesian Information Criterion (BIC) forward selection with the reduced 25 variables. BIC forward selection is a statistical method used for model selection in regression analysis. Forward selection is a greedy optimization approach to find the best subset of predictors that results in the most parsimonious model with a good balance between fit and complexity. The lower the BIC values, the better the model fits the data. The rationale of forward selection is that we start with no predictors in the null model and add variables individually based on their contribution to reducing the BIC. The procedure stops when adding any remaining predictor does not lower the BIC value. The BIC evaluation for regression models of the five selective exposure indices (including six measurements because we use two approaches for the measurement of Identity Diversity) from Level 1 to Level 4 can be found in Figs. \ref{fig:bic_l1}, \ref{fig:bic_l2}, \ref{fig:bic_l3} \& \ref{fig:bic_l4}. We only keep the variables that decrease BIC values in the final regression models.

\begin{figure}[H]
\centering
\includegraphics[width=1.0\linewidth]{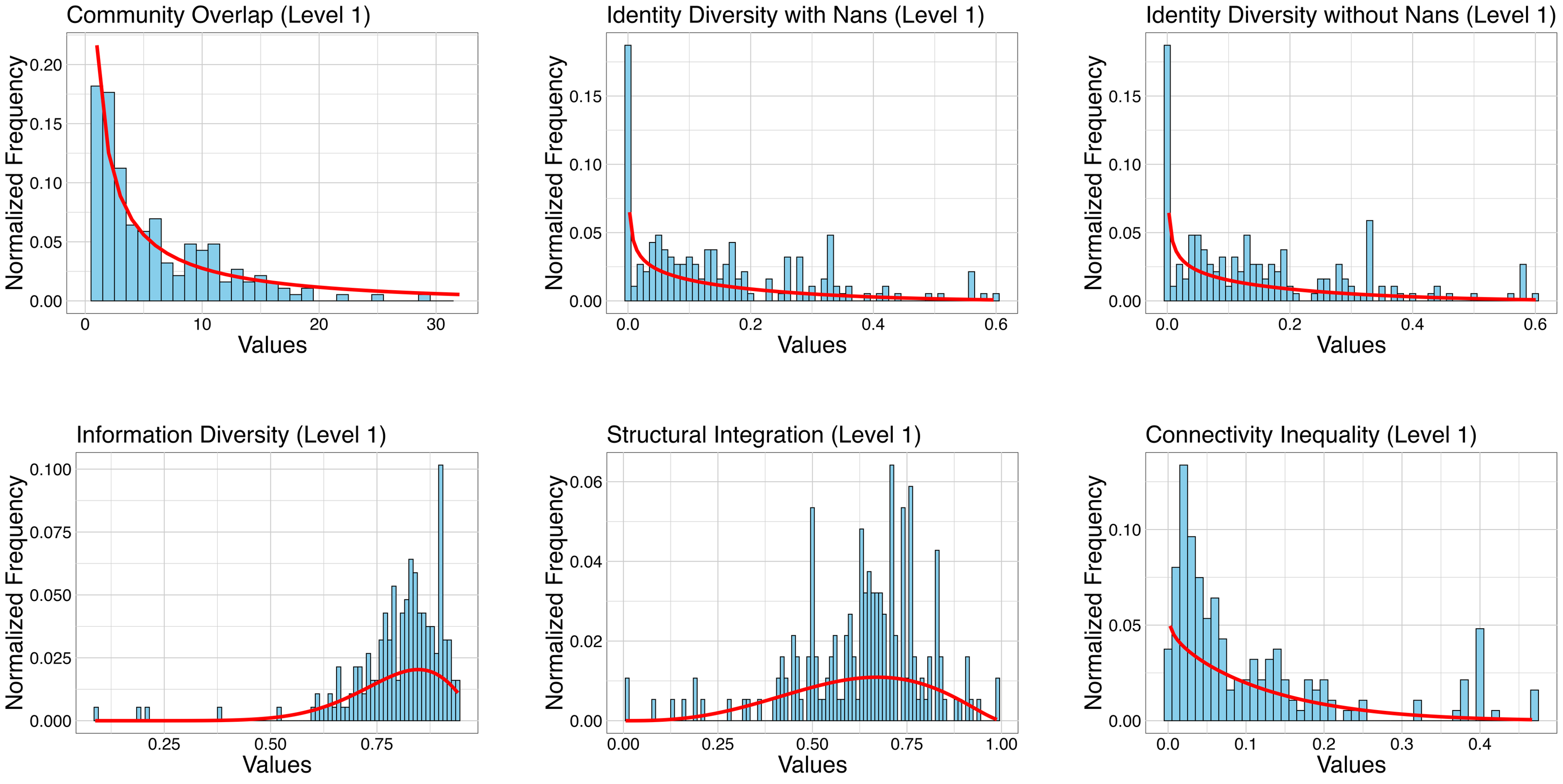}
\caption{Comparison of data distributions between original values and estimated values of the six measurements: Community Overlap, Identity Diversity with unlabeled accounts, Identity Diversity without unlabeled accounts, Information Diversity, Structural Integration, and Connectivity Inequality, at scale level 1.}
\label{fig:distribution_l1}
\end{figure}

\begin{figure}[H]
\centering
\includegraphics[width=1.0\linewidth]{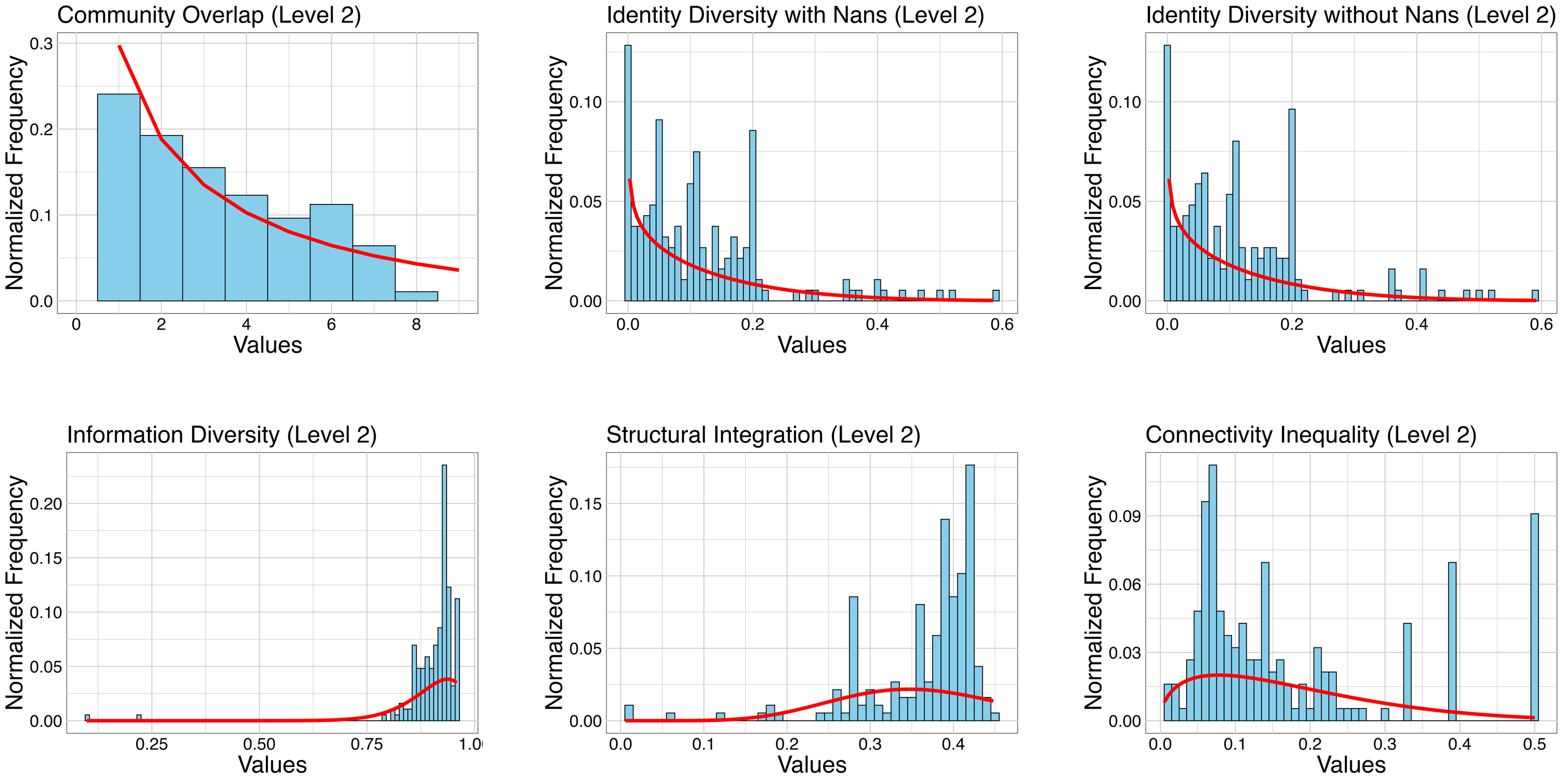}
\caption{Comparison of data distributions between original values and estimated values of the six measurements: Community Overlap, Identity Diversity with unlabeled accounts, Identity Diversity without unlabeled accounts, Information Diversity, Structural Integration, and Connectivity Inequality, at scale level 2.}
\label{fig:distribution_l2}
\end{figure}

\begin{figure}[H]
\centering
\includegraphics[width=1.0\linewidth]{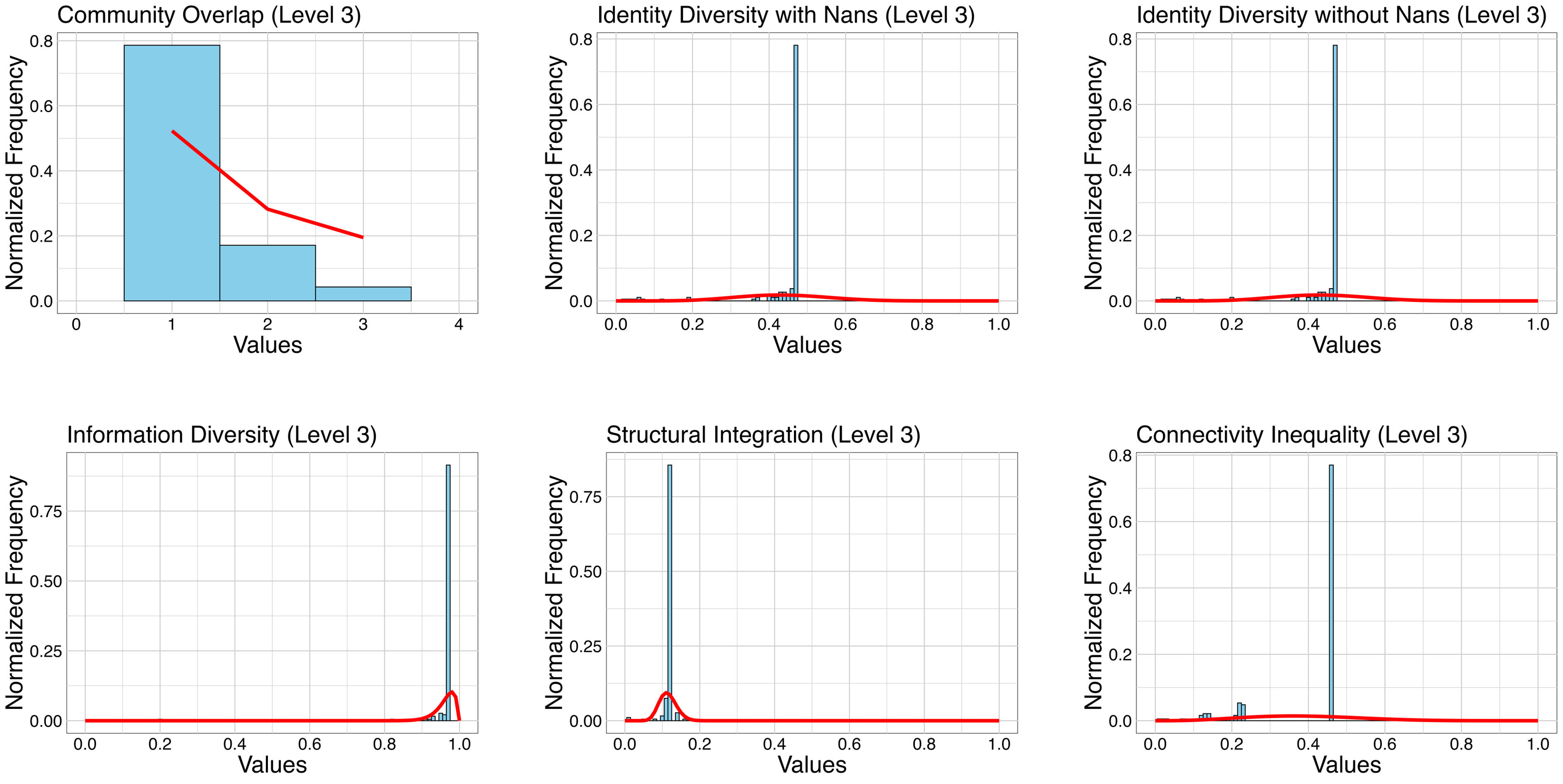}
\caption{Comparison of data distributions between original values and estimated values of the six measurements: Community Overlap, Identity Diversity with unlabeled accounts, Identity Diversity without unlabeled accounts, Information Diversity, Structural Integration, and Connectivity Inequality, at scale level 3.}
\label{fig:distribution_l3}
\end{figure}

\begin{figure}[H]
\centering
\includegraphics[width=1.0\linewidth]{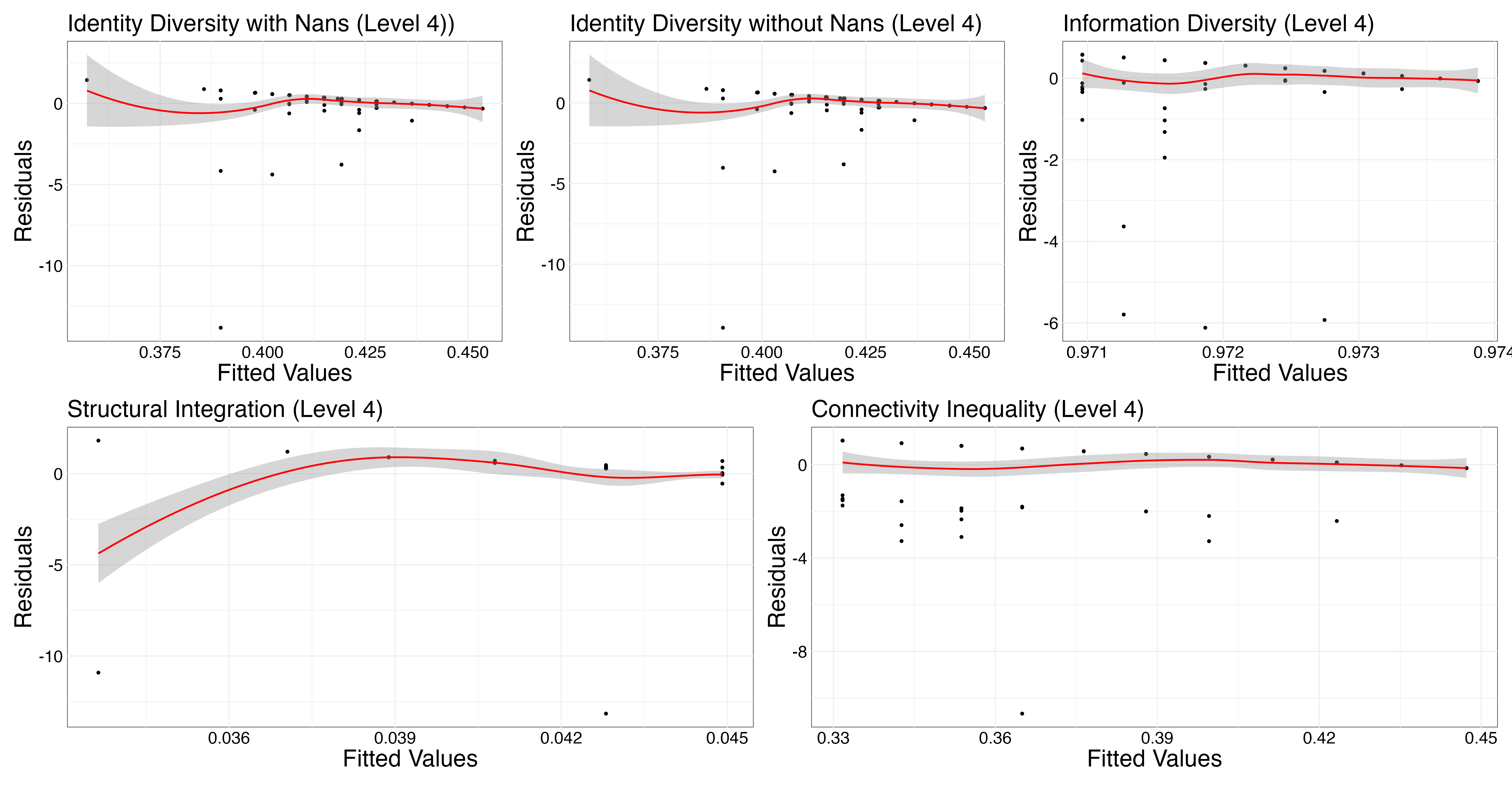}
\caption{Residual check of Beta regression models of five measurements: Identity Diversity with unlabeled accounts, Identity Diversity without unlabeled accounts, Information Diversity, Structural Integration, and Connectivity Inequality, at scale level 4.}
\label{fig:residual_l4}
\end{figure}

To validate the model fit, the distribution of predicted values obtained from the regression models are compared with the actual distribution of values of the six measurements (see \sifigref{fig:distribution_l1}, \sifigref{fig:distribution_l2}, \sifigref{fig:distribution_l3}, \sifigref{fig:residual_l4}). The alignment of fitted distributions and the real value distributions demonstrate a good model fit.

\begin{figure}[H]
\centering
\includegraphics[width=1.0\linewidth]{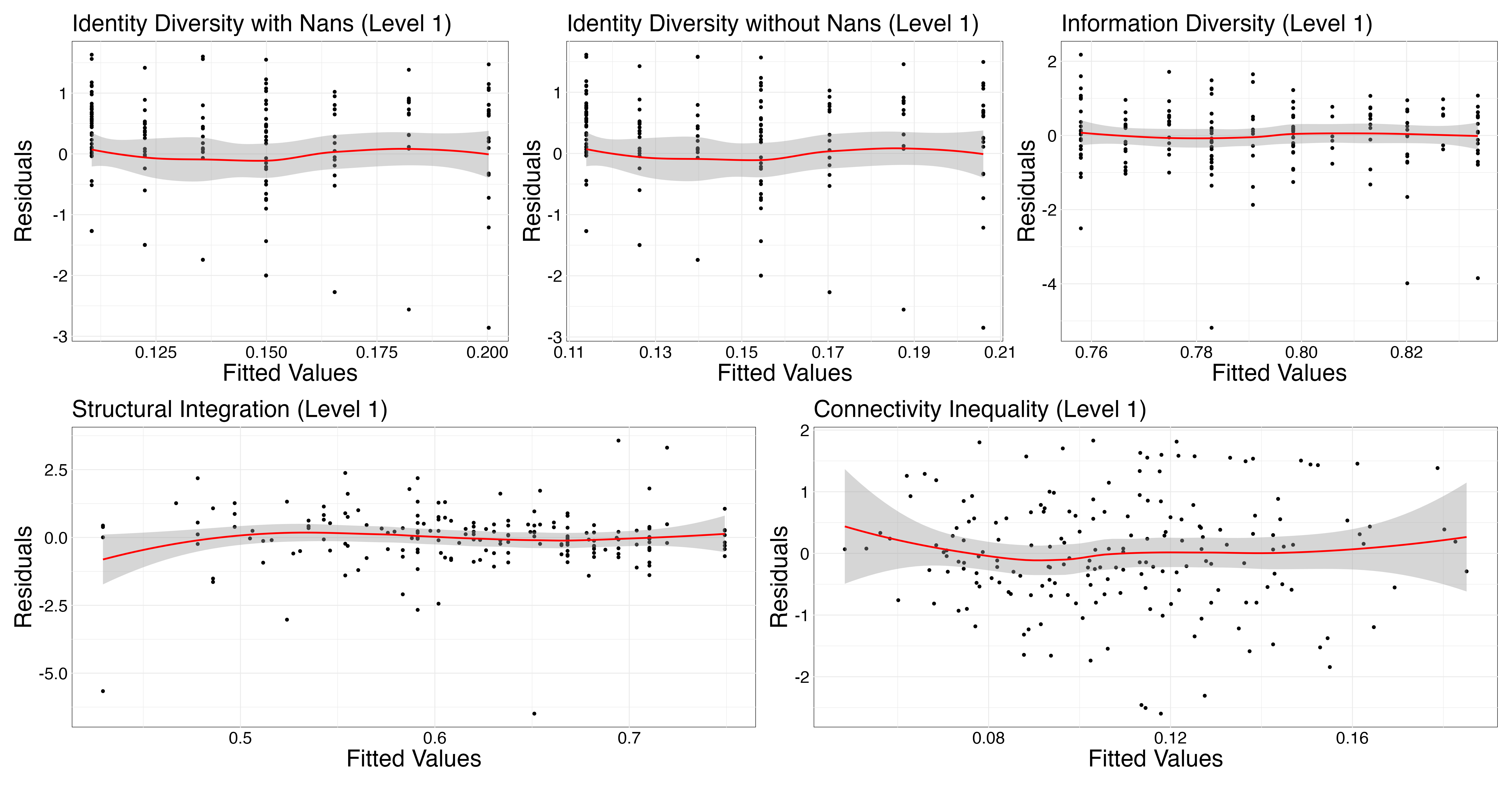}
\caption{Residual check of Beta regression models of five measurements: Identity Diversity with unlabeled accounts, Identity Diversity without unlabeled accounts, Information Diversity, Structural Integration, and Connectivity Inequality, at scale level 1.}
\label{fig:residual_l1}
\end{figure}

\begin{figure}[H]
\centering
\includegraphics[width=1.0\linewidth]{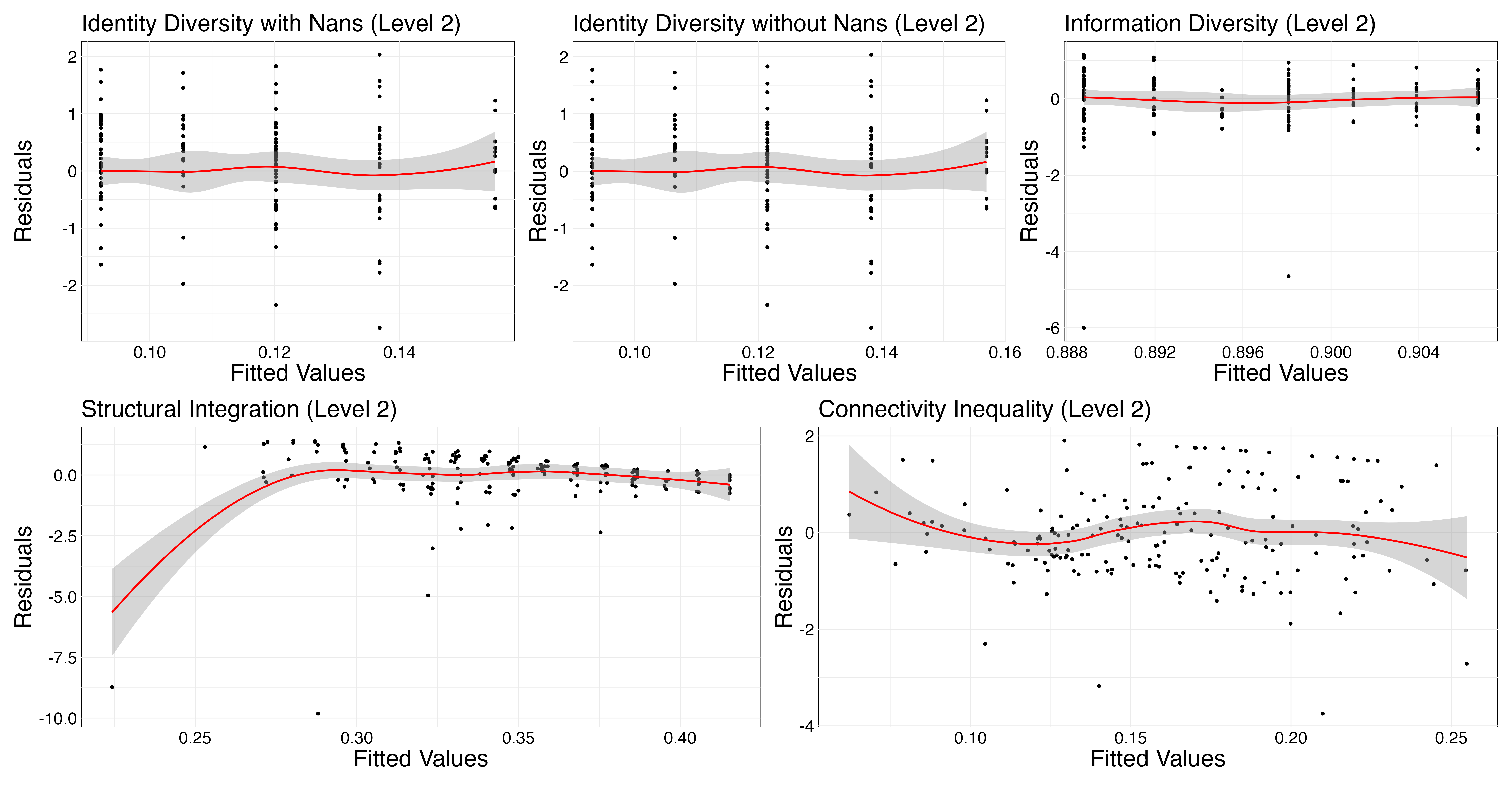}
\caption{Residual check of Beta regression models of five measurements: Identity Diversity with unlabeled accounts, Identity Diversity without unlabeled accounts, Information Diversity, Structural Integration, and Connectivity Inequality, at scale level 2.}
\label{fig:residual_l2}
\end{figure}

\begin{figure}[H]
\centering
\includegraphics[width=1.0\linewidth]{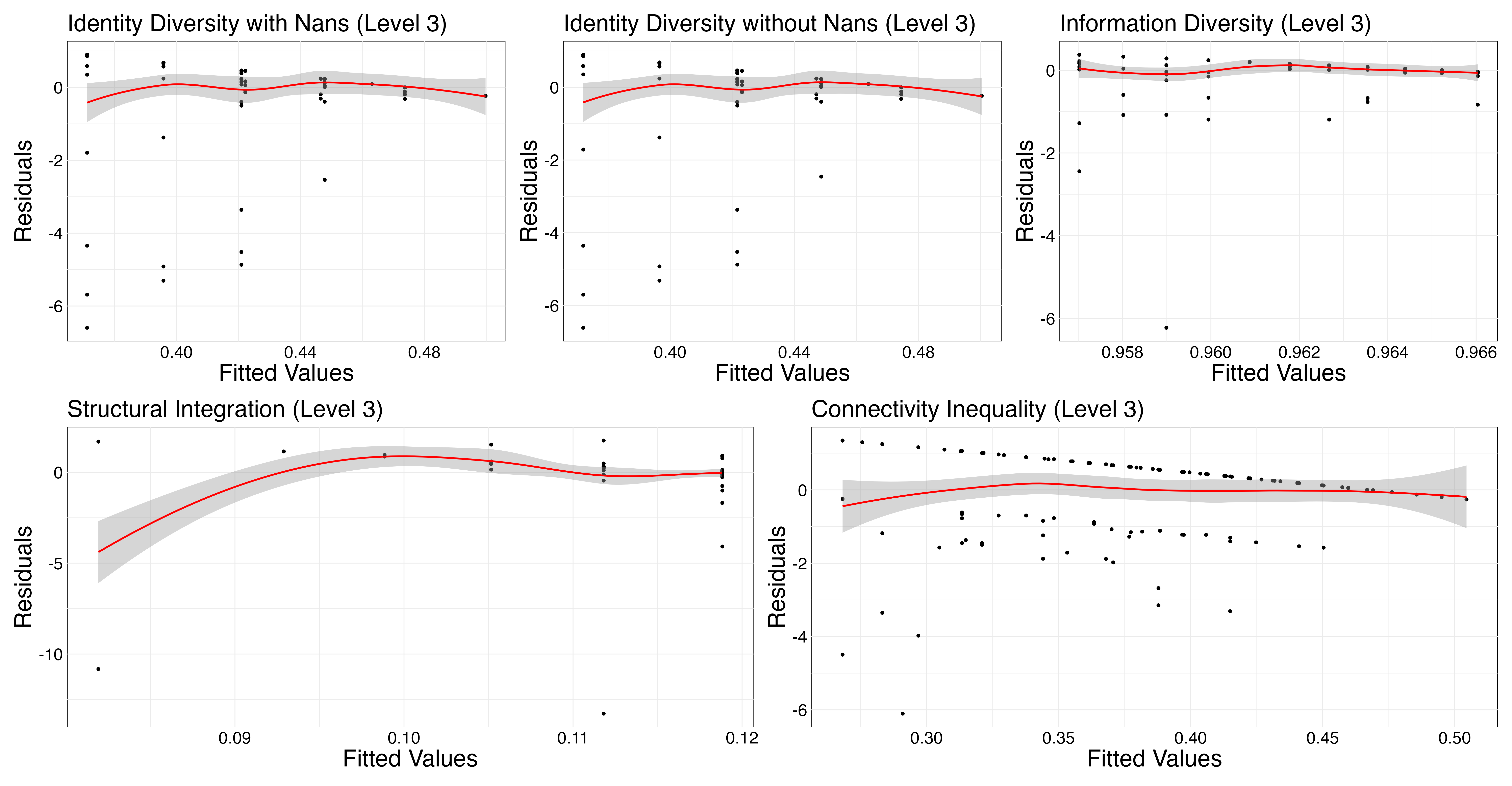}
\caption{Residual check of Beta regression models of five measurements: Identity Diversity with unlabeled accounts, Identity Diversity without unlabeled accounts, Information Diversity, Structural Integration, and Connectivity Inequality, at scale level 3.}
\label{fig:residual_l3}
\end{figure}

\begin{figure}[H]
\centering
\includegraphics[width=1.0\linewidth]{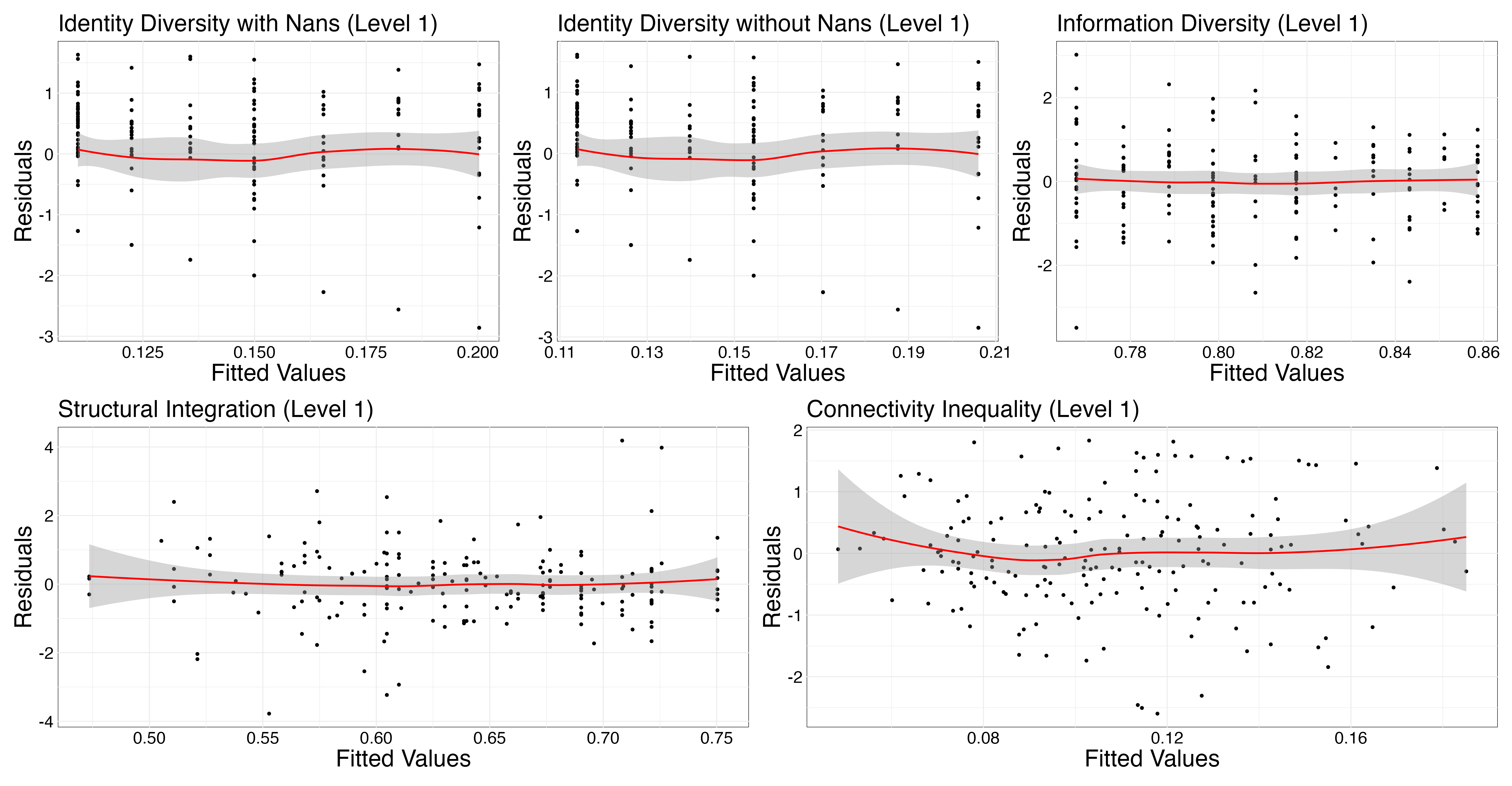}
\caption{Residual check of Beta regression models after removing outliers of five measurements: Identity Diversity with unlabeled accounts, Identity Diversity without unlabeled accounts, Information Diversity, Structural Integration, and Connectivity Inequality, at scale level 1.}
\label{fig:residual_ro_l1}
\end{figure}

\begin{figure}[H]
\centering
\includegraphics[width=1.0\linewidth]{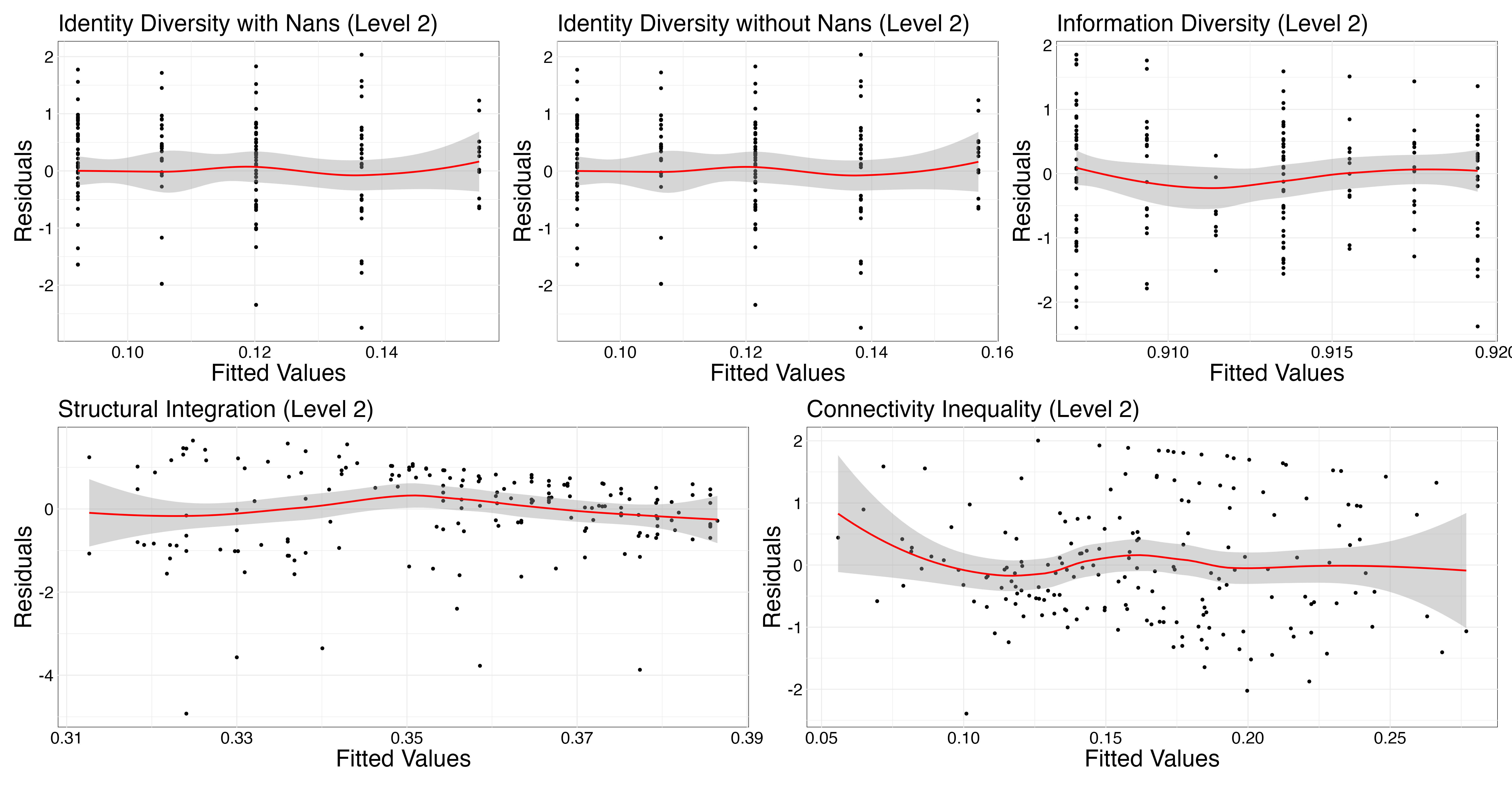}
\caption{Residual check of Beta regression models after removing outliers of five measurements: Identity Diversity with unlabeled accounts, Identity Diversity without unlabeled accounts, Information Diversity, Structural Integration, and Connectivity Inequality, at scale level 2.}
\label{fig:residual_ro_l2}
\end{figure}

\begin{figure}[H]
\centering
\includegraphics[width=1.0\linewidth]{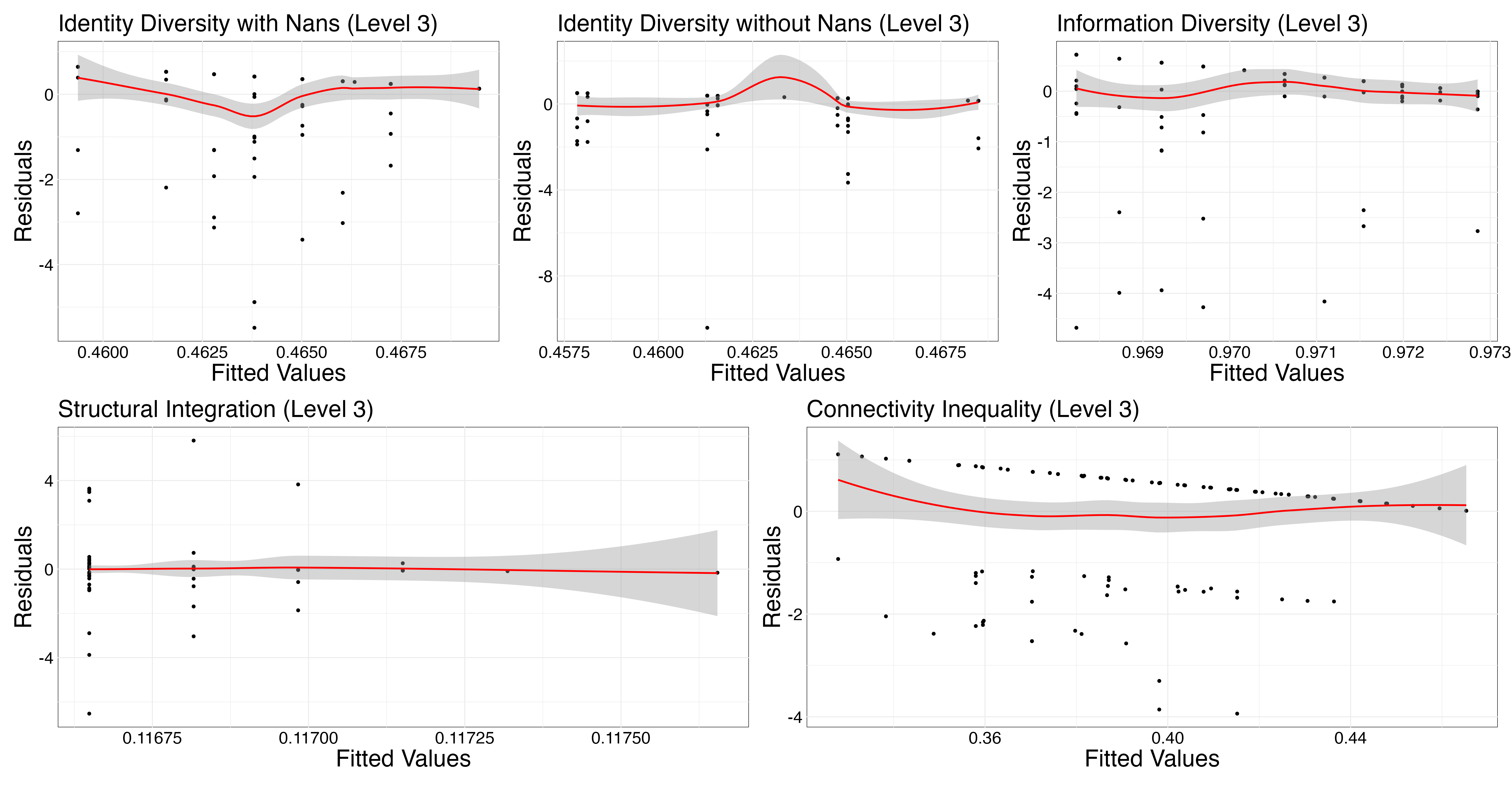}
\caption{Residual check of Beta regression models after removing outliers of five measurements: Identity Diversity with unlabeled accounts, Identity Diversity without unlabeled accounts, Information Diversity, Structural Integration, and Connectivity Inequality, at scale level 3.}
\label{fig:residual_ro_l3}
\end{figure}

\begin{figure}[H]
\centering
\includegraphics[width=1.0\linewidth]{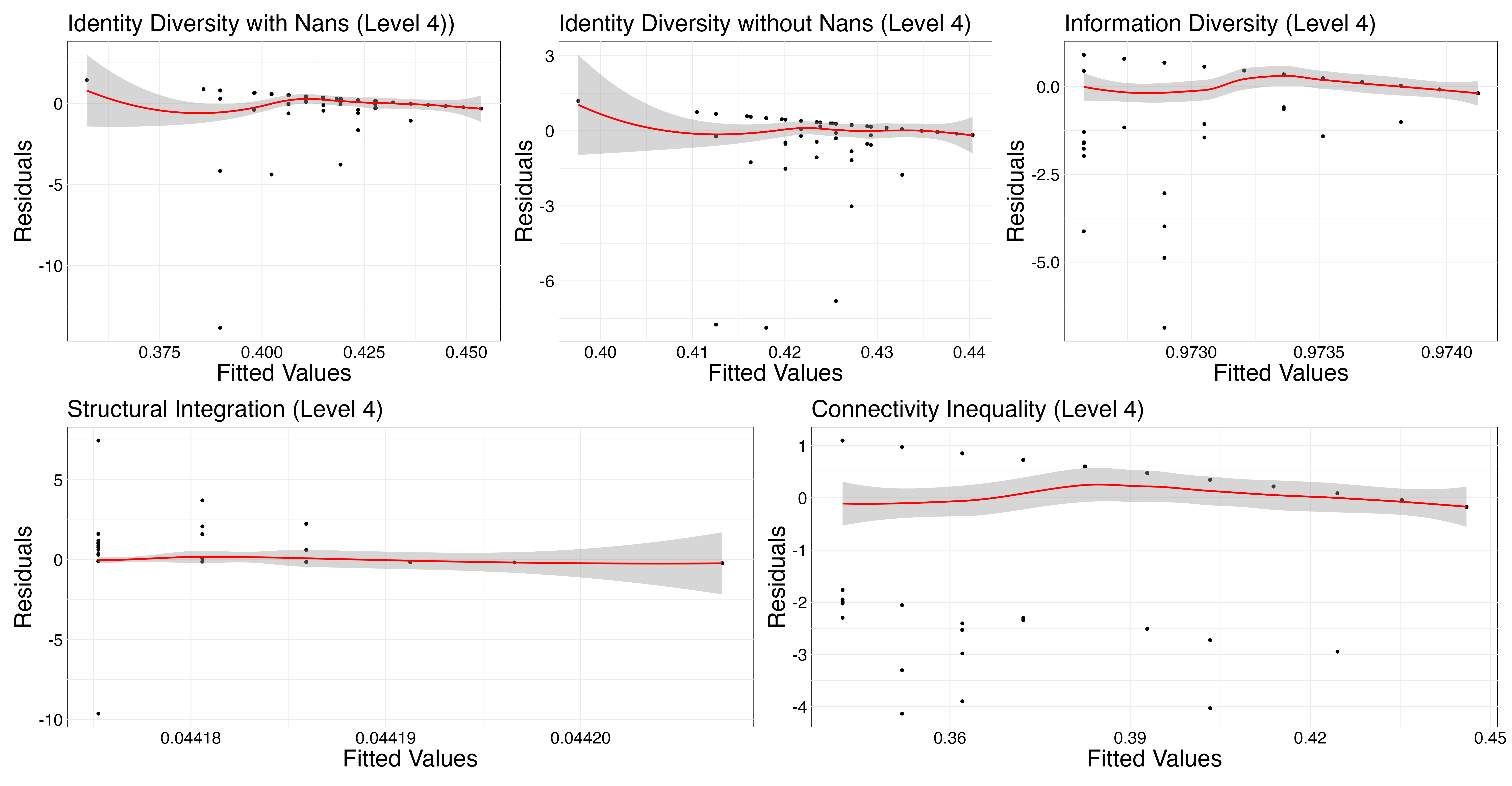}
\caption{Residual check of Beta regression models after removing outliers of five measurements: Identity Diversity with unlabeled accounts, Identity Diversity without unlabeled accounts, Information Diversity, Structural Integration, and Connectivity Inequality, at scale level 4.}
\label{fig:residual_ro_l4}
\end{figure}

\begin{figure}[H]
\centering
\includegraphics[width=0.8\linewidth]{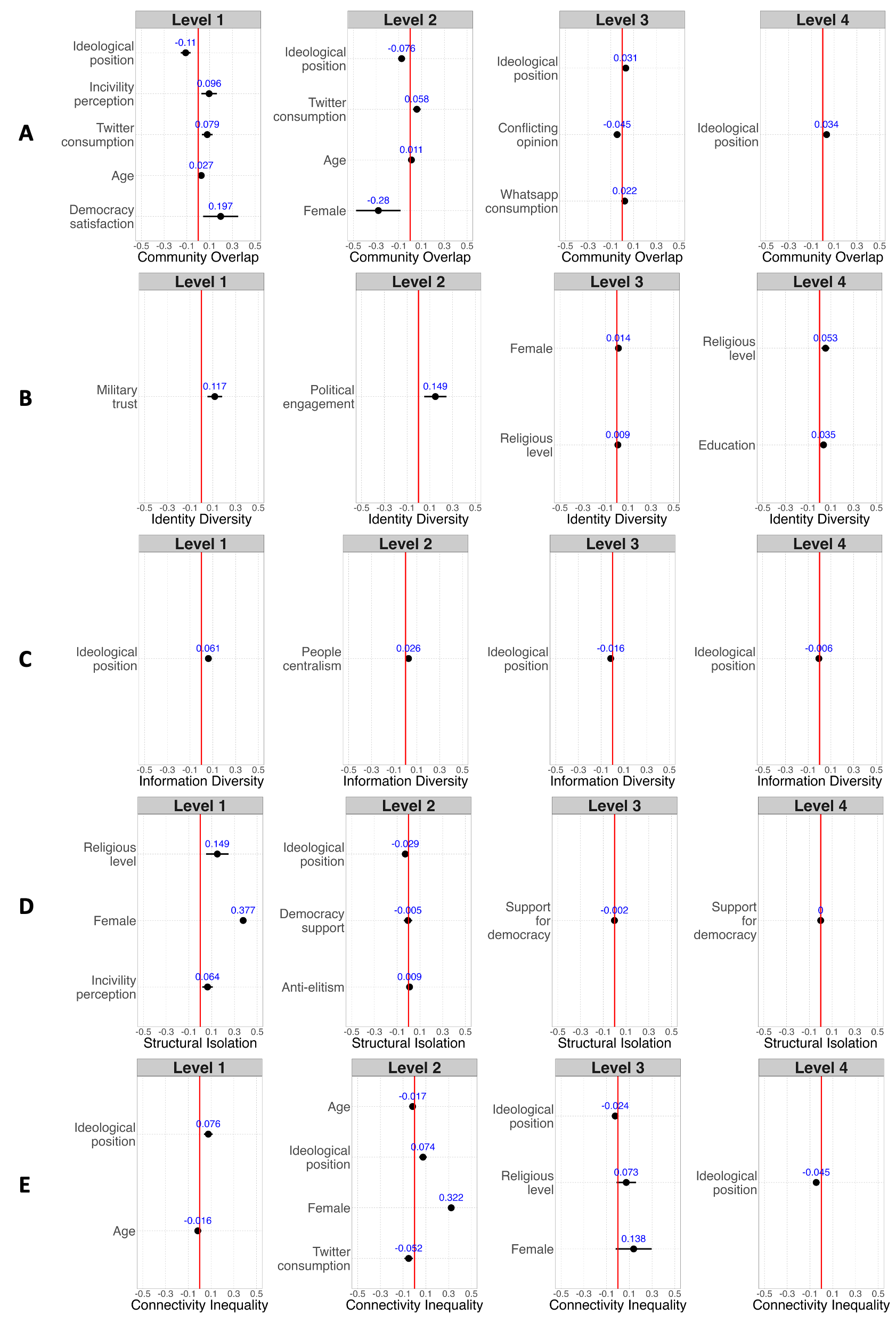}
\caption{Regression plot of sensitivity check. Regression analysis is conducted for the five indices - Community Overlap, Identity Diversity, Information Diversity, Structural Integration, and Connectivity Inequality (only the result of the probability assignment approach for Identity Diversity Index is shown) and at four scale levels, after removing outliers in residual plots. The regression model of Structural Integration Index at scale level 4 does not converge when outliers are removed.}
\label{fig:sensitivity check}
\end{figure}

The residual distributions for the results of Beta Regression are examined at the four scale levels (see \sifigref{fig:residual_l1}, \sifigref{fig:residual_l2}, \sifigref{fig:residual_l3}, \sifigref{fig:residual_l4}). Additionally, we conduct a sensitivity check by manually removing outliers that cause the simulated line to deviate from 0 and then rerun the regression. The residual distributions after removing outliers are shown in \sifigref{fig:residual_ro_l1}, \sifigref{fig:residual_ro_l2}, \sifigref{fig:residual_ro_l3}, \sifigref{fig:residual_ro_l4}. The regression plot after removing outliers is displayed in \sifigref{fig:sensitivity check}. Even though there are small changes in the coefficients, the signs of all the coefficients remain unchanged (except for the regression model of Structural Integration at scale level 4, which does not converge when outliers are removed). This indicates that the results are not biased by outliers.

\newpage

\newpage

\begin{figure}[H]
\centering
\includegraphics[width=1.0\linewidth]{SI_Fig.1_Sample_validation_1.png}
    \caption{Comparison of distributions on Gender, Age, and Area between survey participants and national population. Pearson's chi-squared test is applied for categorical variables, including Gender and Area. Mann–Whitney U test is applied for discrete variables, including Age. The national population statistics are provided by the survey company NetQuest as soft quotas.}
\label{fig:sample validation_1}
\end{figure}

\newpage

\begin{figure}[H]
\centering
\includegraphics[width=0.8\linewidth]{SI_Fig.2_Sample_validation_2.png}
\caption{Comparison of distributions on Age, Gender, Ethnic, Religion, Income, and Education between survey respondents (N = 1,018) and its sub-samples who follow political influencers (N = 204). Pearson's chi-squared test is applied for categorical variables, including Gender, Ethnic, and Religion. Mann–Whitney U test is applied for discrete variables, including Age, Income, and Education.}
\label{fig:sample validation_2}
\end{figure}

\newpage

\begin{figure}[H]
\centering
\includegraphics[width=1.0\linewidth]{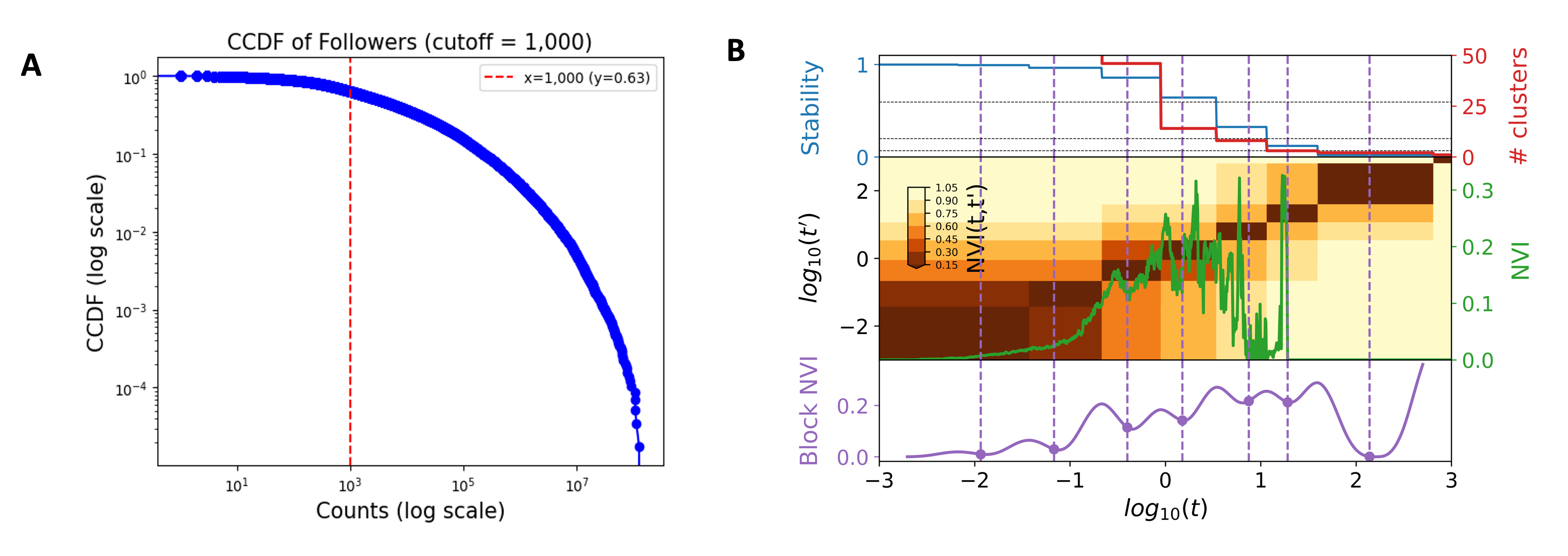}
\caption{(A) displays the complementary cumulative distribution function (CCDF) plots of the number of followers of Twitter accounts followed by survey respondents, with cutoffs at 1,000. (B) shows the corresponding multi-scale community detection outcomes at different scales.}
\label{fig:community detection}
\end{figure}

\newpage

\newpage

\newpage

\newpage

\newpage

\newpage

\newpage

\newpage

\newpage

\newpage

\newpage

\newpage

\newpage

\newpage

\newpage

\begin{figure}[H]
\centering
\includegraphics[width=1.0\linewidth]{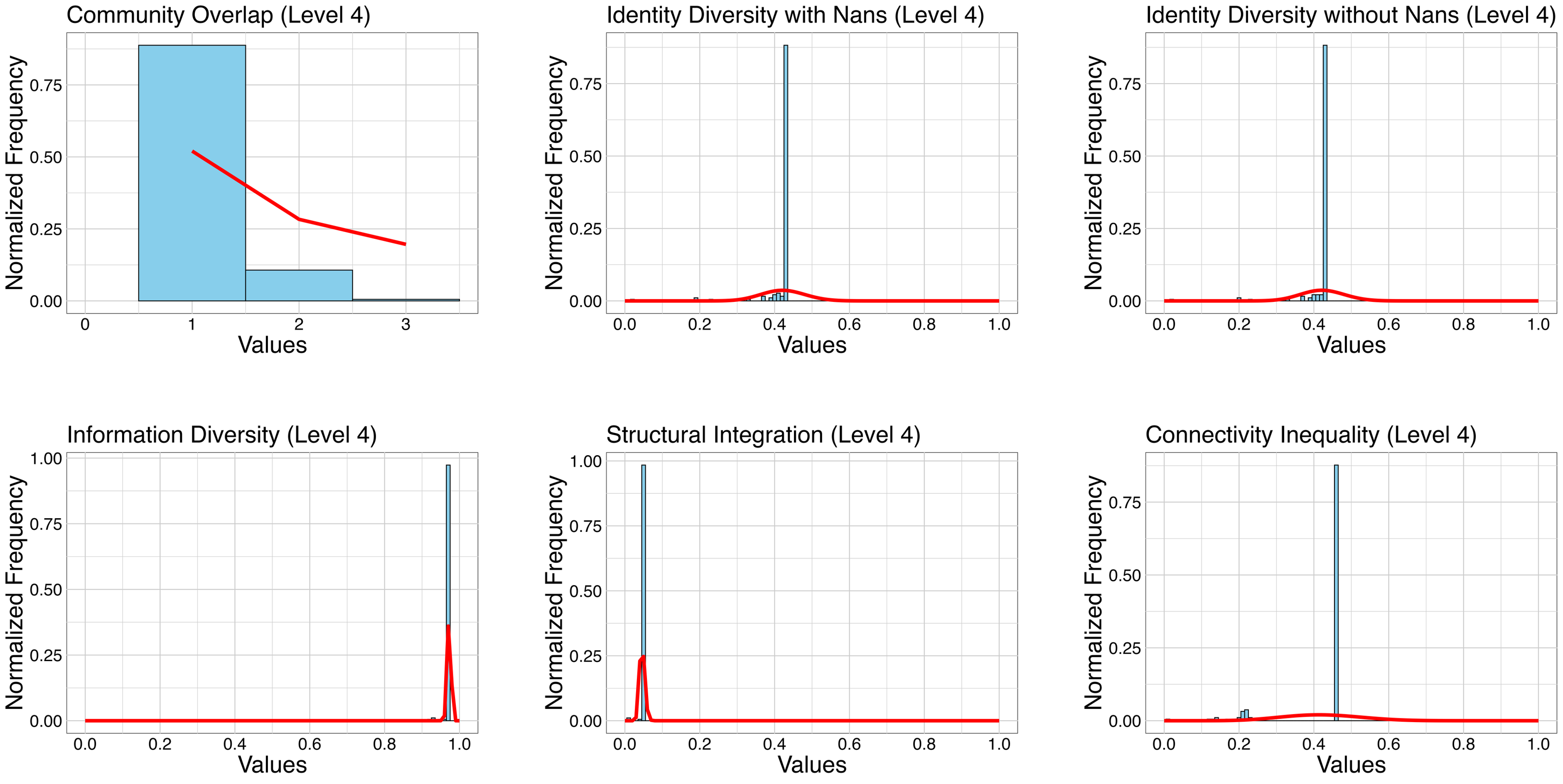}
\caption{Comparison of data distributions between original values and estimated values of the six measurements: Community Overlap, Identity Diversity with unlabeled accounts, Identity Diversity without unlabeled accounts, Information Diversity, Structural Integration, and Connectivity Inequality, at scale level 4.}
\label{fig:distribution_l4}
\end{figure}

\newpage

\newpage

\newpage

\newpage

\newpage

\newpage

\newpage

\newpage

\newpage

\newpage

\begin{table}[H]
\centering
\caption{Soft quota of the national population provided by NetQuest.}
\label{tab:softquota}
\begin{tabular}{llc}
\toprule
\textbf{Variable} & \textbf{Category} & \textbf{Proportion (\%)} \\
\midrule
\multirow{2}{*}{Gender} 
 & Hombre (Male) & 49\% \\
 & Mujer (Female) & 51\% \\
\midrule
\multirow{5}{*}{Age} 
 & 16--24 & 10\% \\
 & 25--34 & 37\% \\
 & 35--44 & 26\% \\
 & 45--54 & 16\% \\
 & $\geq$55 & 11\% \\
\midrule
\multirow{8}{*}{Geographic Area} 
 & Área I & 21\% \\
 & Área II & 14\% \\
 & Área III & 8\% \\
 & Área IV & 15\% \\
 & Área V & 13\% \\
 & Área VI & 17\% \\
 & Área VII & 9\% \\
 & Cap Norte & 3\% \\
\bottomrule
\end{tabular}
\end{table}

\newpage

\begin{table}[H]
\centering
\caption{Demographic distribution of survey participants (N = 1,018).}
\label{tab:survey_demographics}
\begin{tabular}{llcc}
\toprule
\textbf{Variable} & \textbf{Category} & \textbf{Count} & \textbf{Proportion (\%)} \\
\midrule
\multirow{5}{*}{Age} 
 & 16--24 & 102 & 10.0\% \\
 & 25--34 & 370 & 36.3\% \\
 & 35--44 & 269 & 26.4\% \\
 & 45--54 & 164 & 16.1\% \\
 & 55+    & 113 & 11.1\% \\
\midrule
\multirow{2}{*}{Gender} 
 & Homem (Male) & 498 & 48.9\% \\
 & Mulher (Female) & 520 & 51.1\% \\
\midrule
\multirow{6}{*}{Ethnic} 
 & Branca (White) & 546 & 53.7\% \\
 & Preta (Black) & 82 & 8.1\% \\
 & Parda (Mixed) & 355 & 34.9\% \\
 & Amarela (Asian) & 28 & 2.8\% \\
 & Indígena (Indigenous) & 4 & 0.4\% \\
 & Outro (Other) & 1 & 0.1\% \\
\midrule
\multirow{11}{*}{Religion} 
 & Católico (Catholic) & 351 & 36.4\% \\
 & P. Ev. (Prot. Evang.) & 195 & 20.2\% \\
 & P. N-Ev. (Prot. Non-Evang.) & 16 & 1.7\% \\
 & N-Crist. (Non-Christian) & 3 & 0.3\% \\
 & T.J. (Jeova’s Witness) & 12 & 1.2\% \\
 & Afro-Br. (Afro-Brazilian) & 36 & 3.7\% \\
 & Kardecista (Kardecist) & 64 & 6.6\% \\
 & Judeu (Jewish) & 6 & 0.6\% \\
 & Outras (Others) & 30 & 3.1\% \\
 & Agnóstico (Agnostic) & 197 & 20.5\% \\
 & Ateu (Atheist) & 53 & 5.5\% \\
\midrule
\multirow{8}{*}{Income} 
 & $\leq$\$1.2K & 74 & 7.7\% \\
 & \$1.2--2.4K & 184 & 19.2\% \\
 & \$2.4--3.6K & 197 & 20.6\% \\
 & \$3.6--6K & 239 & 25.0\% \\
 & \$6K--12K & 184 & 19.2\% \\
 & \$12--24K & 55 & 5.7\% \\
 & \$24--36K & 18 & 1.9\% \\
 & $>$\$36K & 6 & 0.6\% \\
\midrule
\multirow{10}{*}{Education} 
 & Pre-K & 1 & 0.1\% \\
 & Elem-5 & 5 & 0.5\% \\
 & Elem-9 & 15 & 1.5\% \\
 & HS-1 & 12 & 1.2\% \\
 & HS-2 & 9 & 0.9\% \\
 & HS-3 & 249 & 24.5\% \\
 & Inc.HE & 180 & 17.7\% \\
 & Comp.HE & 345 & 33.9\% \\
 & PG/MSc & 197 & 19.4\% \\
 & Ph.D. & 5 & 0.5\% \\
\bottomrule
\end{tabular}
\end{table}

\newpage

\begin{table}[H]
\centering
\caption{Identification of Political Influencers}
\label{table:political_influencers}
\begin{tabular}{p{2cm} p{3cm} p{10cm}}
\toprule
\textbf{Criteria} & \textbf{Category} & \textbf{Keywords} \\
\midrule
\textbf{Political Keywords} & General & política, político, political, politics, democracia, democracy \\
                           & Election & bolsonaro, bolsonarista, lula, lulista, candidato, partido, presidente \\
                           & Public sector & federal, conselho nacional de, ministro, senador, deputado, governador, prefeito, vereador, secretário \\
                           & Ideology & conservador, conservative, liberal, liberalismo, libertairia, esquerdopata, esquerda, direita, direitista, comunista, comunismo, nacionalista, patriota, globalista, feminista, armamentista, fascista, racist, colonialista, socialista, ativista, progressista \\
                           & Topic (culture) & aborto, mulher, preta, lgbt, gay, bissexualismo, homophobic, catílico, jesus, deus, ambiente, clima, justiça, imigrante, foreigner \\
                           & Topic (economic) & economia, bem-estar, pobre, desigualdade \\
\midrule
\textbf{Political Accounts} & Political party & Partido da Mulher Brasileira, Partido dos Trabalhadores, Partido da Social Democracia Brasileira, Progressistas, Partido Democrático Trabalhista, Partido Trabalhista Brasileiro, União Brasil, Partido Liberal, Partido Socialista Brasileiro, Republicanos, Cidadania, Partido Comunista do Brasil, Partido Social Cristão, Podemos, Partido Social Democrático, Partido Verde, Patriota, Solidariedade, Partido da Mobilização Nacional, Avante, Partido Trabalhista Cristão, Partido Socialismo e Liberdade, Democracia Cristã, Partido Renovador, Trabalhista Brasileiro, Partido Republicano da Ordem Social, Partido da Mulher Brasileira, Partido Novo, Rede Sustentabilidade, Partido Socialista dos Trabalhadores Unificado, Partido Comunista Brasileiro, Partido da Causa Operária, Unidade Popular, Avante, Agir, MDB Nacional \\
                           & Politician & Aldo Rebelo, Soraya Thronicke, Jair Bolsonaro, Luiz Inácio Lula da Silva, Ciro Gomes, Simone Tebet, André Janones, Luiz Felipe D'Avila, José Maria Eymael, Leonardo Péricles, Sofia Manzano, Vera Lúcia Salgado, Luciano Bivar, Pablo Marçal, Wilson Witzel, Janaina Paschoal, José Reguff, Ibaneis Rocha, Renan Filho, Renato Casagrande, Michel Temer, Jorge Kajuru, Padre Kelmon \\
\midrule
\textbf{Media Keywords} & Individual aggregator & jornalista, journalist, correspondent, repórter, comandante, commentator, comentarista, influencer, news, semanal \\
\midrule
\textbf{Media Accounts} & News outlet & Globo News online (incl. G1), UOL online, Record News online (incl. R7.com), O Globo online, Band News online, Folha de S. Paulo online, O Estado de S. Paulo online, BBC News online, Rede TV News online, notícias, Jornal Extra online, TV SBT (incl. SBT Brasil), TV Band News, CNN, TV Brasil (public broadcaster) \\
\bottomrule
\end{tabular}
\end{table}

\newpage

\begin{table}[H]
\centering
\caption{Qualitative interpretation of sub-dimensions of independent variables reflected by PCA projection.}
\label{table:survey attributes}
\begin{tabular}{p{5cm} p{4cm} p{7cm}}
\toprule

\textbf{Defined Groups} & \textbf{Sub-dimensions} & \textbf{Interpretation} \\
\midrule
\multirow{2}{*}{Demographics} & Demographics & Age, Religion, Religious level, Gender, Ethnic, Education, Living, Income \\
\midrule
\multirow{6}{*}{News Consumption} & Information from news & Frequency of news consumption from online sources such as Television, Social Media, and other digital sources \\ & Information from weak ties & Whether they receive political information from people they don't know very well (e.g., colleagues, acquaintances, neighbors) \\
& Information from strong ties & Whether they receive political information from family and friends  \\
\midrule
\multirow{2}{*}{Political Communication} & Discussion conflict & To what extent they enjoy political discussions and conflicts \\ & Opinion Alignment & Frequency of encountering different political opinions \\
\midrule
\multirow{2}{*}{Political Identification} & Ideological position & Ideological spectrum from Left to Right \\ & Partisan degree & To what extent they feel aligned with a political party \\
\midrule
\multirow{2}{*}{Political Engagement} & Political Engagement & Frequency of engaging in political activities, such as electoral campaign, political protest, civic activities, and online discussions \\
\midrule
\multirow{3}{*}{Incivility Perception} 
& Online incivility perception & To what extent they perceive uncivil discourses on social media \\
& Online-Offline incivility comparison & To what extent they perceive uncivil discourses on social media compared to offline experience \\
\midrule
\multirow{1}{*}{Disinformation Perception} 
& Online disinformation perception & To what extent they perceive disinformation on social media \\
\midrule
\multirow{3}{*}{Authority Trust} 
& Political Trust & To what extend they trust the parliament, politicians, political parties, media, and legal system \\ & Military Trust & To what extend they trust the police, military, and government \\
\midrule
\multirow{2}{*}{Populism} & Anti-elitism & To what extent they agree that political and economic elites are harming people's interests\\ & People centralism & To what extent they agree that the majority of the people should be prioritized \\
\midrule
\multirow{1}{*}{Attitudes to Democracy} & Attitudes to democracy & Support for Democracy, Satisfaction with Democracy \\
\bottomrule
\end{tabular}
\end{table}

\newpage

\begin{longtable}[H]{>{\raggedright\arraybackslash}p{4cm} >{\raggedright\arraybackslash}p{8cm} >{\raggedright\arraybackslash}p{4cm}}

    \caption{Results of variable selection}
    \label{table:variable clusters} \\
    \toprule
    \textbf{Sub-dimensions} & \textbf{Variables} & \textbf{Selected Variable} \\
    \midrule
    \endfirsthead
    
    \caption[]{(continued)} \\
    \toprule
    \textbf{Sub-dimensions} & \textbf{Variables} & \textbf{Selected Variable} \\
    \midrule
    \endhead
    
    \midrule \multicolumn{3}{r}{Continued on next page} \\
    \endfoot
    
    \bottomrule
    \endlastfoot
    
    Demographics & Age, Religion, Religious level, Gender, Ethnic, Education, Living, Income & Age, Religion, Religious level, Gender, Ethnic, Education, Living, Income \\
    \midrule
    Information from News & News\_Television news, News\_Online news sources, News\_News via social media, Campaign News\_Television news, Campaign News\_Online news sources, Campaign News\_News via social media, Social Media\_Twitter, Read/Watch\_Twitter & Read/Watch\_Twitter\\
    Information from Weak-ties & News\_National newspapers, News\_Regional newspapers, News\_Radio news, Social Media\_Facebook, 'Social Media\_YouTube, Social Media\_Whatsapp, Social Media\_Telegram, Campaign News\_National newspapers, Campaign News\_Regional newspapers, Campaign News\_Radio news, Read/Watch\_Youtube, Read/Watch\_Facebook, Read/Watch\_Whatsapp, Read/Watch\_Telegram, Share/Liked\_Twitter, Share/Liked\_Youtube, Share/Liked\_Facebook, Share/Liked\_Whatsapp, Share/Liked\_Telegram, Comment/Post\_Twitter, Comment/Post\_Youtube, Comment/Post\_Facebook, Comment/Post\_Whatsapp, Comment/Post\_Telegram, Mess Apps Information\_2, Mess Apps Information\_3, Mess Apps Discussion\_2, Mess Apps Discussion\_3, Mess App Groups & Read/Watch\_Whatsapp\\
    Information from Strong-ties & Mess Apps Information\_1, Mess Apps Information\_97, Mess Apps Discussion\_1, Mess Apps Discussion\_97 & Mess Apps Information\_1\\
    \midrule
    Discussion Conflict & Fatigue\_1, Fatigue\_2, Fatigue\_3, Conflict Orientation\_1, Conflict Orientation\_2, Conflict Orientation\_3, Conflict Orientation\_4, Conflict Orientation\_5, Discussion\_1, Discussion\_2, Discussion\_3, Discussion\_4 & Fatigue\_1\\
    Opinion Alignment & Conflicting Opinion\_1, Conflicting Opinion\_2, Conflicting Opinion\_3, Conflicting Opinion\_4 & Conflicting Opinion\_1\\
    \midrule
    Ideological Position & Ideological Position, Politician Likeability\_5, Politician Likeability\_2, Politician Likeability\_11, Party Likeability\_2, Party Likeability\_5 & Ideological Position \\
    Partisan Degree & Party Alignment, Alignment Measurement, Politician Likeability\_1, Politician Likeability\_3, Politician Likeability\_4, Politician Likeability\_6, Politician Likeability\_7, Politician Likeability\_8, Politician Likeability\_9, Politician Likeability\_10, Politician Likeability\_12, Party Likeability\_1, Party Likeability\_3, Party Likeability\_4, Party Likeability\_6, Party Likeability\_7, Party Likeability\_8, Party Likeability\_9, Party Likeability\_10, Party Likeability\_11, Party Likeability\_12 & Alignment Measurement\\
    \midrule
    Political Engagement & Participation\_Institutional/Electoral campaign\_1, Participation\_Institutional/Electoral campaign\_2, Participation\_Institutional/Electoral campaign\_3, Participation\_Institutional/Electoral campaign\_4, Participation\_Institutional/Electoral campaign\_5, Participation\_Institutional/Electoral campaign\_6, Participation\_Institutional/Electoral campaign\_7, Participation\_Protest\_1, Participation\_Protest\_2, Participation\_Protest\_3, Participation\_Protest\_4, Participation\_Civic engagement\_1, Participation\_Civic engagement\_2, Participation\_Online participation\_1, Participation\_Online participation\_2, Participation\_Online participation\_3, Participation\_Online participation\_4, Participation\_Online participation\_5, Political Interest, Politician/Party Accounts Following\_Twitter, Politician/Party Accounts Following\_Facebook, Distance of Following Accounts & Participation\_Protest\_1\\
    \midrule
    Online Incivility Perception & Incivility Perception\_Impoliteness\_Twitter, Incivility Perception\_Impoliteness\_Youtube, Incivility Perception\_Impoliteness\_Facebook, Incivility Perception\_Impoliteness\_Whatsapp, Incivility Perception\_Impoliteness\_Telegram, Incivility Perception\_Physical harm/violence\_Twitter, Incivility Perception\_Physical harm/violence\_Youtube, Incivility Perception\_Physical harm/violence\_Facebook, Incivility Perception\_Physical harm/violence\_Whatsapp, Incivility Perception\_Physical harm/violence\_Telegram, Incivility Perception\_Negativity\_Twitter, Incivility Perception\_Negativity\_Youtube, Incivility Perception\_Negativity\_Facebook, Incivility Perception\_Negativity\_Whatsapp, Incivility Perception\_Negativity\_Telegram, Incivility Perception\_Personal attack\_Twitter, Incivility Perception\_Personal attack\_Youtube, Incivility Perception\_Personal attack\_Facebook, Incivility Perception\_Personal attack\_Whatsapp, Incivility Perception\_Personal attack\_Telegram, Incivility Perception\_Stereotype/Hate speech/Discrimination\_Twitter, Incivility Perception\_Stereotype/Hate speech/Discrimination\_Youtube, Incivility Perception\_Stereotype/Hate speech/Discrimination\_Facebook, Incivility Perception\_Stereotype/Hate speech/Discrimination\_Whatsapp, Incivility Perception\_Stereotype/Hate speech/Discrimination\_Telegram, Incivility Perception\_Threat to democratic freedoms\_Twitter, Incivility Perception\_Threat to democratic freedoms\_Youtube, Incivility Perception\_Threat to democratic freedoms\_Facebook, Incivility Perception\_Threat to democratic freedoms\_Whatsapp, Incivility Perception\_Threat to democratic freedoms\_Telegram, Experience\_1, Experience\_2, Experience\_3, Experience\_4 & Incivility Perception\_Impoliteness\_Twitter\\
    Online-Offline Incivility Comparison & Online-offline Incivility\_1, Online-offline Incivility\_2, Online-offline Incivility\_3, Online-offline Incivility\_4, Online-offline Incivility\_5, Online-offline Incivility\_6 & Online-offline Incivility\_1 \\
    \midrule
    Online Disinformation Perception & False Information\_Facebook, False Information\_Twitter, False Information\_Youtube, False Information\_Whatsapp, False Information\_Telegram, False Information\_News Media & False Information\_Youtube\\
    \midrule
    Political Trust & Trust\_1, Trust\_2, Trust\_3, Trust\_4, Trust\_5 & Trust\_3 \\
    Military Trust & Trust\_6, Trust\_7, Trust\_8 & Trust\_7 \\
    \midrule
    Anti-elitism & Populism\_1, Populism\_2, Populism\_4, Populism\_6 & Populism\_1\\
    People centralism & Populism\_3, Populism\_5, Populism\_7 & Populism\_7\\
    \midrule
    Attitudes to Democracy & Support for Democracy, Satisfaction with Democracy & Support for Democracy, Satisfaction with Democracy\\
\end{longtable}

\newpage






\bibliographystyle{unsrt}  
\bibliography{reference_SI}

\begin{thebibliography}{10}

\bibitem{khamis2017self}
Susie Khamis, Lawrence Ang, and Raymond Welling.
\newblock Self-branding, ‘micro-celebrity’ and the rise of social media influencers.
\newblock {\em Celebrity Studies}, 8(2):191--208, 2017.

\bibitem{harff2023influencers}
Dennis Harff and Desirée Schmuck.
\newblock Influencers as empowering agents? following political influencers, internal political efficacy and participation among youth.
\newblock {\em Political Communication}, 40(2):147--172, 2023.

\bibitem{soares2018influencers}
Felipe~Bonow Soares, Raquel Recuero, and Gabriela Zago.
\newblock Influencers in polarized political networks on twitter.
\newblock In {\em Proceedings of the 9th international conference on social media and society}, pages 168--177. ACM, July 2018.

\bibitem{dubois2014multiple}
Elizabeth Dubois and Devin Gaffney.
\newblock The multiple facets of influence: Identifying political influentials and opinion leaders on twitter.
\newblock {\em American Behavioral Scientist}, 58(10):1260--1277, 2014.

\bibitem{flamino2023political}
J~Flamino, J~Blackburn, T~Caulfield, G~Stringhini, S~Zannettou, and E~De~Cristofaro.
\newblock Political polarization of news media and influencers on twitter in the 2016 and 2020 us presidential elections.
\newblock {\em Nature Human Behaviour}, pages 1--13, 2023.

\bibitem{arnaudon2023pygenstability}
A.~Arnaudon, D.~J. Schindler, R.~L. Peach, A.~Gosztolai, M.~Hodges, M.~T. Schaub, and M.~Barahona.
\newblock Pygenstability: Multiscale community detection with generalized markov stability.
\newblock 2023.

\bibitem{lambiotte2014random}
R~Lambiotte, J.~C Delvenne, and M~Barahona.
\newblock Random walks, markov processes and the multiscale modular organization of complex networks.
\newblock {\em IEEE Transactions on Network Science and Engineering}, 1(2):76--90, 2014.

\bibitem{arnaudon_algorithm_2024}
Alexis Arnaudon, Dominik~J Schindler, Robert~L Peach, Adam Gosztolai, Maxwell Hodges, Michael~T Schaub, and Mauricio Barahona.
\newblock Algorithm xxx: Pygenstability, a multiscale community detection with generalized markov stability.
\newblock {\em ACM Transactions on Mathematical Software}, page 3651225, March 2024.

\bibitem{smithson2006better}
Michael Smithson and Jay Verkuilen.
\newblock A better lemon squeezer? maximum-likelihood regression with beta-distributed dependent variables.
\newblock {\em Psychological Methods}, 11(1):54, 2006.

\bibitem{cribari2010beta}
Francisco Cribari-Neto and Achim Zeileis.
\newblock Beta regression in r.
\newblock {\em Journal of statistical software}, 34:1--24, 2010.

\end{thebibliography}


\begin{thebibliography}{99}

\bibitem{rainie2012social}
Rainie L, Smith A, Schlozman KL, Brady H, Verba S. Social media and political engagement. Pew Internet \& American Life Project. 2012;19(1):2--13.

\bibitem{bright2018explaining}
Bright J. Explaining the emergence of political fragmentation on social media: The role of ideology and extremism. Journal of Computer-Mediated Communication. 2018;23(1):17--33.

\bibitem{pildes2021age}
Pildes RH. The age of political fragmentation. Journal of Democracy. 2021;32(4):146--159.

\bibitem{terren2021echo}
Terren L, Borge-Bravo R. Echo chambers on social media: A systematic review of the literature. Review of Communication Research. 2021;9:99--118.

\bibitem{cinelli2021echo}
Cinelli M, De Francisci Morales G, Galeazzi A, Quattrociocchi W, Starnini M. The echo chamber effect on social media. Proceedings of the National Academy of Sciences. 2021;118(9):e2023301118.

\bibitem{GonzalezBailon2023}
González-Bailón S, Lazer D, Barberá P, Zhang M, Allcott H, Brown T, Tucker JA. Asymmetric ideological segregation in exposure to political news on Facebook. Science. 2023;381(6656):392--398.

\bibitem{van2005global}
Van Alstyne M, Brynjolfsson E. Global village or cyber-balkans? Modeling and measuring the integration of electronic communities. Management Science. 2005;51(6):851--868.

\bibitem{garcia2023influence}
Garcia D. Influence of Facebook algorithms on political polarization tested. Nature. 2023:39--41.

\bibitem{santos2021link}
Santos FP, Lelkes Y, Levin SA. Link recommendation algorithms and dynamics of polarization in online social networks. Proceedings of the National Academy of Sciences. 2021;118(50):e2102141118.

\bibitem{huszar2022algorithmic}
Huszár F, Ktena SI, O'Brien C, Belli L, Schlaikjer A, Hardt M. Algorithmic amplification of politics on Twitter. Proceedings of the National Academy of Sciences. 2022;119(1):e2025334119.

\bibitem{haroon2023auditing}
Haroon M, Wojcieszak M, Chhabra A, Liu X, Mohapatra P, Shafiq Z. Auditing YouTube's recommendation system for ideologically congenial, extreme, and problematic recommendations. Proceedings of the National Academy of Sciences. 2023;120(50):e2213020120.

\bibitem{guess2023social}
Guess AM, Malhotra N, Pan J, Barberá P, Allcott H, Brown T, Tucker JA. How do social media feed algorithms affect attitudes and behavior in an election campaign? Science. 2023;381(6656):398--404.

\bibitem{allcott2024effects}
Allcott H, Gentzkow M, Mason W, Wilkins A, Barberá P, Brown T, Tucker JA. The effects of Facebook and Instagram on the 2020 election: A deactivation experiment. Proceedings of the National Academy of Sciences. 2024;121(21):e2321584121.

\bibitem{ohara2015echo}
OHara K, Stevens D. Echo chambers and online radicalism: Assessing the Internet's complicity in violent extremism. Policy \& Internet. 2015;7(4):401--422.

\bibitem{messing2014selective}
Messing S, Westwood SJ. Selective exposure in the age of social media: Endorsements trump partisan source affiliation when selecting news online. Communication Research. 2014;41(8):1042--1063.

\bibitem{delvicario2016echo}
Del Vicario M, Vivaldo G, Bessi A, Zollo F, Scala A, Caldarelli G, Quattrociocchi W. Echo chambers: Emotional contagion and group polarization on Facebook. Scientific Reports. 2016;6(1):37825.

\bibitem{bessi2016users}
Bessi A, Zollo F, Del Vicario M, Puliga M, Scala A, Caldarelli G, Quattrociocchi W. Users polarization on Facebook and YouTube. PloS ONE. 2016;11(8):e0159641.

\bibitem{bail2018exposure}
Bail CA, Argyle LP, Brown TW, Bumpus JP, Chen H, Hunzaker MF, Volfovsky A. Exposure to opposing views on social media can increase political polarization. Proceedings of the National Academy of Sciences. 2018;115(37):9216--9221.

\bibitem{yang2020exposure}
Yang T, Majó-Vázquez S, Nielsen RK, González-Bailón S. Exposure to news grows less fragmented with an increase in mobile access. Proceedings of the National Academy of Sciences. 2020;117(46):28678--28683.

\bibitem{levin2021dynamics}
Levin SA, Milner HV, Perrings C. The dynamics of political polarization. Proceedings of the National Academy of Sciences. 2021;118(50):e2116950118.

\bibitem{balietti2021reducing}
Balietti S, Getoor L, Goldstein DG, Watts DJ. Reducing opinion polarization: Effects of exposure to similar people with differing political views. Proceedings of the National Academy of Sciences. 2021;118(52):e2112552118.

\bibitem{tornberg2022how}
Törnberg P. How digital media drive affective polarization through partisan sorting. Proceedings of the National Academy of Sciences. 2022;119(42):e2207159119.

\bibitem{abramowitz2008is}
Abramowitz AI, Saunders KL. Is polarization a myth? The Journal of Politics. 2008;70(2):542--555.

\bibitem{barbera2015birds}
Barberá P. Birds of the same feather tweet together: Bayesian ideal point estimation using Twitter data. Political Analysis. 2015;23(1):76--91.

\bibitem{johnson2020issues}
Johnson BK, Neo RL, Heijnen ME, Smits L, van Veen C. Issues, involvement, and influence: Effects of selective exposure and sharing on polarization and participation. Computers in Human Behavior. 2020;104:106155.

\bibitem{zucco2021fragmentation}
Zucco C, Power TJ. Fragmentation without Cleavages? Endogenous Fractionalization in the Brazilian Party System. Comparative Politics. 2021;53(3):477--500.

\bibitem{cinelli2020selective}
Cinelli M, Morales GDF, Galeazzi A, Quattrociocchi W, Starnini M. Selective exposure shapes the Facebook news diet. PloS ONE. 2020;15(3):e0229129.

\bibitem{flamino2023political}
Flamino J, Galeazzi A, Feldman S, Macy MW, Cross B, Zhou Z, Szymanski BK. Political polarization of news media and influencers on Twitter in the 2016 and 2020 US presidential elections. Nature Human Behaviour. 2023;7(6):904--916.

\bibitem{efstratiou2023non}
Efstratiou A, Blackburn J, Caulfield T, Stringhini G, Zannettou S, De Cristofaro E. Non-Polar Opposites: Analyzing the Relationship Between Echo Chambers and Hostile Intergroup Interactions on Reddit. Proceedings of the International AAAI Conference on Web and Social Media. 2023;17:197--208.

\bibitem{perez2019political}
Pérez-Curiel C, Limón-Naharro P. Political influencers. A study of Donald Trump's personal brand on Twitter and its impact on the media and users. Communication \& Society. 2019:57--75.

\bibitem{riedl2021rise}
Riedl M, Schwemmer C, Ziewiecki S, Ross LM. The rise of political influencers—Perspectives on a trend towards meaningful content. Frontiers in Communication. 2021;6:752656.

\bibitem{garcia2023political}
García-Sánchez E, Benetti PR, Higa GL, Alvarez MC, Gomez-Nieto E. Political discourses, ideologies, and online coalitions in the Brazilian Congress on Twitter during 2019. New Media \& Society. 2023;25(5):1130--1152.

\bibitem{bartels2018partisanship}
Bartels LM. Partisanship in the Trump era. The Journal of Politics. 2018;80(4):1483--1494.

\bibitem{fortunato2007resolution}
Fortunato S, Barthelemy M. Resolution limit in community detection. Proceedings of the National Academy of Sciences. 2007;104(1):36--41.

\bibitem{delvenne2010stability}
Delvenne JC, Yaliraki SN, Barahona M. Stability of graph communities across time scales. Proceedings of the National Academy of Sciences. 2010;107(29):12755--12760.

\bibitem{clauset_hierarchical_2008}
Clauset A, Moore C, Newman MEJ. Hierarchical structure and the prediction of missing links in networks. Nature. 2008;453(7191):98--101.

\bibitem{ahn2010link}
Ahn YY, Bagrow JP, Lehmann S. Link communities reveal multiscale complexity in networks. Nature. 2010;466(7307):761--764.

\bibitem{luo2019multiscale}
Luo W, Zhang D, Ni L, Lu N. Multiscale local community detection in social networks. IEEE Transactions on Knowledge and Data Engineering. 2019;33(3):1102--1112.

\bibitem{reichardt2006statistical}
Reichardt J, Bornholdt S. Statistical mechanics of community detection. Physical Review E. 2006;74(1):016110.

\bibitem{lancichinetti2009detecting}
Lancichinetti A, Fortunato S, Kertész J. Detecting the overlapping and hierarchical community structure in complex networks. New Journal of Physics. 2009;11(3):033015.

\bibitem{Rosvall2011}
Rosvall M, Bergstrom CT. Multilevel Compression of Random Walks on Networks Reveals Hierarchical Organization in Large Integrated Systems. PLoS ONE. 2011;6(4):e18209.

\bibitem{peixoto_hierarchical_2014}
Peixoto TP. Hierarchical Block Structures and High-Resolution Model Selection in Large Networks. Physical Review X. 2014;4(1):011047.

\bibitem{arnaudon_algorithm_2024}
Arnaudon A, Schindler DJ, Peach RL, Gosztolai A, Hodges M, Schaub MT, Barahona M. Algorithm 1044: PyGenStability, a multiscale community detection framework with generalized Markov stability. ACM Transactions on Mathematical Software. 2024;50(2):1--8.

\bibitem{lambiotte2014random}
Lambiotte R, Delvenne JC, Barahona M. Random walks, Markov processes and the multiscale modular organization of complex networks. IEEE Transactions on Network Science and Engineering. 2014;1(2):76--90.

\bibitem{newman2004fast}
Newman ME. Fast algorithm for detecting community structure in networks. Physical Review E. 2004;69(6):066133.

\bibitem{Traag2019}
Traag VA, Waltman L, Van Eck NJ. From Louvain to Leiden: guaranteeing well-connected communities. Scientific Reports. 2019;9(1):1--12.

\bibitem{khamis2017self}
Khamis S, Ang L, Welling R. Self-branding, 'micro-celebrity' and the rise of social media influencers. Celebrity Studies. 2017;8(2):191--208.

\bibitem{harff2023influencers}
Harff D, Schmuck D. Influencers as empowering agents? Following political influencers, internal political efficacy and participation among youth. Political Communication. 2023;40(2):147--172.

\bibitem{bakshy2015exposure}
Bakshy E, Messing S, Adamic LA. Exposure to ideologically diverse news and opinion on Facebook. Science. 2015;348(6239):1130--1132.

\bibitem{tajfel1979integrative}
Tajfel H, Turner JC. An integrative theory of intergroup conflict. In: Austin WG, Worchel S, editors. Intergroup Relations: Essential Readings. Psychology Press; 2001. p. 94--109.

\bibitem{turner1987rediscovering}
Turner JC, Hogg MA, Oakes PJ, Reicher SD, Wetherell MS. Rediscovering the social group: A self-categorization theory. Oxford: Blackwell; 1987.

\bibitem{huddy2001from}
Huddy L. From social to political identity: A critical examination of social identity theory. Political Psychology. 2001;22(1):127--156.

\bibitem{del2017modeling}
Del Vicario M, Scala A, Caldarelli G, Stanley HE, Quattrociocchi W. Modeling confirmation bias and polarization. Scientific Reports. 2017;7(1):40391.

\bibitem{knobloch2020confirmation}
Knobloch-Westerwick S, Mothes C, Polavin N. Confirmation bias, ingroup bias, and negativity bias in selective exposure to political information. Communication Research. 2020;47(1):104--124.

\bibitem{simpson1949measurement}
Simpson EH. Measurement of diversity. Nature. 1949;163(4148):688--688.

\bibitem{shi2000normalized}
Shi J, Malik J. Normalized cuts and image segmentation. IEEE Transactions on Pattern Analysis and Machine Intelligence. 2000;22(8):888--905.

\bibitem{gastwirth1972estimation}
Gastwirth JL. The estimation of the Lorenz curve and Gini index. The Review of Economics and Statistics. 1972:306--316.

\bibitem{cribari2010beta}
Cribari-Neto F, Zeileis A. Beta regression in R. Journal of Statistical Software. 2010;34:1--24.

\bibitem{himelboim2013tweeting}
Himelboim I, Smith M, Shneiderman B. Tweeting apart: Applying network analysis to detect selective exposure clusters in Twitter. Communication Methods and Measures. 2013;7(3-4):195--223.

\bibitem{conover2011political}
Conover M, Ratkiewicz J, Francisco M, Gonçalves B, Menczer F, Flammini A. Political polarization on Twitter. Proceedings of the International AAAI Conference on Web and Social Media. 2011;5(1):89--96.

\bibitem{martin2023multipolar}
Martin-Gutierrez S, Losada JC, Benito RM. Multipolar social systems: Measuring polarization beyond dichotomous contexts. Chaos, Solitons \& Fractals. 2023;169:113244.

\bibitem{peralta2024multidimensional}
Peralta AF, Ramaciotti P, Kertész J, Iñiguez G. Multidimensional political polarization in online social networks. Physical Review Research. 2024;6(1):013170.

\bibitem{phillips2024why}
Phillips SC, Carley KM, Joseph K. Why do people think liberals drink lattes? How social media afforded self-presentation can shape subjective social sorting. arXiv preprint arXiv:2404.02338. 2024.

\bibitem{Manza1999}
Manza J, Brooks C. Social Cleavages and Political Change: Voter Alignments and US Party Coalitions. Oxford: OUP Oxford; 1999.

\bibitem{golder2016far}
Golder M. Far right parties in Europe. Annual Review of Political Science. 2016;19(1):477--497.

\bibitem{layton2021demographic}
Layton ML, Smith AE, Moseley MW, Cohen MJ. Demographic polarization and the rise of the far right: Brazil's 2018 presidential election. Research \& Politics. 2021;8(1):2053168021990204.

\bibitem{yarchi2021political}
Yarchi M, Baden C, Kligler-Vilenchik N. Political polarization on the digital sphere: A cross-platform, over-time analysis of interactional, positional, and affective polarization on social media. In: Dissonant Public Spheres. Routledge; 2024. p. 185--226.

\bibitem{neal2021comparing}
Neal ZP, Domagalski R, Sagan B. Comparing alternatives to the fixed degree sequence model for extracting the backbone of bipartite projections. Scientific Reports. 2021;11(1):23929.

\end{thebibliography}


\end{document}